\documentclass[]{rsos_biblatex}
\usepackage{pgfplots}
\pgfplotsset{compat=1.16}
\usepackage{subcaption}
\usepackage{dsfont}
\usepackage{amssymb}
\usepackage{amsmath}
\usepackage{bm}
\usepackage{csquotes}
\usepackage{cancel}
\usepackage{booktabs}
\usepackage{epigraph}
\usepackage{fancyref}
\usepackage{etoolbox}
\let\bbordermatrix\bordermatrix
\patchcmd{\bbordermatrix}{8.75}{4.75}{}{}
\patchcmd{\bbordermatrix}{\left(}{\left[}{}{}
\patchcmd{\bbordermatrix}{\right)}{\right]}{}{}
\usepackage{hyperref}
\definecolor{myred}{RGB}{234,51,35}
\hypersetup{
         colorlinks,
         urlcolor = myred,
         citecolor = myred,
         linkcolor = myred,
}
\usepackage{soul}
\newcommand{\ve}[1]{\boldsymbol{\mathbf{#1}}}
\usepackage[
    backend=biber,
    style=authoryear-comp,
    sortcites=true,
    sorting=ynt,
    uniquename=mininit,
    uniquelist=false,
    natbib=true,
    firstinits=true,
]{biblatex}
\begin{document}

\title{Astronomia ex machina: a history, primer, and outlook on neural networks in astronomy}

\author{Michael J. Smith and James E. Geach}

\address{Department of Physics, Astronomy \& Mathematics, School of Physics, Engineering and Computer Science, University of Hertfordshire, Hatfield, AL10 9AB}

\subject{Astrophysics, Computer Science}

\keywords{neural networks, astrophysics, machine learning}

\corres{M. J. Smith\\\email{mike@mjjsmith.com}\\
J. E. Geach\\\email{j.geach@herts.ac.uk}}

\begin{abstract}
In this review, we explore the historical development and future prospects of artificial intelligence (AI) and deep learning in astronomy. We trace the evolution of connectionism in astronomy through its three waves, from the early use of multilayer perceptrons, to the rise of convolutional and recurrent neural networks, and finally to the current era of unsupervised and generative deep learning methods. With the exponential growth of astronomical data, deep learning techniques offer an unprecedented opportunity to uncover valuable insights and tackle previously intractable problems. As we enter the anticipated fourth wave of astronomical connectionism, we argue for the adoption of GPT-like foundation models fine-tuned for astronomical applications. Such models could harness the wealth of high-quality, multimodal astronomical data to serve state-of-the-art downstream tasks. To keep pace with advancements driven by Big Tech, we propose a collaborative, open-source approach within the astronomy community to develop and maintain these foundation models, fostering a symbiotic relationship between AI and astronomy that capitalizes on the unique strengths of both fields.
\end{abstract}

\begin{fmtext}

\section{Introduction}

The concept of artificial intelligence (AI) can be traced back at least 350
years to Leibniz's {\it Dissertation on the Art of Combinations}
\citep{ref_leibniz1989}. Inspired by Descartes and Llull, Leibniz posited that,
through the development of a `universal language,' all ideas could be
represented by the combination of a small set of fundamental concepts, and that
{\it new} concepts could be generated in a logical fashion, potentially by some
computing machine.  Leibniz's ambitious vision (`{\it let us calculate}') has
not yet been realised, but the quest to emulate human reasoning, or at least to
build a machine to mimic the computational and data processing capabilities of
the human brain, has persisted to this day.

\end{fmtext}

\maketitle

{
\hypersetup{linkcolor=black}
\tableofcontents
}

It might be fair to say that the roots of AI stretch even as far back
as Llull's medieval philosophy that inspired Leibniz
\citep{ref_fidora2011,ref_gray2016}. However, if we now consider AI to be a
{\it bona fide} scientific discipline, then that discipline clearly emerged in
the post-war years of the twentieth century, following Turing's simple enquiry
`{\it can machines think?}'  \citep{ref_turing1950}. Somewhat philosophical in
nature, Turing's 1950 question succinctly articulates the ambition of AI, but
from a nuts and bolts standpoint it took a further five years from Turing's
query for what one might call the first AI program---the so-called `Logic
Theorist'---to be developed by Allen Newell, Cliff Shaw, and Herbert Simon.
Funded by the Research and Development (RAND) Corporation, the Logic Theorist
was designed, in part, to emulate the role of a human mathematician, in that it
could automate the proof of mathematical theorems. This was a breakthrough in
computer science and the Logic Theorist  was presented at the seminal Dartmouth
Summer Research Project on Artificial Intelligence (DSRPAI) conference in 1956,
now regarded as the true birth of AI as a field. Indeed, it was DSRPAI
organiser John McCarthy who is credited with coining the term `artificial
intelligence' \citep{ref_moor2006}. 

\begin{figure}[htbp]
    \centering
    \input{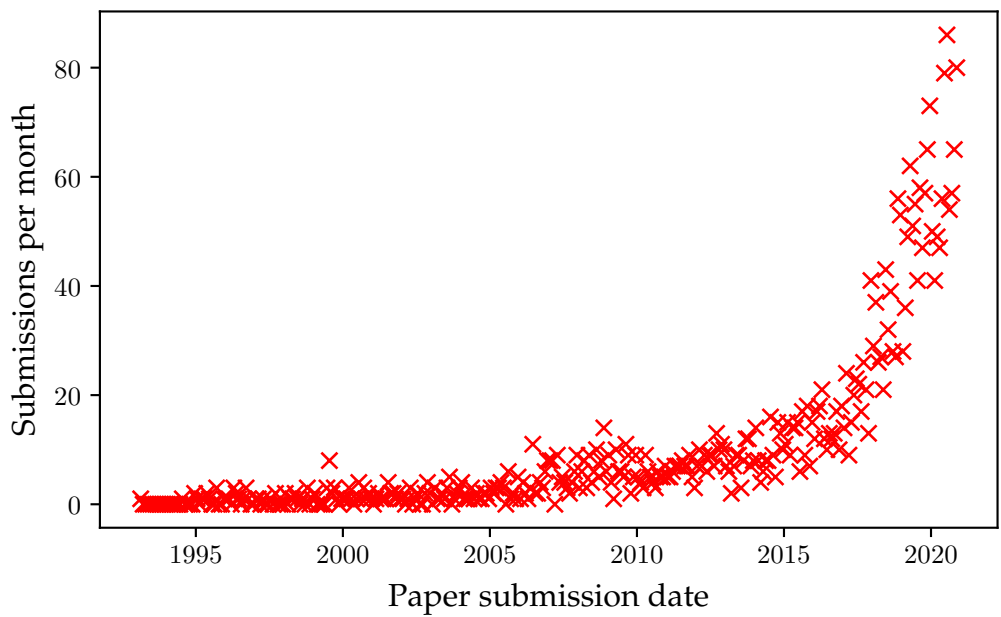}
    \caption[The increasing rate of machine learning papers submitted to
    \emph{astro-ph}]{Here we see the number of arXiv:astro-ph submissions
    per month that have abstracts or titles containing one or more of the
    strings: `machine learning', `ML', `artificial intelligence', `AI', `deep
    learning', or `neural network'. The raw data is in the public domain and is
    available at \url{https://www.kaggle.com/Cornell-University/arxiv}.}
    \label{fig_arxivgrowth}
\end{figure}

Natural intrigue---and clearly a good deal of fear---of the idea of AI has
inspired popular culture no end, from Dick's {\it Do Androids Dream Of Electric
Sheep?} to Crichton's {\it Westworld}, {\it Terminator's `Skynet'} and beyond.
Iain M. Banks's Galactic civilisation known as  `The Culture' imagines a
society run by powerful `Minds' whose intelligence and wisdom far exceeds that
of humans, and where biological beings and machines of equivalent sentience
generally co-exist peacefully,  cooperatively, and equitably. Science fiction
notwithstanding, if these dreams are even possible, we are still years away
from a machine that can genuinely think for itself
\citep{ref_bostrom2014,ref_mitchell2019}.  Nevertheless, the question of how
one mathematically (and algorithmically) models the workings and
inter-relationships of biological neurons---neural networks---and the
subsequent exploration of how they can find utility as tools in the data
analyst's workshop is really what is being referred to when most people use the
term `AI' today\footnote{And the term is regularly misused, not only
erroneously, but often cynically.}.  While we must always be wary of hype and
buzzwordism, it is the {\it application} of neural networks---and the
possibility of tackling hitherto intractable problems---that offers genuine
reason for excitement across many disparate fields of enquiry, including
astronomy.

Astronomers have made use of artificial neural networks (ANNs) for over three
decades. In 1994, Ofer Lahav, an early trailblazer, wryly identified the `{\it
neuro-skeptics}'---those resistant to the use of such techniques in serious
astrophysics research---and argued that ANNs `{\it should be  viewed as a
general statistical framework, rather than as an estoteric approach}'
\citep{ref_lahav1994}.
Unfortunately, this skepticism has persisted. This is despite
the recent upsurge in the use of neural networks (and machine learning in
general) in the field, as illustrated in Fig.~\ref{fig_arxivgrowth}. This
skepticism also stands contrary to achievements within astronomy that would
not be possible without the use of ANNs, such as photometric redshift
estimation \citep[e.g.][]{ref_tagliaferri2003,ref_firth2003}, astronomical object
identification and clustering at scale \citep[e.g.][]{ref_hayat2021}, and entirely
data-driven simulation \citep[e.g.][]{ref_bretonniere2021,ref_smith2022}.
Most of the criticism of machine learning techniques, and deep
learning\footnote{Deep learning referring to the use of a network constructed
of many layers of artificial neurons.} in particular, is levelled at the
perceived `black box' nature of the methodology. In this review we provide a
primer on how deep neural networks are constructed, and the mathematical rules
governing their learning, which we hope will serve as a useful resource for
neuro-skeptics.  Nevertheless, we must recognise that a unified theoretical
picture of how deep neural networks work does not yet exist. This remains a
point of debate even within the deep learning community. For example, Yann
LeCun responding to Ali Rahimi's `Test of Time' award talk at the 31st
Conference on Neural Information Processing Systems (NIPS) remarked:
\blockquote[]{ \small
    \emph{Ali gave an entertaining and well-delivered talk. But I fundamentally disagree with the message. The main message was, in essence, that the current practice in machine learning is akin to `alchemy' (his word). It's insulting, yes. But never mind that: It's wrong! Ali complained about the lack of (theoretical) understanding of many methods that are currently used in ML, particularly in deep learning ... Sticking to a set of methods just because you can do theory about it, while ignoring a set of methods that empirically work better just because you don't (yet) understand them theoretically is akin to looking for your lost car keys under the street light knowing you lost them someplace else.
    Yes, we need better understanding of our methods. But the correct attitude is to attempt to fix the situation, not to insult a whole community for not having succeeded in fixing it yet. This is like criticizing James Watt for not being Carnot or Helmholtz.} \citep{ref_lecun2017}
}
Philosophical concerns aside, LeCun's fundamental point is that deep learning
`works' and therefore we should use it, even if we do not fully understand it.
If one were being uncharitable, we could make similar arguments about the
$\Lambda$CDM paradigm.

\begin{figure}[thbp]
    \centering
    \input{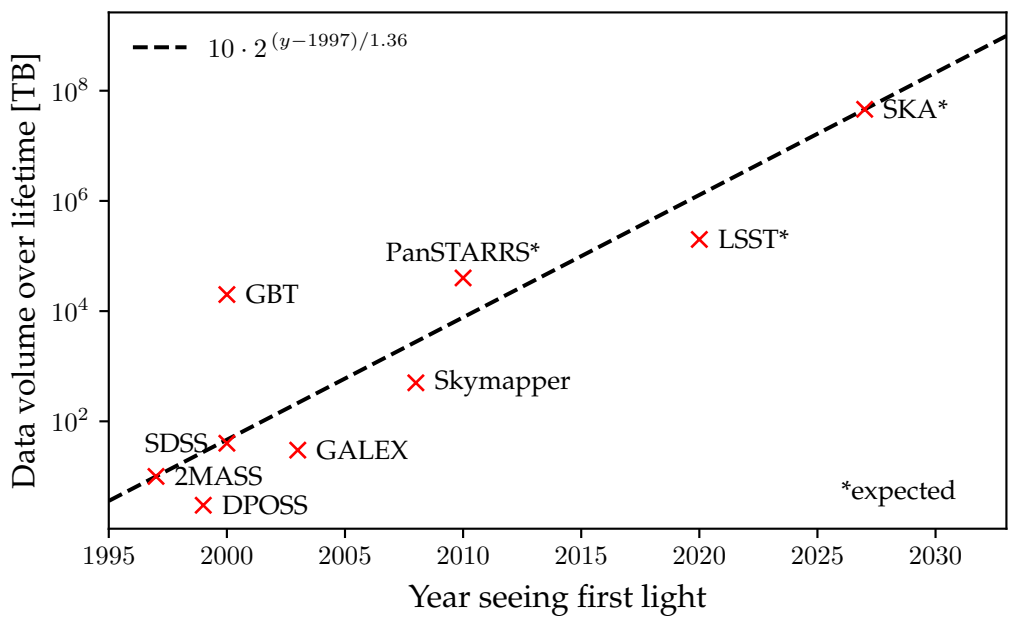}
    \caption[Astronomical survey data volume doubles every 16 months]{The data
    volume output of a selection of astronomical surveys over their lifetimes.
    We can see the astronomical survey data volume doubles every 16 months. Data
    is taken from \citet{ref_zhang2015}.}
    \label{fig_surveys}
\end{figure}

It is clear that in every field that deep learning has infiltrated we have seen
a reduction in the use of specialist knowledge, to be replaced with knowledge
automatically derived from data. We have already seen this process play out in
many `applied deep learning' fields such as computer Go \citep{ref_silver2016},
protein folding \citep{ref_jumper2021}, natural language processing
\citep{ref_brown2020gpt3}, and computer vision \citep{ref_dosovitskiy2020vit}.
We argue that astronomy's data abundance corrals it onto a path no different to
that trodden by other applied deep learning fields. This abundance is not a
passing phase; the total astronomical data volume is already large and will
increase exponentially in the coming years. We illustrate this in
Fig.~\ref{fig_surveys}, where we present a selection of astronomical surveys
and their estimated data volume output over their lifetimes
\citep{ref_zhang2015}. And this is not even considering data associated with
ever larger and more detailed numerical simulations
\citep[e.g.][]{ref_springel2018,ref_vogelsberger2020,ref_angulo2022}.
The current scale of the data volume already poses an issue for astronomy as
many classical methods rely on human supervision and specialist expertise, and
the increasing data volume will make exploring and exploiting these surveys
through traditional human supervised and semi-supervised means an intractable
problem. Of serious concern is the possibility that we will miss---or
substantially delay---interesting and important discoveries simply due to our
inability to accurately and consistently interrogate astronomical data at
scale. Deep learning has shown great promise in automating information
extraction in various data intensive fields, and so is ideally poised as a
solution to the challenge of processing ultra-large scale astronomical data.
But we do not need to stop there. This review's outlook ventures a step
further, and argues that astronomy's wealth of data should be considered a
unique opportunity, and not merely an albatross.

Since astronomical connectionism's\footnote{
    Since its inception, AI research can be broadly categorised
    into two schools: `symbolic' and `connectionist'.  Symbolists see the mind
    as a collection of fully-formed representations, and attempt to mimic human
    reasoning through a logical rule-based processing of these symbols.  This
    approach contrasts with connectionist (or neural network-based) AI, which
    takes a bottom-up approach and simulates cognition by mimicking the way
    neurons in the human brain work.
}
humble beginnings in the late 1980s, there have
been numerous excellent reviews on the application of artificial neural
networks to astronomy
\citep[e.g.][]{ref_miller1993,ref_ball2010,ref_huertas2022}.
We take an alternative approach to previous literature reviews and survey
the field holistically, in an attempt to paint astronomical
connectionism's `Big Picture' with broad strokes.  While we cannot possibly
include all works within astronomical connectionism\footnote{
    We refer the reader to Fig.~\ref{fig_arxivgrowth}!
}, we hope that this review serves as a historical background on
astronomy's `three waves' of increasingly automated connectionism, as well as
presenting a general primer on neural networks that may assist those seeking to
explore this fascinating topic for the first time. 

In \S\ref{sec_artificalneurons} and
\S\ref{sec_mlpinastro} we explore initial work on multilayer perceptrons within
astronomy, where models required manually selected emergent properties as
input. In \S\ref{sec_condeeplearn} and \S\ref{sec_cnnrnn_apps} we explore the
second wave, which coincided with the dissemination of convolutional neural
networks and recurrent neural networks---models where the multilayer
perceptron's manually selected inputs are replaced
with raw data ingestion. In the third wave that is happening now we are seeing the
removal of human supervision altogether with deep learning methods inferring
labels and knowledge directly from the data, and we explore this wave in
\S\ref{sec_gen_modelling}--\S\ref{sec_athirdera}.
Finally, in \S\ref{ch_conclusions}, we look to the future and predict that we will soon
enter a fourth wave of astronomical connectionism. We argue that if astronomy follows the pattern of other applied deep learning fields we will see the removal of
expertly crafted deep learning models, to be replaced with fine-tuned versions
of an all-encompassing `foundation' model. As part of this fourth wave we argue
for a symbiosis between astronomy and connectionism, a symbiosis
predicated on astronomy's relative data wealth and deep
learning's insatiable data appetite. Many ultra-large datasets in machine
learning are proprietary or of poor quality, and so there is an opportunity for
astronomers as a community to develop and provide a high quality multimodal
public dataset. In turn, this dataset could be used to train an astronomical
foundation model to serve state-of-the-art downstream tasks. Due to foundation
models' hunger for data and compute, a single astronomical research group could
not bring about such a model alone. Therefore, we conclude that astronomy as a
discipline has slim chance of keeping up with a research pace set by the Big
Tech goliaths---that is, unless we follow the examples of EleutherAI and
HuggingFace and pool our resources in a grassroots open source fashion.

Before moving on, we must first admit to our readers that we have not been
entirely honest with them. The abstract of this review has not been written by
us. It was generated by prompting OpenAI's generative pretrained transformer 4
(`GPT-4') neural network-based foundation model with this paper's introduction
\citep{ref_openai2023gpt4,ref_bubeck2023}. To be precise, we prompted the GPT-4 engine provided by
`ChatGPT Plus' with all the text in \S1 up until this paragraph in raw {\LaTeX}
format. We then appended the following prompt to the introduction text:
\SetBlockThreshold{1}
\blockquote[]{\small\emph{Write an abstract for the above text that will catch the reader's eye, and make them interested in the paper. Make the abstract 160 words or less, and touch on the value of GPT-like models in astronomy.}}
We did not alter the GPT generated output whatsoever.
We explore these foundation models and their possible astronomical uses in
more detail in \S\ref{ch_conclusions}.

\section{A primer on artificial neurons} \label{sec_artificalneurons}

In \citeyear{ref_mcculloch1943} \citeauthor{ref_mcculloch1943} proposed the
first computational model of a biological neuron \citep[MP neuron;][]{ref_mcculloch1943}.
Their model consisted of a set of binary inputs $x_i \in \{0, 1\}$ and
a single binary output $y \in \{0, 1\}$. Their model also defines a single `inhibitory'
input $\mathcal{I} \in \{0, 1\}$ that blocks output if $\mathcal{I} = 1$. 
If the sum of the inputs exceeds a threshold value $\Theta$, the MP neuron
`fires' and outputs $y = 1$. Mathematically we can write the MP neuron function
as
\begin{equation*}
    \text{MP}(\mathbf{x}) = \begin{cases}
        \mbox{1} & \mbox{if } \sum^n_{i=1} x_i > \Theta \mbox{ and } \mathcal{I} = 0, \\
        \mbox{0} & \mbox{otherwise.}
    \end{cases}
\end{equation*}
The MP neuron is quite a powerful abstraction. Single MP neurons can calculate
simple boolean functions, and more complicated functions can be calculated when
many MP neurons are chained together. However, there is one show-stopping issue:
the MP neuron is missing the capacity to learn.
\citet{ref_perceptron} addressed this by combining the MP neuron with Hebb's
neuronal wiring theory\footnote{Also known by the mantra `cells that fire
together wire together'.} \citep{ref_hebb1949}, and we will explore a related
training formulation in the next subsection.

\subsection{The perceptron} \label{sec_AN}

This subsection aims to provide the reader a foundation and intuition for the
gradient-based learning that dominates contemporary neural network
architectures. Therefore, we diverge from \citeauthor{ref_perceptron}'s
original learning algorithm and instead describe a gradient-based training
algorithm.  The interested reader will find an analysis of
\citeauthor{ref_perceptron}'s original learning algorithms in the `Mathematical
analysis of learning in the perceptron' section of \citet{ref_perceptron}.

\begin{figure}[htbp]
    \centering
    \includegraphics{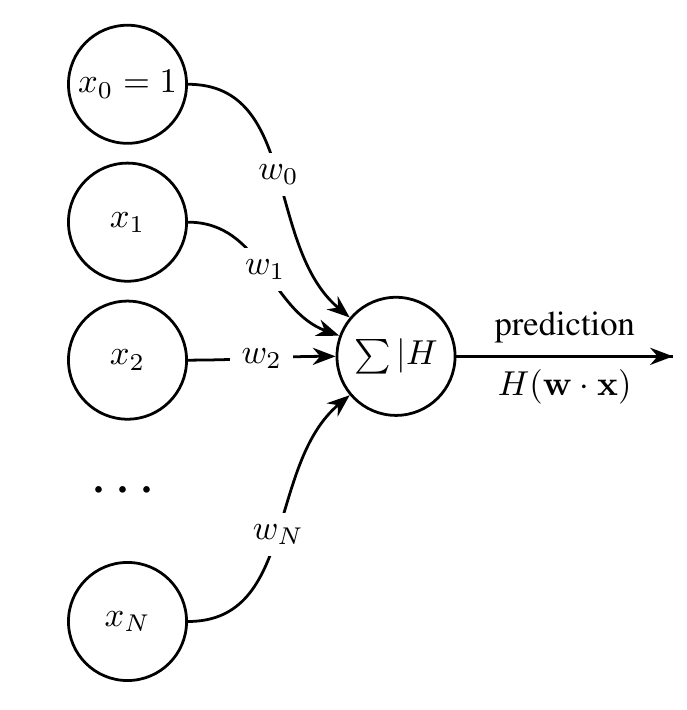}
    \caption[\citeauthor{ref_perceptron}'s perceptron]{A single neuron (or
    perceptron) with a bias $w_0$, inputs $x_1, x_2, \ldots, x_N$, and weights
    $w_1, w_2, \ldots, w_N$.} 
    \label{fig_neuron}
\end{figure}

Like the MP neuron, the perceptron takes a number of numeric inputs ($x_i$).
However, unlike the MP neuron each one of these inputs is multiplied by a
corresponding weight ($w_i$) signifying the importance the perceptron assigns
to a given input. As shown in Fig.~\ref{fig_neuron}, we can then sum this list
of products and pass it into an `activation function'.  Let us use the
Heaviside step function as our activation function:

\vspace{1em}
\noindent%
\begin{minipage}[c]{0.38\textwidth}
    \centering
    \begin{tikzpicture}[declare function={sigma(\x)=1/(1+exp(-\x));
                        sigmap(\x)=sigma(\x)*(1-sigma(\x));}]
       \begin{axis}%
       [
           xmin=-6,
           xmax=6,
           axis x line=bottom,
           ytick={0,.5,1},
           ymax=1.1,
           axis y line=center,
           domain=-6:6,
           scale=0.62,
           xlabel=$\mathbf{w}\cdot\mathbf{x}$,
           legend style={at={(0.3,0.85)},anchor=east},
       ]
           \addplot[red,ultra thick,mark=none] coordinates {(-6,0) (0,0) (0,1) (6,1)};
       \end{axis}
    \end{tikzpicture}
\end{minipage}
\begin{minipage}[c]{0.61\textwidth}
    \centering
    \begin{equation}
        \begin{aligned}
            \text{prediction} = H(\mathbf{w} \cdot \mathbf{x}) = \begin{cases}
                \mbox{0} & \mbox{if } \mathbf{w} \cdot \mathbf{x} < 0, \\
                \mbox{1} & \mbox{if } \mathbf{w} \cdot \mathbf{x} \ge 1,
            \end{cases}
            \label{eqn_step}
        \end{aligned}
    \end{equation}
\end{minipage}
\vspace{1em}

\noindent where $\mathbf{x}$ is a set of inputs, and $\mathbf{w}$
is a set of `weights' that represent the importance of each input.

To concretise how we could train our perceptron we will use an example.
Let us say that we want to automatically label a set of
galaxy images as either `spiral' or `elliptical'. To do this we first need
to compile a training dataset of galaxy images.
This training set would consist of spiral and elliptical galaxies, and each
image would have a ground truth label $y$---say `0' for a spiral galaxy and `1'
for an elliptical. To train our perceptron
we randomly choose one image from the training set, and feed it to the
perceptron, with the numerical value of each pixel corresponding to an input
$\{x_1,\ldots,x_N\}$.  These inputs are multiplied by their corresponding
weight $\{w_1,\ldots,w_N\}$. A bias term $(b = w_0\,x_0, \text{where}\, x_0 = 1)$
is also added to the inputs, which allows the neuron to shift its activation
function linearly.
Since we do not want our perceptron to have any
prior knowledge of the task, we initialise the weights at random. The resulting
products are then summed. Finally, our activation function $H$ transforms
$\mathbf{w} \cdot \mathbf{x}$ and produces a prediction $p$.  We then compare
$p$ to $y$ via a `loss function,' which is a function that measures the
difference between $p$ and $y$. The loss can be any differentiable
function, so for illustration purposes we will define it here as the L1 loss:
$\mathcal{L}(y, p) = \lvert y - p \rvert$.  Now that we can compare to the ground
truth, we need to work out how a change in one of our weights affects
the loss (that is, we want to find $\partial \mathcal{L} / \partial \mathbf{w}$).
We can calculate this change with the chain rule
\begin{equation}
    \frac{\partial \mathcal{L}}{\partial \mathbf{w}} = \frac{\partial \mathcal{L}}{\partial p} 
                                              \frac{\partial p}{\partial \mathbf{w}},
    \label{eqn_bp1}
\end{equation}
and since $p = H(\mathbf{w}\cdot\mathbf{x})$ and $\partial p / \partial
\mathbf{w} = H' \mathbf{x}^T$ we get 
\begin{equation*}
    \frac{\partial \mathcal{L}}{\partial \mathbf{w}} = \frac{\partial \mathcal{L}}{\partial p}\, 
                                                       \odot (H' \mathbf{x}^T)
\end{equation*}
where $\odot$ is the distributive Hadamard product. 
Thus we can update the weights to decrease the loss function:
\begin{equation*}
    \begin{split}
        \textbf{w}_{\text{next}} &= \mathbf{w} - \eta \frac{\partial \mathcal{L}}{\partial \mathbf{w}}\\
                                 &= \mathbf{w} - \eta \frac{\partial \mathcal{L}}{\partial p}\odot(H' \mathbf{x}^T),
    \end{split}
\end{equation*}
where $\eta$ is the learning rate\footnote{
    The eagle-eyed reader may have noticed that since the derivative of the
    Heaviside step function is the Dirac delta function, we will only update
    the perceptron's weights on an incorrect prediction. If we want to also
    learn from positive examples, we need to use a smoothly differentiable
    activation function. This is explored in the next subsection.
}.
If we repeat this process our perceptron will get better and better at
classifying our galaxies!

While we provide the above example for illustrative purposes, we
will need a more powerful algorithm to produce a useful classifier of galaxy
morphology. This need is perhaps most famously discussed in
\emph{Perceptrons: An Introduction to Computational Geometry} 
\citep[e.g. \S13.0;][]{ref_minsky1969}. \citeauthor{ref_minsky1969} show that 
the single layer perceptron is only able to calculate linearly separable
functions, among other limitations. Their book (alongside a consensus that AI
had failed to deliver on its early grandiose promises) delivered a big blow to
the connectionist school of artificial intelligence\footnote{
    See \citet{ref_olazaran1996} and \citet{ref_metz2021} for a closer look at
    the conflicts and personalities that shaped AI.
}.  
In the years following \citet{ref_minsky1969} governmental
and industry funding was pulled from connectionist research laboratories,
ushering in the first `AI winter'\footnote{
    At least, in the Western world. Connectionism continued in earnest in
    the Soviet Union \citep[][]{ref_ivakhnenko1965,ref_ivakhnenko1971}.
}.

\begin{figure}[htbp]
    \centering
    \includegraphics{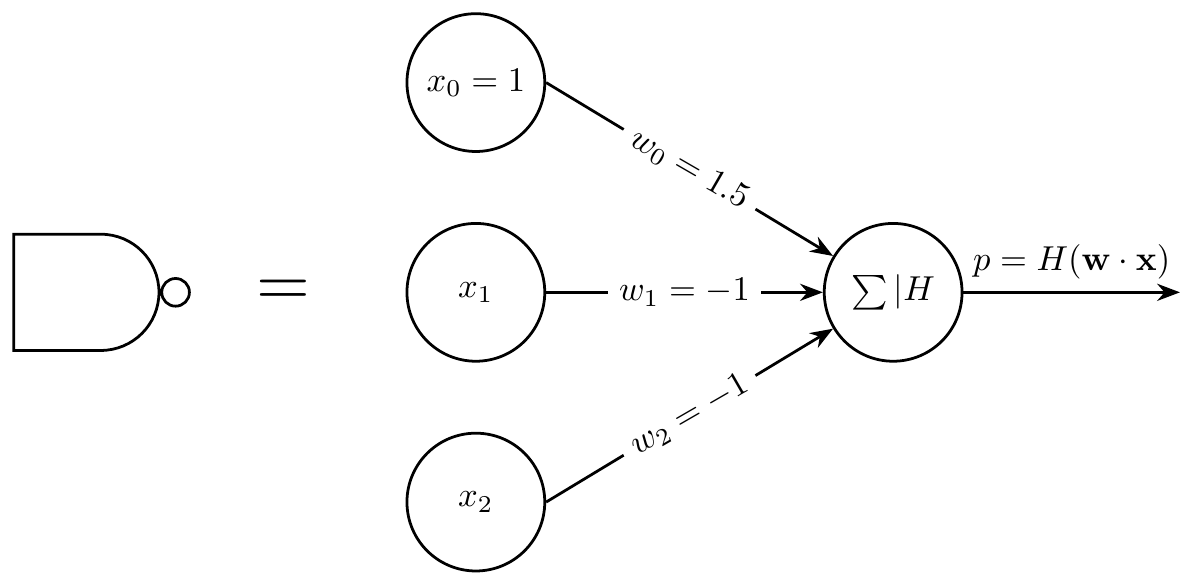}
    \begin{tabular*}{\textwidth}{l@{\extracolsep{\fill}}llll} \\ \toprule
        $x_1$ & $x_2$ & $\neg(x_1 \land x_2)$ & $p = H(\mathbf{w}\cdot\mathbf{x}$) \\
        \midrule
        0 & 0 & 1 & $H(1.5 + (-1)\cdot0 + (-1)\cdot0) = 1$ \\
        0 & 1 & 1 & $H(1.5 + (-1)\cdot0 + (-1)\cdot1) = 1$ \\
        1 & 0 & 1 & $H(1.5 + (-1)\cdot1 + (-1)\cdot0) = 1$ \\
        1 & 1 & 0 & $H(1.5 + (-1)\cdot1 + (-1)\cdot1) = 0$ \\
        \bottomrule
    \end{tabular*}
    \caption[\texttt{NAND} is computed with a single perceptron]{If we define
    $H(\mathbf{w}\cdot\mathbf{x})$ as in Eq.~\ref{eqn_step} we can set a
    perceptron's weights so that it is equivalent to the \texttt{NAND} gate.}
    \label{fig_nandeqneuron}
\end{figure}

Yet, as exemplified in \citet[\S5.2,~theorem~1;][]{ref_rosenblatt1962}
it was known at the time that multilayer perceptrons could calculate
non-linearly separable functions (such as the `exclusive or').  We can prove
intuitively that a set of neurons can calculate \emph{any} function: a
perceptron can perfectly emulate a \texttt{NAND} gate
(Fig.~\ref{fig_nandeqneuron}), and the singleton set $\{\texttt{NAND}\}$ is
functionally complete. Since we can combine a set of \texttt{NAND} gates to
calculate any function, \emph{we must also be able to combine a set of neurons to
calculate any function}. This result is also explored in a more
formal proof by both \citet{ref_cybenko1989} and \citet{ref_kurt1991}. They
show that an infinitely wide neural network can calculate any function.
Similarly, \citet{ref_lu2017width} show that an infinitely deep neural network
is a universal approximator.
Such a group of neurons is known as the multilayer perceptron (MLP).
Unfortunately, we cannot simply stack perceptrons together as we are missing
one vital ingredient: a way to train the network! At the time of
\citeauthor{ref_minsky1969}'s treatise on perceptrons there was no widely known
algorithm \citep[in the West; see][]{ref_ivakhnenko1965} that could train such
a multilayer network.  In \citeauthor{ref_minsky1969}'s own words:
\blockquote[\S13.2; \citet{ref_minsky1969}, on MLPs]{\small
    \emph{Nevertheless, we consider it to be an important research problem to
    elucidate (or reject) our intuitive judgment that the extension [from one
    layer to many] is sterile.  Perhaps some powerful convergence theorem will
    be discovered, or some profound reason for the failure to produce an
    interesting `learning theorem' for the multilayered machine will be
    found.}
}
\noindent The field had to wait almost two decades for such an algorithm to
become widespread. In the next subsection we will explore backpropagation, the
algorithm that ultimately proved \citeauthor{ref_minsky1969}'s intuition wrong.

\subsection{The multilayer perceptron} \label{sec_MLP}

\begin{figure}[htbp]
    \centering
    \includegraphics{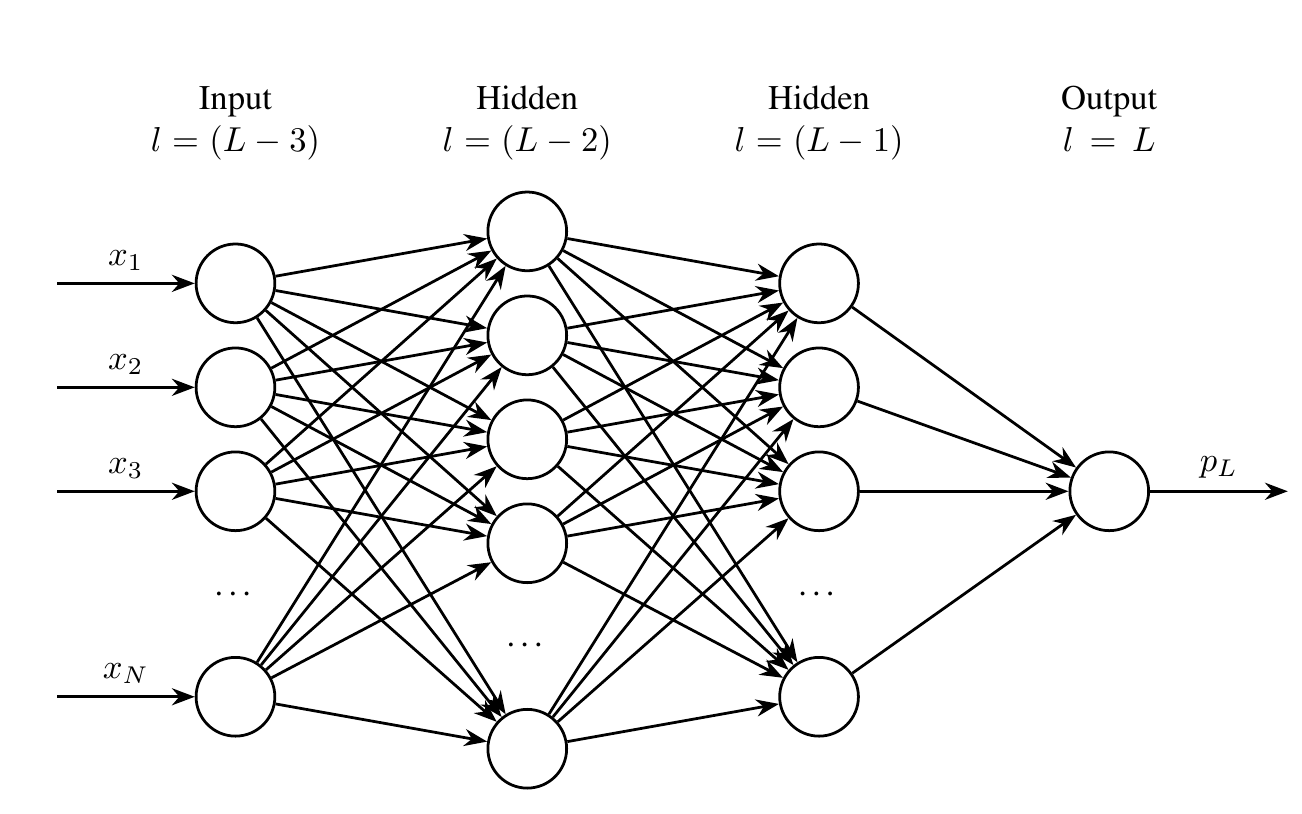}
    \caption[Multi layer perceptron]{The multilayer perceptron, or artificial
    neural network. The depicted network has two hidden layers. It takes $N$
    inputs $x_1, x_2, \ldots, x_N$, and outputs a prediction $p_L$. Note that
    here we omit the explicit bias terms (i.e. $w_0$).}
    \label{fig_MLP}
\end{figure}

Grouping many artificial neurons together may result in something resembling
Fig.~\ref{fig_MLP}. This network consists of an input layer, two intermediate
`hidden' layers, and an output layer.  As in the previous section, let us say
that we want a classifier that can classify a set of galaxy images into
elliptical and spiral types.  In an MLP similar to Fig.~\ref{fig_MLP} a neuron
would be assigned to each pixel in a galaxy image.  Each neuron would take the
numeric value of that pixel, and propagate that signal forward into the
network. The next layer of neurons does the same, with the input being the
previous layer's output. This process continues until we reach the output
layer. In a binary classification task like our galaxy classifier this layer
outputs a value between zero and one. Thus, if we define a spiral galaxy as zero,
and an elliptical galaxy as one, we would want the network output to be near
zero for a spiral galaxy input (and vice versa).

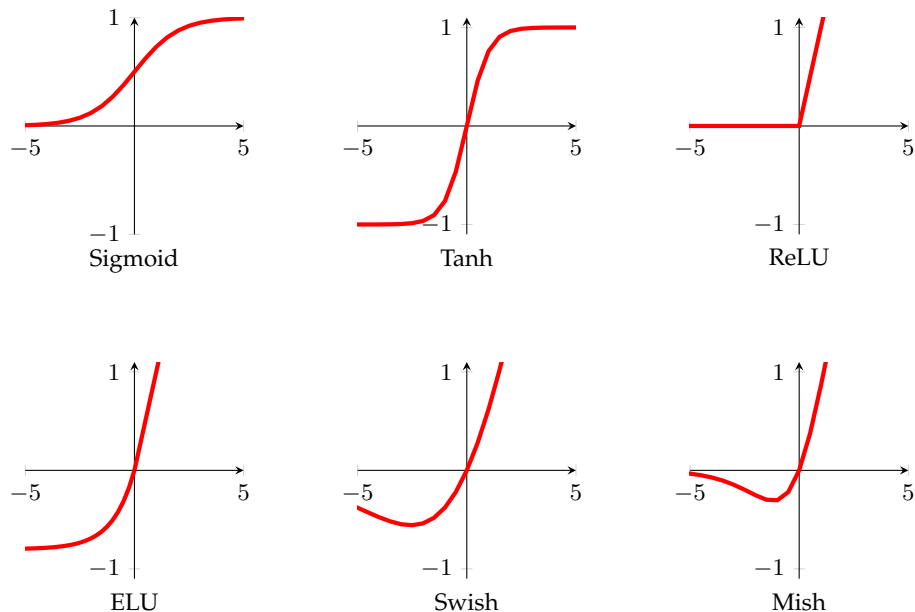
\begin{figure}[htbp]
    \centering
    \begin{tikzpicture}[
             declare function={
                             sigma(\x)=1/(1+exp(-\x));
                             relu(\x)=(\x>=0)*\x;
                             elupos(\x)=\x;
                             eluneg(\x)=0.8*(exp(\x)-1);
                             swish(\x)=\x/(1+exp(-0.5*\x));
                             mish(\x)=\x*tanh(ln(1+exp(\x)));
                         }]

        \begin{axis}%
        [
            width=0.33\linewidth,
            height=0.33\linewidth,
            xmin=-5,
            xmax=5,
            axis x line=center,
            ytick={-1,0,1},
            xtick={-5,0,5},
            ymax=1,
            ymin=-1,
            axis y line=center,
            domain=-6:6,
            name=plot0,
            title=Sigmoid,
            title style={at={(0.5,-0.1)},anchor=north,yshift=-0.1},
        ]
            \addplot[red,ultra thick,mark=none]   (x,{sigma(x)});
        \end{axis}

        \begin{axis}%
        [
            width=0.33\linewidth,
            height=0.33\linewidth,
            xmin=-5,
            xmax=5,
            axis x line=center,
            ytick={-1,0,1},
            xtick={-5,0,5},
            ymax=1.1,
            ymin=-1.1,
            axis y line=center,
            domain=-6:6,
            name=plot1,
            at=(plot0.right of north east), anchor=left of north west,
            xshift=1cm,
            title=Tanh,
            title style={at={(0.5,-0.1)},anchor=north,yshift=-0.1},
        ]
            \addplot[red,ultra thick,mark=none]   (x,{tanh(x)});
        \end{axis}

        \begin{axis}%
        [
            width=0.33\linewidth,
            height=0.33\linewidth,
            xmin=-5,
            xmax=5,
            axis x line=center,
            ytick={-1,0,1},
            xtick={-5,0,5},
            ymax=1.1,
            ymin=-1.1,
            axis y line=center,
            domain=-6:6,
            name=plot2,
            at=(plot1.right of north east), anchor=left of north west,
            xshift=1cm,
            title=ReLU,
            title style={at={(0.5,-0.1)},anchor=north,yshift=-0.1},
        ]
            \addplot[red,ultra thick,mark=none]   (x,{relu(x)});
        \end{axis}

        \begin{axis}%
        [
            width=0.33\linewidth,
            height=0.33\linewidth,
            xmin=-5,
            xmax=5,
            axis x line=center,
            ytick={-1,0,1},
            xtick={-5,0,5},
            ymax=1.1,
            ymin=-1.1,
            axis y line=center,
            name=plot3,
            at=(plot0.below south west), anchor=above north west,
            title=ELU,
            title style={at={(0.5,-0.1)},anchor=north,yshift=-0.1},
            yshift=-1cm,
        ]
            \addplot[red,ultra thick,mark=none,domain=-6:0]   (x,{eluneg(x)});
            \addplot[red,ultra thick,mark=none,domain=0:6]   (x,{elupos(x)});
        \end{axis}

        \begin{axis}%
        [
            width=0.33\linewidth,
            height=0.33\linewidth,
            xmin=-5,
            xmax=5,
            axis x line=center,
            ytick={-1,0,1},
            xtick={-5,0,5},
            ymax=1.1,
            ymin=-1.1,
            axis y line=center,
            domain=-6:6,
            name=plot4,
            at=(plot3.right of north east), anchor=left of north west,
            title=Swish,
            title style={at={(0.5,-0.1)},anchor=north,yshift=-0.1},
            xshift=1cm,
        ]
            \addplot[red,ultra thick,mark=none]   (x,{swish(x)});
        \end{axis}

        \begin{axis}%
        [
            width=0.33\linewidth,
            height=0.33\linewidth,
            xmin=-5,
            xmax=5,
            axis x line=center,
            ytick={-1,0,1},
            xtick={-5,0,5},
            ymax=1.1,
            ymin=-1.1,
            axis y line=center,
            domain=-6:6,
            name=plot5,
            at=(plot4.right of north east), anchor=left of north west,
            title=Mish,
            title style={at={(0.5,-0.1)},anchor=north,yshift=-0.1},
            xshift=1cm,
        ]
            \addplot[red,ultra thick,mark=none]   (x,{mish(x)});
        \end{axis}
    \end{tikzpicture}

    \caption[A curated selection of activation functions.]
    {A curated selection of activation functions. In all plots, the x axis is
    the input, and the y axis is the output. The rectified linear unit (ReLU)
    activation function was first introduced in the context of neural networks
    in \citet{ref_neocognitron} and later rediscovered, named, and
    popularised in \citet{ref_nair2010relu}. The exponential linear unit (ELU),
    Swish and Mish activations were respectively introduced in
    \citet{ref_clevert2015}, \citet{ref_ramachandran2017swish}, and
    \citet{ref_mish}.} 
    \label{fig_nonlinearities}
\end{figure}

In \S\ref{sec_AN} we found the change we needed to apply to a single neuron's
weights to make it learn from a training example. We can train an MLP in a
similar way by employing the reverse mode of automatic differentiation (or
backpropagation) to learn from our galaxy training data set
\citep{ref_backpropenglish,ref_werbos1981,ref_rumelhart1986}\footnote{
    Some controversy surrounds backpropagation's discovery.
    The Finnish computer scientist \citeauthor{ref_backpropenglish} proposed
    the reverse mode of automatic differentiation and adapted the algorithm to
    run on computers in their 1970 (Finnish language) thesis
    \citep{ref_backpropfinnish}. They first published their findings in
    English in \citeyear{ref_backpropenglish}.  \citeauthor{ref_werbos1981}
    then proposed applying an adaptation of \citeauthor{ref_backpropenglish}'s
    method to artificial neural networks.  \citet{ref_rumelhart1986} showed
    experimentally that backpropagation can generate meaningful internal
    representations within a neural network, and popularised the method. Here
    we will err on the side of caution and cite all three manuscripts. For
    further reading we recommend \citet{ref_schmidhuber2014} and
    \citet{ref_baydin2018backprop}.
    \label{ftn_backprop}
}.  
We want our network to learn when it makes both a correct and incorrect
prediction, so we define our activation function as a smoothed version of
\citeauthor{ref_perceptron}'s perceptron activation. This ensures that a signal
is present in the derivative no matter which values are input. This activation
function is known as the `sigmoid' function, and is shown in
Fig.~\ref{fig_nonlinearities}. As in \S\ref{sec_AN} we define a loss function
$\mathcal{L}(y, p)$ that describes the similarity between a ground truth ($y$)
and a prediction ($p$). We also define a neuron's activation function as
$\varphi(\mathbf{w}\cdot\mathbf{x})$ where $\mathbf{w}\cdot\mathbf{x}$ is the
weighted sum of a neuron's inputs. Following from Eq.~\ref{eqn_bp1}:
\begin{equation*}
    \frac{\partial \mathcal{L}}{\partial \mathbf{w}_l} = \frac{\partial \mathcal{L}}{\partial \mathbf{p}_l}
                                                         \frac{\partial \mathbf{p}_l}{\partial \mathbf{w}_l}
\end{equation*}
where $l$ is a layer in the MLP. In the same way as in \S\ref{sec_AN} we can
calculate an MLP's final layer's ($l=L$) weight updates in terms of known
values:
\begin{equation}
    \frac{\partial \mathcal{L}}{\partial \mathbf{w}_{L}} = \frac{\partial \mathcal{L}}{\partial p_{L}}\odot
                                                           \left(\varphi_L' \mathbf{p}^T_{L-1}\right),
    \label{eqn_bp_L}
\end{equation}
where $\mathbf{p}_{L-1}$ are the outputs from the previous layer. To calculate
the $(L-1)$th layer's weight updates we use the chain rule:
\begin{equation*}
    \frac{\partial \mathcal{L}}{\partial \mathbf{w}_{L-1}} = \frac{\partial \mathcal{L}}{\partial p_{L}}
                                                             \frac{\partial p_{L}}{\partial \mathbf{p}_{L-1}}
                                                             \frac{\partial \mathbf{p}_{L-1}}{\partial \mathbf{w}_{L-1}}.
\end{equation*}
Likewise for the $(L-n)$th layer:
\begin{equation*}
    \frac{\partial \mathcal{L}}{\partial \mathbf{w}_{L-n}} = \frac{\partial \mathcal{L}}{\partial p_{L}}
                                                             \left(\prod^{n}_{i=1} \frac{\partial \mathbf{p}_{L+1-i}}{\partial \mathbf{p}_{L-i}}\right)
                                                             \frac{\partial \mathbf{p}_{L-n}}{\partial \mathbf{w}_{L-n}}.
\end{equation*}
Now we can start plugging in some known values. Since $\mathbf{p}_{l} =
\varphi_l(\mathbf{w}_l \cdot \mathbf{p}_{l-1})$, it follows that 
$\partial \mathbf{p}_l / \partial \mathbf{p}_{l-1} =  \varphi'_l \mathbf{w}^T_l$, and 
$\partial \mathbf{p}_{l} / \partial \mathbf{w}_{l} = \varphi'_{l} \mathbf{p}^T_{l-1}$. So:
\begin{equation}
    \frac{\partial \mathcal{L}}{\partial \mathbf{w}_{L-n}} = \frac{\partial \mathcal{L}}{\partial p_{L}}\odot
                                                             \left(\prod^{n}_{i=1} \varphi'_{L-i} \mathbf{w}^T_{L-i}\right)\,
                                                             \left(\varphi'_{L-n} \mathbf{p}^T_{L-n-1}\right).
    \label{eqn_bp_n}
\end{equation}
Combining Eq.~\ref{eqn_bp_L} with Eq.~\ref{eqn_bp_n} we get the
weight update algorithm for the $(L-n)$th layer of the MLP:
\begin{equation}
    \mathbf{w}_{\text{next}} = \mathbf{w} - \eta 
        \begin{cases}
            \frac{\partial \mathcal{L}}{\partial p_{L}} \odot
            \left(\varphi_L' \mathbf{p}^T_{L-1}\right), & \text{for } n = 0,\\

            \frac{\partial \mathcal{L}}{\partial p_{L}} \odot
            \left(\prod^{n}_{i=1} \varphi'_{L-i} \mathbf{w}^T_{L-i}\right)\,
            \left(\varphi'_{L-n} \mathbf{p}^T_{L-n-1}\right), & \text{for } n > 0.
        \end{cases}
    \label{eqn_bp}
\end{equation}
With this equation\footnote{
    If we examine Eq.~\ref{eqn_bp} carefully, we
    can see why we add nonlinearities between the MLP layers; without
    activation functions Eq.~\ref{eqn_bp} collapses to the equivilent of a
    single layer MLP!
}
in hand we can use the same technique described earlier in this section and in
\S\ref{sec_AN} to update the network's weights with each galaxy image to
decrease the loss function $\mathcal{L}$. Again, as $\mathcal{L}$ is
minimised, our MLP will classify our elliptical and spiral galaxy images
with increasing accuracy.

\section{Astronomy's first wave of connectionism} \label{sec_mlpinastro}

Connectionism was first discussed within astronomy in the late 1980s, after the
popularisation of backpropagation (see footnote~\ref{ftn_backprop}) and the
consequent passing of the first `AI winter'. Two radical studies emerged in
1988 that recognised areas where astronomy could benefit from the use of
ANNs
\citep{ref_rappaport1988,ref_adorf1988}. 
Together, they identified that astronomical object classification\footnote{
    Specifically, galaxies were discussed in \citet{ref_rappaport1988} and
    point sources observed with the Infra-Red Astronomical Satellite (IRAS)
    were discussed in \citet{ref_adorf1988}.
}
and telescope scheduling could be solved through the use of an ANN.  These studies were followed by a rapid broadening of the
field, and the application of connectionism to many disparate astronomical use
cases \citep[][and references therein]{ref_miller1993}.  In this section, we
will outline areas where MLPs found an early use in astronomy.

\subsection{Classification problems}

\citet{ref_odewahn1992} classified astronomical objects into star and galaxy
types. These
were taken from the Palomar Sky Survey Automated Plate Scanner catalogue
\citep{ref_pennington1993}. To compile their dataset, they first extracted a set of
emergent image parameters from the scanned observations.
These parameters included the diameter, ellipticity, area, and plate transmission.
The parameters were then used to train both a linear perceptron and a
feedforward MLP to classify the objects into stars or galaxies.
\citet{ref_odewahn1992} found that their best performing model could
classify galaxies with a completeness of $95\%$ for objects down to a magnitude $<
19.5$. This work was followed by many more studies on the star/galaxy classification
problem \citep[e.g.][]{ref_odewahn1993,ref_bazell1998,ref_sextractor,ref_andreon2000}. Galaxy morphological type classification was explored in the early 1990s.
\citet{ref_storrie1992} describe an MLP that takes an input a selected set
of thirteen galaxy summary statistics, and uses this information to classify
a galaxy into one of five morphological types. \citet{ref_storrie1992}
report a top one accuracy of 64\%, and a top two accuracy of 90\%. This pilot
study was followed by several studies from the same group that confirmed that
MLPs are effective automatic galaxy morphological classifiers
\citep[][see \S\ref{sec_cnnrnn_apps} for a continuation of this line of
research]{ref_lahav1995,ref_lahav1996,ref_naim1995,ref_naim1995compar,ref_odewahn1996,ref_ball2004}.

MLPs were also used in other classification tasks; here we highlight
a few further areas where MLPs were applied.  
\citet{ref_von1994} classified stellar spectra into temperature types, and
\citet{ref_klusch1993} did the same for Morgan-Keenan System types.
\citet{ref_chon1998} described the use of an MLP to search for and classify
muon events (and therefore neutrino observations) in the Sudbury Neutrino
Observatory. Quasar classification has been explored in several studies
\citep{ref_carballo2004,ref_claeskens2006,ref_carballo2008}.
Seminally, \citet{ref_carballo2004} used an MLP to select quasar candidates
given their radio flux, integrated-to-peak flux ratio, photometry and point
spread function in the red and blue bands, and their radio-optical position
separation. They found good agreement between their model and that of the
decision tree described in \citet{ref_white2000}, confirming MLPs as a
competitive alternative to more traditional machine learning.
As part of the Supernova photometric Classification Challenge
\citep[SPCC;][]{ref_kessler2010}, \citet{ref_karpenka2013} proposed the
use of a neural network to classify supernovae into Type-1a/non-Type-1a
classes. To classify their light curves, they first used a hand-crafted fitting
function, and then trained their MLP on the fitted coefficients. They found
that their model was competitive with other, more complex models trained on the
SPCC dataset.
From the studies discussed in this section we can safely conclude that MLPs are effective
classifiers of astronomical data, when given important parameters extracted by
an expert guide.

\subsection{Regression problems}

MLPs have also been used in regression problems. \citet{ref_angel1990} applied
them first to adaptive telescope optics. They trained their MLP on $250\,000$
simulated in focus and out of focus observations of stars as seen by the
Multiple Mirror Telescope (MMT). From the flattened $13 \times 13$ pixel
observations, their network predicted the piston position and tilt required for
each of the MMT's mirrors to bring the stars into focus. After the application
of these corrections, the authors were able to recover the original profile.
In follow up studies, \citet{ref_sandler1991} and
\citet{ref_lloydhart1992} proved that \citeauthor{ref_angel1990}'s MLP
worked on the real MMT.

Photometric redshift estimation was explored in many concurrent studies
\citep[e.g.][]{ref_firth2003,ref_tagliaferri2003,ref_vanzella2004,ref_collister2004,ref_ball2004}.
\citet{ref_firth2003} train a neural network to predict the redshift
of galaxies contained in the Sloan Digital Sky Survey (SDSS) early data release
\citep{ref_stoughton2002}. The galaxies were input to the neural network
as a set of summary parameters, and the output was a single float representing
 the galaxy redshift. They found their network attained a performance comparable
to classical techniques. Extending and confirming the work by
\citet{ref_firth2003}, \citet{ref_ball2004} used an MLP to predict
the redshift of galaxies contained in the SDSS's first data release
\citep{ref_sdss}. They also showed that MLPs were capable of predicting the
galaxies' spectral types and morphological classifications.

Of course, MLPs have been used more widely in astronomical regression tasks.
Here we will cherry pick a few studies to show the MLP's early breadth of use.
Sunspot maxima prediction was carried out by \citet{ref_koons1990}. They
found their MLP based method was capable of predicting the number of sunspots
when trained on previous cycles.  \citet{ref_bailer1997} predicted the
effective temperature of a star from its spectrum. 
\citet{ref_auld2007} and \citet{ref_auld2008} applied MLPs to
cosmology, demonstrating that MLPs are capable of predicting the cosmic
microwave background power spectra and matter power spectra when given a set of
cosmological parameters.
\citet{ref_norgaardnielsen2008} used an MLP to remove the foreground from
microwave temperature maps.  From the studies discussed in this section we can
see that MLPs are effective regressors of astronomical data, when given
significant parameters extracted by an expert guide.

\section{Contemporary supervised deep learning} \label{sec_condeeplearn}

There are some issues with MLPs. Primarily they do not scale well to high
dimensional datasets. For example, if our dataset consists of images with a
$128 \times 128$ pixels, we will need $16\,384$ neurons in the MLP's input
layer alone! As we move into the hidden layers, this scaling issue only gets
worse.  Also, since MLPs must take an unrolled image as an input, they
disregard any spatial properties of their training images, and so either need a
substantial amount of training data to classify or generate large
images\footnote{
    At the height of the convolutional neural network architecture's
    popularity in the mid 2010s these were real problems. However, with the growth
    of computing power and data in recent years we are seeing a resurgence of the more
    general MLP model
    \citep[e.g.][]{ref_tolstikhin2021,ref_touvron2021,ref_liu2021,ref_melaskyriazi2021}.
    This follows the prevailing trend in AI where the removal of human-crafted
    features and biases ultimately results in more expressive models that learn
    such features and biases dieectly from data \citep{ref_sutton2019,ref_gwern2022}.
    \label{ftn_mlpresurgence}
}, or an expert to extract descriptive features from the data in a preprocessing step. 
We can see this issue writ large in the previous section---most of the MLP
applications described in \S\ref{sec_mlpinastro} require an expert to extract
features from the data for the network to then train on! This drawback is not
ideal; what if there are features within the raw data that are not present in
these cherry picked statistics? In that case, it would be preferable to let the
neural network take in the raw data as input, and then learn which features are
the most descriptive. We will discuss neural network architectures that solve
both the MLP scaling problem and the expert reliance problem in
this section. After we have explored these architectures in general, we will
discuss their application to astronomical problems in \S\ref{sec_cnnrnn_apps}.

\subsection{Convolutional neural networks} \label{sec_CNN}

\begin{figure}[htbp]
    \centering
    \includegraphics[width=\linewidth]{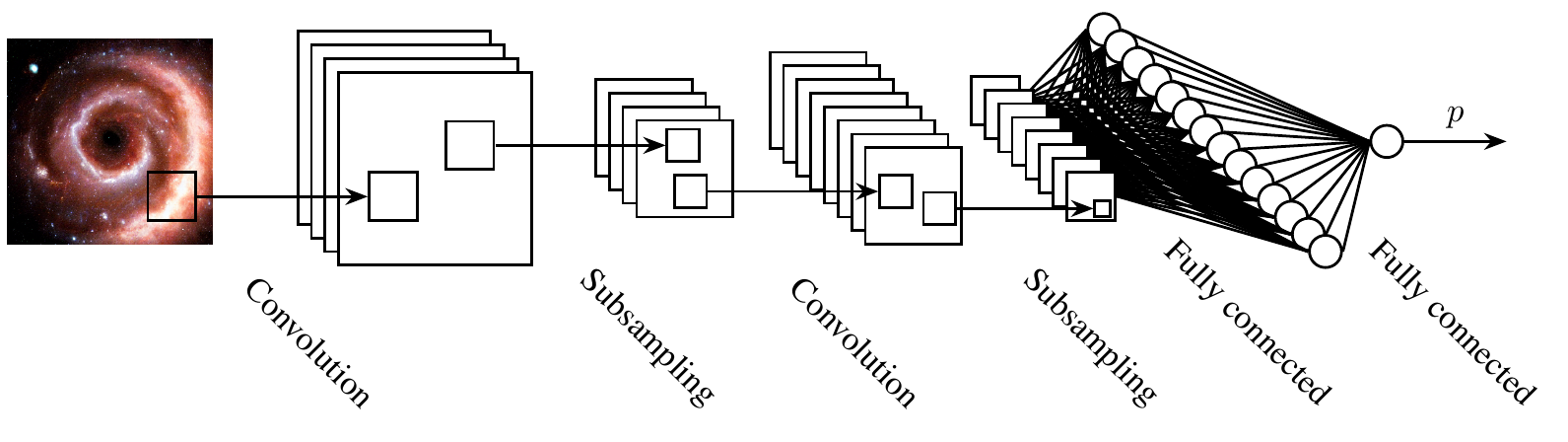}
    \caption[Convolutional neural network]{A convolutional neural network
    classifying a spiral galaxy image\protect\footnotemark.}
    \label{fig_CNN}
\end{figure}

Unlike the MLP described in the previous section, convolutional neural networks
(CNNs; introduced in \citet{ref_neocognitron} and first combined with
backpropagation in \citet{ref_lecun1989}) do not entirely consist of fully
connected layers, where each neuron is connected to every neuron in the
previous and subsequent layers. Instead, the CNN (such as the one depicted in
Fig.~\ref{fig_CNN}) uses convolutional layers in place of the majority (or all)
of the dense layers.
 
\footnotetext{All astronomical objects shown in the neural network diagrams
within this manuscript are generated
via text prompts fed into a latent diffusion neural network model
\citep{ref_rombach2021}.}

We can think of a convolutional layer as a set of learnt `feature filters'.
These feature filters perform a local transform on input imagery. In classical
computer vision, these filters are hand crafted, and perform a
predetermined function, such as edge detection or blurring. In contrast, a CNN
learns the optimal set of filters for its task (say, galaxy classification).
Eq.~\ref{eqn.conv} shows two different convolution\footnote{ 
    We must note that in Eq.~\ref{eqn.conv} we follow
    most deep learning libraries and perform a cross-correlation and \emph{not} a
    convolution. However, since the weights are learnt, this does not matter; the
    neural network will simply learn a flipped representation of the
    cross-correlation.
}
operators being performed on an array.
\begin{equation}
    \begin{split}
        \begin{bmatrix}
            39  &  57  &  86  &  9  &  26 \\
            90  &  74  &  63  &  87  &  98 \\
            79  &  34  &  26  &  16  &  46 \\
            67  &  61  &  96  &  1  &  79 \\
            33  &  47  &  15  &  49  &  29
        \end{bmatrix}
        \star
        \begin{bmatrix}
            0 & 0 & 0 \\
            0 & 0 & 0 \\
            0 & 0 & 1 \\
        \end{bmatrix}
        &=
        \begin{bmatrix}
            26  &  16  &  46 \\
            96  &  1  &  79 \\
            15  &  49  &  29
        \end{bmatrix}\\\\
        \begin{bmatrix}
            39  &  57  &  86  &  9  &  26 \\
            90  &  74  &  63  &  87  &  98 \\
            79  &  34  &  26  &  16  &  46 \\
            67  &  61  &  96  &  1  &  79 \\
            33  &  47  &  15  &  49  &  29
        \end{bmatrix}
        \star
        \begin{bmatrix}
            1 & 0 & 0 \\
            0 & 1 & 0 \\
            0 & 0 & 1 \\
        \end{bmatrix}
        &=
        \begin{bmatrix}
            139  &  136  &  219 \\
            220  &  101  &  158 \\
            155  &  179  &  56
        \end{bmatrix}
    \end{split}
    \label{eqn.conv}
\end{equation}
In the above equation the operation is represented as a matrix. In a CNN the
matrix is a set of neuronal weights. As seen in Fig.~\ref{fig_CNN} there are
multiple feature maps in a convolutional layer, each containing a set of
weights independent to the other feature maps, and learning to extract a
different feature. 
Due to the convolution operator's inbuilt translational
equivarience, these features can be detected by the convolutional layer no
matter where they are in the image.
As in the MLP described in the previous section, the weights are updated using
backpropagation to minimise a loss function.  We will discuss astronomical
applications of CNNs in \S\ref{sec_cnnrnn_apps}, after we introduce modern CNN
architectures.

\subsection{Recurrent neural networks} \label{sec_rnn}

\begin{figure}[htbp]
    \centering
    \input{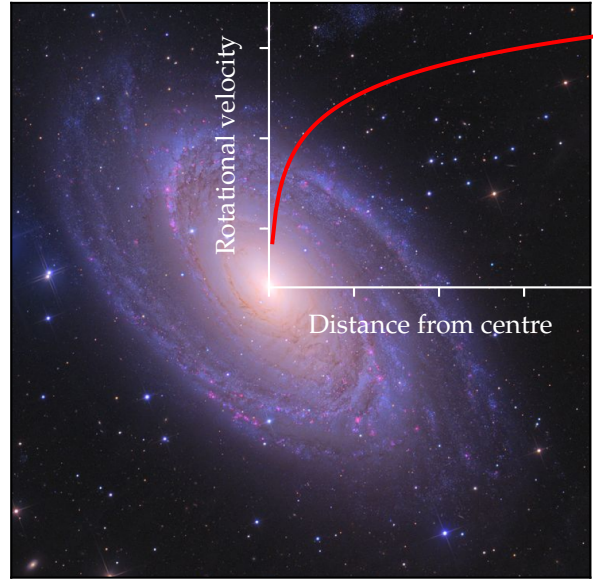}
    \caption[A galaxy rotation curve example]{An example of a galaxy rotation curve, 
    plotted over an image of Messier 81 \mbox{\citep{ref_m81}}.}
    \label{fig_rcurve}
\end{figure}

Standard feedforward neural networks like the MLP (\S\ref{sec_MLP}) and CNN
(\S\ref{sec_CNN}) generate a fixed size vector given a fixed size
input\footnote{
    As with any rule there are exceptions, such as CNNs containing a global
    average pooling layer \citep{ref_lin2013}.
}. 
But, what if we want to
classify or generate a variably sized vector? For example, we might want to
classify a galaxy's morphology given its rotation curve. A rotation curve
describes the velocity of a galaxy's visible stars versus their distance from the
galaxy's centre.  Fig.~\ref{fig_rcurve} shows a possible rotation curve for
Messier 81.  A rotation curve's length depends on the size of its galaxy, and due
to this variable length, and the fact that MLPs take a fixed size input, we
cannot easily use an MLP for classification. Recurrent neural networks (RNNs),
however, can take a variable length input and produce a variable length
output.
An RNN differs from a feed forward MLP by having a hidden state that acts as a
`memory' store of previously seen information. As the RNN encounters new data,
its weights are altered through the backpropagation through time algorithm
\citep[BPTT;][and references therein. Also see footnote
\ref{ftn_backprop}]{ref_werbos1990}.

\begin{figure}[htbp]
    \centering
    \includegraphics[width=\textwidth]{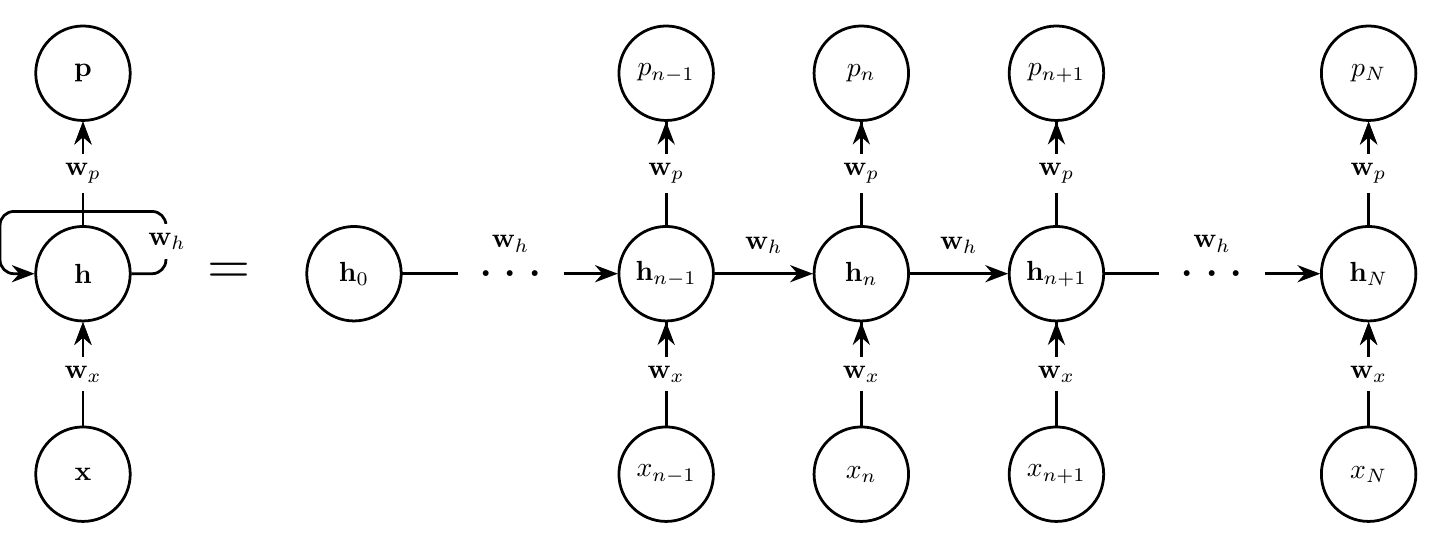}
    \caption[Recurrent neural network]{A recurrent neural network
    with weights $\{\mathbf{w}_x, \mathbf{w}_p, \mathbf{w}_h\}$, a hidden state
    $\mathbf{h}_n$, inputs $\mathbf{x}$, and a prediction $p_{n=N}$ is unrolled into its
    constituent processes.} 
    \label{fig_RNN}
\end{figure}

We can use an RNN similar to Fig.~\ref{fig_RNN} to classify
our rotation curves. We express the rotation curve as a list
$\{x_1,x_2,\ldots,x_N\}$, with each $x$ being a measurement of the rotational
velocity at a certain radius. Then we feed this list into the RNN
sequentially in the same way as shown in Fig.~\ref{fig_RNN}. The RNN will
produce an output for each $x$ fed to it, but we ignore those until we feed in
$x_N$, the rotational velocity furthest from the galaxy's centre. When we feed
in $x_N$, the RNN produces a prediction $p_N$, which we can then compare to a
ground truth $y_N$ via a loss function $\mathcal{L}_N$. In our case, $y$ is an
integer label representing the galaxy's morphological class. The comparison
$\mathcal{L}_N(y_N, p_N)$ is a function that represents the distance between the
RNN prediction and the ground truth. We can then reduce $\mathcal{L}_N(y_N, p_N)$
by updating the RNN's weights through BPTT so that the weights
$\{\mathbf{w}_x,\mathbf{w}_p,\mathbf{w}_h\}$ follow $\nabla \mathcal{L}_N$ 
downwards. As we do this, our RNN will improve its galaxy classifications.

BPTT's mathematical derivation is akin to the one we explored in
\S\ref{sec_MLP}, and we will quickly derive it here for posterity. Let us first
look at the forward propagation equations:
\begin{align*}
    \mathcal{L}_{n} &= \lvert y_n - p_n \rvert,\\
    p_n &= \varphi(\mathbf{w}_p \cdot \mathbf{h}_n),\,\text{and}\\
    \mathbf{h}_n &= \phi(\mathbf{w}_h \cdot \mathbf{h}_{n-1} + \mathbf{w}_x \cdot \mathbf{x}_n).
\end{align*}
From these we see that we need to express
$\partial\mathcal{L}_n / \partial\mathbf{w}_p$,
$\partial\mathcal{L}_n / \partial\mathbf{w}_h$, and
$\partial\mathcal{L}_n / \partial\mathbf{w}_x$ as known values to train
the network.  $\partial\mathcal{L}_n / \partial\mathbf{w}_p$ is
relatively easy; via the chain rule, and the fact that $\partial p_n / \partial \mathbf{w}_p = \varphi' \mathbf{h}_n^T$:
\begin{align}
    \frac{\partial\mathcal{L}_n}{\partial\mathbf{w}_p} &= \frac{\partial\mathcal{L}_n}{\partial p_n}\frac{\partial p_n}{\partial\mathbf{w}_p},\nonumber\\
    &= \frac{\partial\mathcal{L}_n}{\partial p_n} \odot \varphi' \mathbf{h}_n^T.
\end{align}
$\partial\mathcal{L}_n/\partial\mathbf{w}_h$ is more tricky, so we will
go step by step. We already know that
\begin{equation}
    \frac{\partial\mathcal{L}_n}{\partial\mathbf{w}_h} = \frac{\partial\mathcal{L}_n}{\partial p_n}\frac{\partial p_n}{\partial\mathbf{h}_n}\frac{\partial \mathbf{h}_n}{\partial\mathbf{w}_h}. \label{eqn_dldwh}
\end{equation}
However, we see in Fig.~\ref{fig_RNN} that $\mathbf{h}_n$ depends on
$\mathbf{h}_{n-1}$, which depends on $\mathbf{h}_{n-2}$ (and so on). We also
notice that all the hidden states depend on $\mathbf{w}_h$. We
therefore rewrite Eq.~\ref{eqn_dldwh} to make this explicit:
\begin{align*}
    \frac{\partial\mathcal{L}_n}{\partial\mathbf{w}_h} &= \frac{\partial\mathcal{L}_n}{\partial p_n}\frac{\partial p_n}{\partial\mathbf{h}_n}
    \sum^n_{j=1} \frac{\partial \mathbf{h}_{n}}{\partial\mathbf{h}_j}\frac{\partial \mathbf{h}_{j}}{\partial\mathbf{w}_h},\\
    &= \frac{\partial\mathcal{L}_n}{\partial p_n}\frac{\partial p_n}{\partial\mathbf{h}_n}
    \sum^n_{j=1} \left(\prod_{i=j + 1}^n\frac{\partial \mathbf{h}_i}{\partial \mathbf{h}_{i-1}}\right)\frac{\partial \mathbf{h}_{j}}{\partial\mathbf{w}_h}.
\end{align*}
We can now substitute in some known values:
\begin{equation}
    \frac{\partial\mathcal{L}_n}{\partial\mathbf{w}_h} = \frac{\partial\mathcal{L}_n}{\partial p_n} \odot \varphi' \mathbf{h}_n^T \sum^n_{j=1} \left(\prod_{i=j + 1}^n\phi'\mathbf{w}_{h, i}^T\right){\phi'\mathbf{h}_{j-1}^T}.
    \label{eqn_rnn_vanishinggradient}
\end{equation}
Finally, $\partial \mathcal{L}_n/\partial \mathbf{w}_x$ is derived in
the same way as $\partial\mathcal{L}_n/\partial\mathbf{w}_h$:
\begin{align}
    \frac{\partial\mathcal{L}_n}{\partial\mathbf{w}_x} &= \frac{\partial\mathcal{L}_n}{\partial p_n}\frac{\partial p_n}{\partial\mathbf{h}_n}
    \sum^n_{j=1}\left(\prod_{i=j + 1}^n\frac{\partial \mathbf{h}_i}{\partial \mathbf{h}_{i-1}}\right)\frac{\partial \mathbf{h}_{j}}{\partial\mathbf{w}_x},\nonumber\\
    &= \frac{\partial\mathcal{L}_n}{\partial p_n} \odot \varphi' \mathbf{h}_n^T \sum^n_{j=1}\left(\prod_{i=j + 1}^n\phi'\mathbf{w}_{h, i}^T\right){\phi'\mathbf{x}_{j}^T}. \label{eqn_rnn_vanishinggradientx}
\end{align}
With $\partial\mathcal{L}_n / \partial\mathbf{w}_p$,
$\partial\mathcal{L}_n / \partial\mathbf{w}_h$, and
$\partial\mathcal{L}_n / \partial\mathbf{w}_x$ in hand we can apply
the same update rule shown in Eq.~\ref{eqn_bp}.

Aside from many-to-one encoding, RNNs can produce many predictions given many
inputs, or act similarly to an MLP and produce one or many outputs given a single input.
We will discuss the application of recurrent neural networks to astronomical
data in \S\ref{sec_cnnrnn_apps}, after we introduce gated recurrent neural
networks.

\subsection{Sidestepping the vanishing gradient problem} \label{sec_vanishinggradient}

In the early 1990s, researchers identified a major issue with the
training of deep neural networks through backpropagation.
\citeauthor{ref_hochreiter1991} first formally examined the `vanishing
gradient' problem in their diploma thesis (\citet{ref_hochreiter1991}, see
also later work by \citet{ref_bengio1994}).  Due to the vanishing gradient
problem, it was widely believed that training very deep artificial neural
networks from scratch via backpropagation was impossible. In this section we
will explore what the vanishing gradient problem is, and how contemporary
end-to-end trained neural networks sidestep this issue.  

First let us remind ourselves of the sigmoid activation function introduced in
Fig.~\ref{fig_nonlinearities}: 

\vspace{1em}
\noindent%
\begin{minipage}[c]{0.48\textwidth}
    \centering
    \begin{tikzpicture}[declare function={sigma(\x)=1/(1+exp(-\x));
                        sigmap(\x)=sigma(\x)*(1-sigma(\x));}]
       \begin{axis}%
       [
           xmin=-6,
           xmax=6,
           axis x line=bottom,
           ytick={0,.5,1},
           ymax=1.05,
           axis y line=center,
           domain=-6:6,
           scale=0.62,
           xlabel=$\mathbf{x}$,
           legend style={at={(0.5,1.32)},anchor=north,draw=none},
           legend columns=-1,
       ]
           \addplot[red,ultra thick,mark=none]   (x,{sigma(x)});
           \addlegendentry{$\varphi(\mathbf{x})\quad$};
           \addplot[red,ultra thick,dotted,mark=none]   (x,{sigmap(x)});
           \addlegendentry{$\varphi'(\mathbf{x})$};
       \end{axis}
    \end{tikzpicture}
\end{minipage}
\hfill
\begin{minipage}[c]{0.48\textwidth}
    \noindent
    \begin{equation}
        \varphi(\mathbf{x}) = 1/(1 - e^{-\mathbf{x}}).
        \label{eqn_sigmoidder}
    \end{equation}
\end{minipage}
\vspace{1em}

\noindent Eq.~\ref{eqn_sigmoidder} and its accompanying plot shows the output
of a sigmoid function $\varphi$ and its derivative $\varphi'$, when given an
input $\mathbf{x}$.

Now, let us revisit the weight update rule for the $(L - n)$th layer of a
feedforward MLP (Eq.~\ref{eqn_bp_n}):
\begin{equation}
    \frac{\partial \mathcal{L}}{\partial \mathbf{w}_{L-n}} = \frac{\partial \mathcal{L}}{\partial p_{L}}\odot
                                                            \underbrace{\left(\prod^{n}_{i=1} \varphi'_{L-i}\mathbf{w}^T_{L-i}\right)}_{\lim\limits_{n \to \infty}\prod^{n}_{i=1} \varphi'_{L-i}\mathbf{w}^T_{L-i} =\,0}\,
                                                             \left(\varphi'_{L-n} \mathbf{p}^T_{L-n-1}\right).
    \label{eqn_vanishing_gradients_mlp}
\end{equation}
If $\varphi'$ is typically less than one (as in Eq.~\ref{eqn_sigmoidder} and
most other saturating nonlinearities) the product term in the above
equation becomes an issue. In that case, we can see that the product rapidly
goes to zero as $n$ (the number of layers) becomes large\footnote{
    Likewise, if $\varphi'$ is typically greater than one, the product term
    rapidly `explodes' to infinity. This is known as the `exploding gradient'
    problem, also first identified in \citet{ref_hochreiter1991}.
    \label{foot_explodinggradient}}. 
If we study Eq.~\ref{eqn_rnn_vanishinggradient}, we can see the same problem
also plagues RNNs as we backpropagate through hidden states:
\begin{equation}
    \frac{\partial\mathcal{L}_n}{\partial\mathbf{w}_h} = \frac{\partial\mathcal{L}_n}{\partial p_n} \odot \varphi' \mathbf{h}_n^T \sum^n_{j=1} \underbrace{\left(\prod_{i=j + 1}^n\phi'\mathbf{w}_{h, i}^T\right)}_{\lim\limits_{n \to \infty}\prod^{n}_{i=j+1} \phi'\mathbf{w}^T_{h,i} =\,0}{\phi'\mathbf{h}_{j-1}^T}.
    \label{eqn_vanishing_gradients_rnn}
\end{equation}

Let us solidify this issue by reminding ourselves of Eq.~\ref{eqn_bp}---the
weight update rule for a network trained through backpropagation:
\begin{equation}
    \mathbf{w}_{\text{next}} = \mathbf{w} - \eta \frac{\partial \mathcal{L}}{\partial \mathbf{w}}.
    \label{eqn_updaterule}
\end{equation}
Combining Eq.~\ref{eqn_updaterule} and the limits defined in
Eq.~\ref{eqn_vanishing_gradients_mlp} and Eq.~\ref{eqn_vanishing_gradients_rnn}
results in the below weight update rule in the limit $n \to \infty$.
\begin{equation}
    \lim\limits_{n \to \infty} \mathbf{w}_{\text{next}} = \mathbf{w}.
    \label{eqn_gradientgone}
\end{equation}
Eq.~\ref{eqn_gradientgone} shows that learning via backpropagation slows
as we move deeper into the network.  This problem once again
caused a loss of faith in the connectionist model, ushering in the second AI
winter.  It took until 2012 for a new boom to begin. In the following three
subsections we will explore some of the proposed partial solutions to the
vanishing gradient problem and show how they came together to contribute
to the current deep learning boom.

\subsubsection{Non-saturating activation functions}

We can see in Eq.~\ref{eqn_vanishing_gradients_rnn} and
Eq.~\ref{eqn_vanishing_gradients_mlp} that if $\varphi' = 1$ 
then the product term does not automatically go to zero or infinity. If this is
the case, why not simply design our activation function around this property?
The rectified linear unit
\citep[ReLU;][]{ref_neocognitron,ref_nair2010relu} is an activation
function that does precisely this\footnote{
    $\text{ReLU}'$ is always zero if its inputs are $<0$,
    removing any signal for further training.  This is known as the `dying ReLU'
    problem, but is not as big of an issue as it first seems. Since contemporary
    deep neural networks are greatly overparameterised
    (see for example \citet{ref_frankle2018} and other work on the `lottery ticket
    hypothesis') backpropagation through the ReLU activation
    function can act as a pruning mechanism, creating sparse representations
    within the neural network and thus reducing training time even further
    \citep{ref_glorot2011}. 
}:

\vspace{1em}
\noindent%
\begin{minipage}[c]{0.48\textwidth}
    \centering
    \begin{tikzpicture}[declare function={relu(\x)=(\x>=0)*\x;}]
       \begin{axis}%
       [
           xmin=-1.5,
           xmax=1.5,
           axis x line=bottom,
           ytick={0,.5,1},
           ymax=1.05,
           axis y line=center,
           domain=-6:6,
           scale=0.62,
           xlabel=$\mathbf{x}$,
           legend style={at={(0.5,1.32)},anchor=north,draw=none},
           legend columns=-1,
       ]
           \addplot[red,ultra thick,mark=none]   (x,{relu(x)});
           \addlegendentry{$\text{ReLU}(\mathbf{x})\quad$};
           \addplot[red,ultra thick,dotted,mark=none] coordinates {(-6,0) (0,0) (0,1) (6,1)};
           \addlegendentry{$\text{ReLU}'(\mathbf{x})$};
       \end{axis}
    \end{tikzpicture}
\end{minipage}
\hfill
\begin{minipage}[c]{0.48\textwidth}
    \noindent
    \begin{equation}
        \text{ReLU}(\mathbf{x}) = \max(\mathbf{x}, 0).
        \label{eqn_reluder}
    \end{equation}
\end{minipage}
\vspace{1em}

The gradient of ReLU is unity if the inputs are above
zero, exactly the property we needed to mitigate the vanishing gradient
problem.  Similar non-saturating activation functions also share the ReLU
gradient's useful property, see for example the Exponential Linear Unit, Swish,
and Mish functions in Fig.~\ref{fig_nonlinearities}.

\subsubsection{Graphics processing unit acceleration} \label{sec_gpu_acceleration}

If we can speed up training, we can run an inefficient algorithm (such as
backpropagation through saturating activations) to completion in less time. One
way to speed up training is by using hardware that is specifically suited to the
training of neural networks.  Graphics processing units (GPUs) were originally
developed to render video games and other intensive graphical processing tasks.
These rendering tasks require a processor capable of massive parallelism. We
have seen in the previous sections that neural networks trained through
backpropagation also require many small weight update calculations.  With this
in mind, it is natural to try to accelerate deep neural networks using GPUs. 

\begin{figure}[htbp]
    \centering
    \input{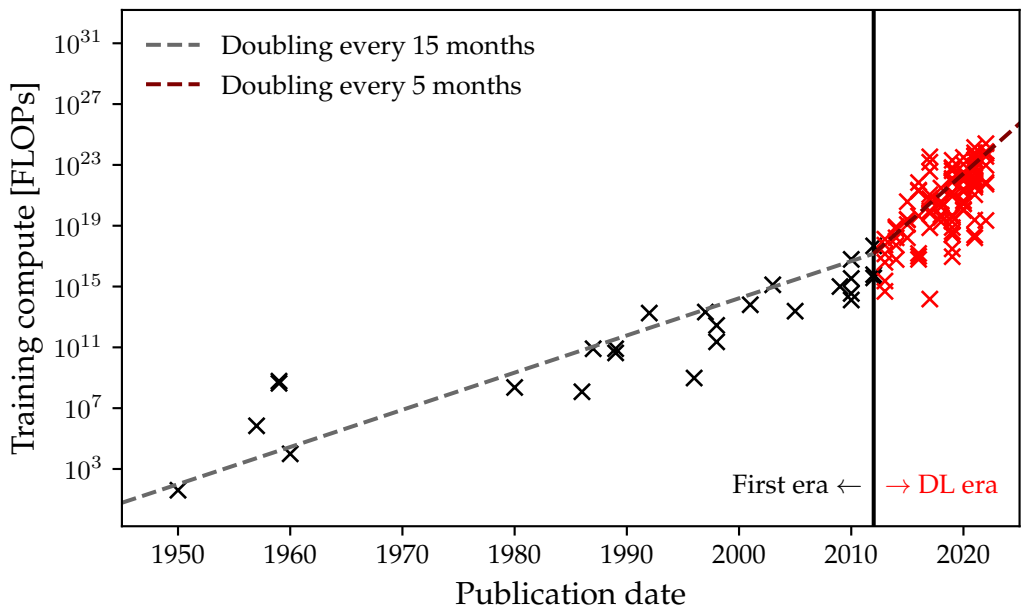}
    \caption[Demarcating the `Deep Learning Era']{If we plot the total number of
    floating point operations (FLOPs) required to train a neural network model,
    and compare it to the model's publication date, we can see a change
    in trend at around \citeyear{ref_alexnet}. This corresponds to the
    popularisation of GPU accelerated training of very deep neural networks,
    with \citeyear{ref_alexnet} demarcating AI's `Deep Learning Era' and 
    the beginning of astronomy's second wave of connectionism
    (\S\ref{sec_cnnrnn_apps}).  Data is taken from \citet{ref_sevilla2022}.}
    \label{fig_mleras}
\end{figure}

In 2004, \citet{ref_oh2004anngpu} were the first to use GPUs to accelerate an MLP model,
reporting a $20\times$ performance increase on inference with an `ATI RADEON 9700 PRO' GPU
accelerated neural network. Shortly after, \citet{ref_steinkrau2005} showed
that backpropagation can also benefit from GPU acceleration, reporting a
three-fold performance increase in both training and inference. These two
breakthroughs were followed by a flurry of activity in the area
\citep[e.g.][]{ref_chellapilla2006,ref_raina2009,ref_dannet2010mlp,ref_dannet},
culminating in a milestone victory for GPU accelerated neural networks at
ImageNet 2012.  AlexNet \citep{ref_alexnet} won the ImageNet
classification and localisation challenges \citep{ref_russakovsky2015},
scoring an unprecedented top-5 classification error of $16.4\%$, and a single
object localisation error of $34.2\%$. In both challenges AlexNet scored over 
$10\%$ better than the models in second place.  
\citeauthor{ref_alexnet}'s
winning network was a CNN \citep{ref_neocognitron} trained through
backpropagation \citep{ref_backpropenglish,ref_lecun1989}, with ReLU
activation \citep{ref_nair2010relu}, and dropout \citep{ref_dropout} as
a regulariser\footnote{
    Dropout reduces the amount of neural network overfitting---where a network performs well on the training set at the expense of performance on data it has not yet seen.
    One performs dropout by randomly removing a set of neurons at each training step, and using all neurons at test time. This set up essentially trains a large ensemble of sub-models, whose average prediction outperforms that inferred by a single model.
}. 
The performance increase afforded by GPU accelerated training
enabled the network to be trained from scratch via backpropagation in a
reasonable amount of time. The discovery that it is possible to train a neural
network from scratch by using readily available hardware ultimately resulted in
the end of connectionism's second winter, and ushered in the Cambrianesque deep
learning explosion of the mid-to-late 2010s and the 2020s (Fig.~\ref{fig_mleras}).

\subsubsection{Gated recurrent neural networks and residual networks} \label{sec_LSTM}

\begin{figure}[htbp]
    \centering
    \includegraphics[width=\textwidth]{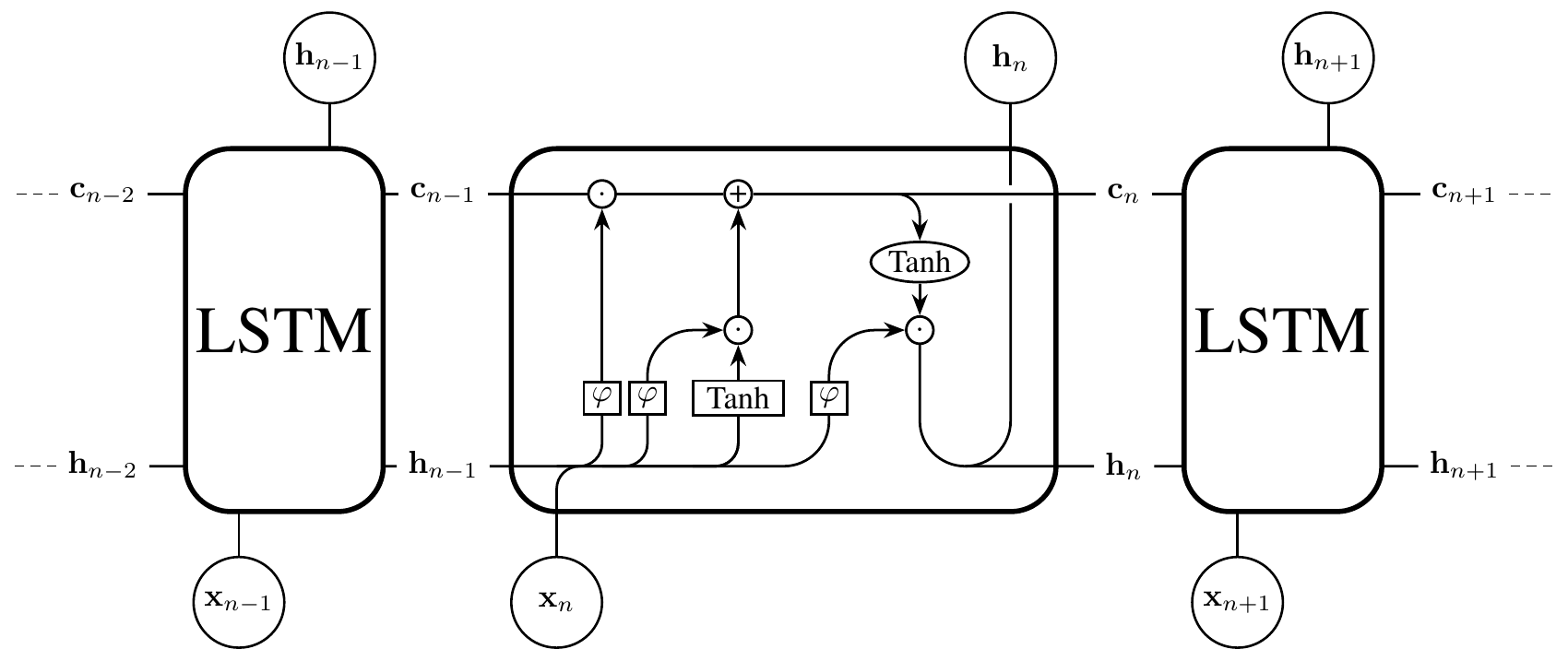}
    \caption[Long-short term memory]{A set of sequential data $\mathbf{x}_n$ is
    input into an LSTM network. Inside the cell
    \tikz\draw[black,thick,fill=white] (0,0) circle (.6ex); 
    denotes elementwise operations and 
    \tikz\draw[black,thick,fill=white] (0,0) rectangle (1.2ex,1.2ex);
    denotes neuronal layers. $\varphi$ is the sigmoid activation function,
    and Tanh is the hyperbolic tangent activation function.
    $\oplus$ is an elementwise addition, $\odot$ is the Hadamard product, and
    line mergers are concatenations.  $\mathbf{c}_n$ is the cell state, and
    $\mathbf{h}_n$ is the hidden state.}
    \label{fig_LSTM}
\end{figure}

The long short-term memory unit \citep[LSTM;][]{ref_lstm,ref_gers2000}\footnote{
    Compare also the gated recurrent unit \citep[GRU;][]{ref_gru}.
}
mitigates the vanishing gradient problem by introducing a new hidden state, the
`cell state' $(\mathbf{c}_n)$, to the standard RNN
architecture.  This cell state allows the network to learn long range
dependencies, and we will show why this is the case via a
brief derivation\footnote{
    Here we loosely follow \citet[\S1.3.4]{ref_bayer2015}.
}. First, as always, let us study Fig.~\ref{fig_LSTM} and write down the
forward pass equation for updating the cell state:
\begin{equation*}
    \mathbf{c}_n = f(\mathbf{c}_{n-1}, \mathbf{h}_{n-1}, \mathbf{x}_n) + g(\mathbf{h}_{n-1}, \mathbf{x}_n)
\end{equation*}
where $f(\mathbf{c}_{n-1}, \mathbf{h}_{n-1}, \mathbf{x}_n) = \mathbf{c}_{n-1} \odot \varphi(\mathbf{h}_{n-1},
\mathbf{x}_n)$. For brevity we define $\varphi_n = \varphi(\mathbf{h}_{n-1},
\mathbf{x}_n)$.

Like the RNN case (Eq.~\ref{eqn_rnn_vanishinggradient} and
Eq.~\ref{eqn_rnn_vanishinggradientx}), we will need to find $\partial
\mathbf{c}_n/\partial \mathbf{c}_{n-1}$ to calculate $\nabla \mathcal{L}$.
Therefore, 
\begingroup
\addtolength{\jot}{0.5em}
\begin{align*}
    \frac{\partial \mathbf{c}_n}{\partial \mathbf{c}_{n-1}} &= \frac{\partial f(\mathbf{c}_{n-1}, \mathbf{h}_{n-1}, \mathbf{x}_n)}{\partial \mathbf{c}_{n-1}} + \cancelto{0}{\frac{\partial g(\mathbf{h}_{n-1}, \mathbf{x}_n)}{\partial \mathbf{c}_{n-1}}},\\
                                          &= \frac{\partial \mathbf{c}_{n-1} \odot \varphi_n}{\partial \mathbf{c}_{n-1}},\\
                                          &= \mathbf{c}_{n-1}\,\cancelto{0}{\frac{\partial \varphi_n}{\partial \mathbf{c}_{n-1}}} + \cancelto{1}{\frac{\partial \mathbf{c}_{n-1}}{\partial \mathbf{c}_{n-1}}}\,\varphi_n,\\
                                          &= \varphi_n.
\end{align*}
\endgroup
Thus, if we want to backpropagate to a cell state deep in the network we must calculate
\begin{equation}
    \frac{\partial \mathbf{c}_n}{\partial \mathbf{c}_N} = 
    \prod_{i=1}^{n - N} \varphi_i, \quad n > N.
    \label{eqn_lstm_backprop}
\end{equation}
The product term above does not depend on the derivative of a
saturating activation function, and so does not automatically vanish as $N$
goes to $\infty$. This means that a gradient signal can be carried through the
LSTM cell state without losing amplitude and vanishing\footnote{
    Which is great in theory. In practice, LSTMs still have trouble learning
    very long range dependencies due to their reliance on recurrent processing
    \citep{ref_seq2seq}. Transformer networks \citep{ref_aiayn} are an
    architecture that uses the concept of attention to address this issue. We
    will discuss transformer networks in \S\ref{sec_transformers}.
}.

\begin{figure}[htbp]
    \centering
    \begin{subfigure}[b]{0.35\textwidth}
        \centering
        \includegraphics[width=\textwidth]{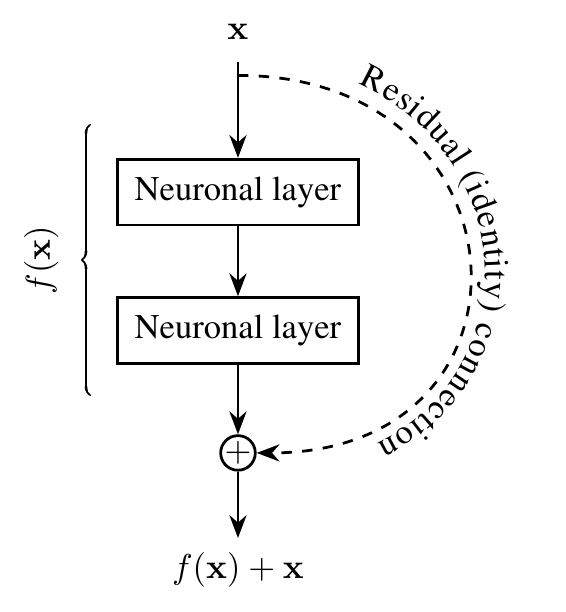}
        \caption[The residual connection]{A single residual connection is
        applied within a neural network.}
        \label{fig_residual_connection}
    \end{subfigure}
    \hfill
    \begin{subfigure}[b]{0.60\textwidth}
        \centering
        \includegraphics[width=\textwidth]{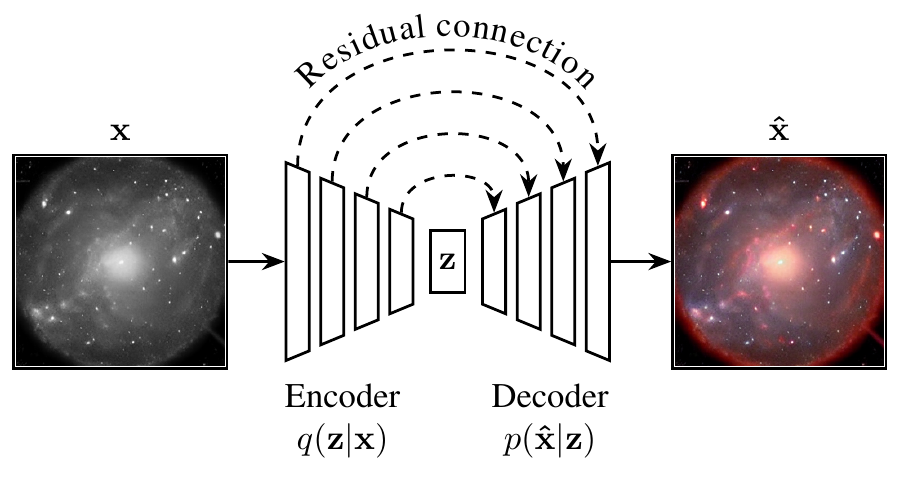}
        \caption[The U-Net]{The U-Net, a network that was originally developed to
        segment biological imagery uses the residual connection.}
        \label{fig_unet}
    \end{subfigure}
    \caption[The residual connection]{The left subfigure shows the residual
    connection as originally introduced in \citet{ref_resnet}. The right
    subfigure shows an application of the residual connection to an
    autoencoder like architecture \citep{ref_unet}, in this case
    colourising an astronomical object. Here, $\mathbf{z}$
    is a compressed shared representation of $\mathbf{x}$ and
    $\mathbf{\hat{x}}$.}
    \label{fig_residual}
\end{figure}

We can use a technique derived from the LSTM to solve our vanishing gradient
problem for deep feedforward neural networks (as studied in \S\ref{sec_MLP}).
\citet{ref_highway} do this by applying the concept of the LSTM's
cell state to their deep convolutional `highway network'. The highway network
uses gated connections to modulate the gradient flow back through neuronal
layers.  Later work by \citet{ref_resnet} introduces the residual network (ResNet) by taking
a highway network and simplifying its connections. They apply an elementwise
addition (or `residual connection') in place of the highway network's gated
connection (Fig.~\ref{fig_residual_connection}).  One can go even further with
residual connections, as \citet{ref_unet} demonstrate with their U-Net model.
The U-Net combines residual connections with an autoencoder-like architecture
(Fig.~\ref{fig_unet}). The U-Net has gone on to become the \emph{de facto} network
for many tasks that require an input and output of the same size (such as
segmentation, colourisation, and style transfer).

\subsection{Translation, attention, and transformers} \label{sec_transformers}

\begin{figure}[htbp]
    \centering
    \includegraphics[width=\textwidth]{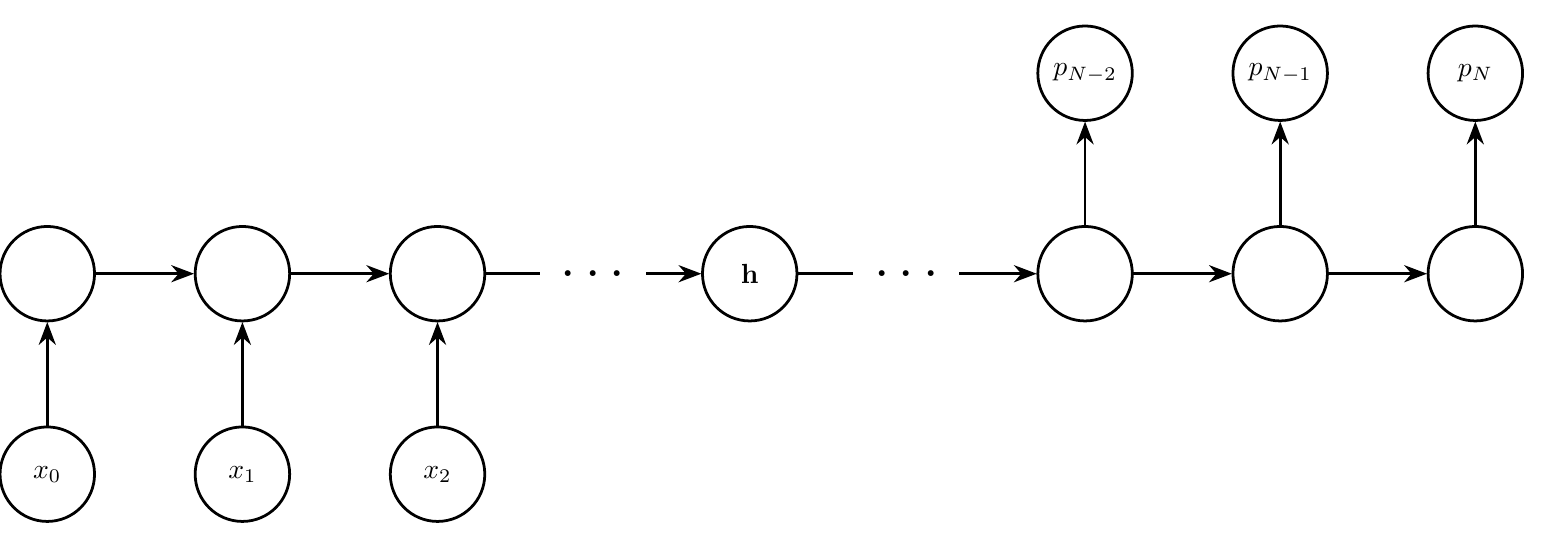}
    \caption[Neural machine translation via RNNs]{A sequence to sequence
    \citep[Seq2Seq;][]{ref_seq2seq} model. A sequence $\mathbf{x}$ is input into
    a GRNN. The final hidden state ($\mathbf{h}$) of the input network is then
    passed into a second GRNN. The second GRNN then unrolls to predict an
    output sequence $\mathbf{p}$. Due to the hidden state acting as a intermediary,
    $\mathbf{x}$ and $\mathbf{p}$ need not be of equal length. 
    }
    \label{fig_seq2seq}
\end{figure}

Theoretically, gated RNNs (GRNNs) such as the LSTM can learn very long
range dependencies (see Eq.~\ref{eqn_lstm_backprop} and its accompanying text).
In practice, GRNNs tend to forget information about
distant inputs.  This is because the GRNN lacks unmediated access to
inputs beyond the immediate antecedent as a consequence of its recurrent
architecture. The problem is especially apparent in neural machine translation
tasks that require knowledge of an entire sequence to produce an output, such
as language to language translation.  Fig.~\ref{fig_seq2seq} shows such a
sequence to sequence \citep[Seq2Seq;][]{ref_seq2seq} model. Seq2Seq
translates between two sets of sequential data by sharing a hidden state
between two GRNN units.  In Fig.~\ref{fig_seq2seq} we can see that the shared
information is bottlenecked by the hidden state.  Therefore, to resolve the
GRNN `forgetting problem' we must find a way to avoid any recursion, or serial
processing of input and output.  We can do this by providing the neural
network access to all input while it is calculating an output.
This was the primary motivation behind the transformer architecture
\citep{ref_bahdanau2014,ref_aiayn}.

\begin{figure}[htbp]
    \centering
    \includegraphics[width=\textwidth]{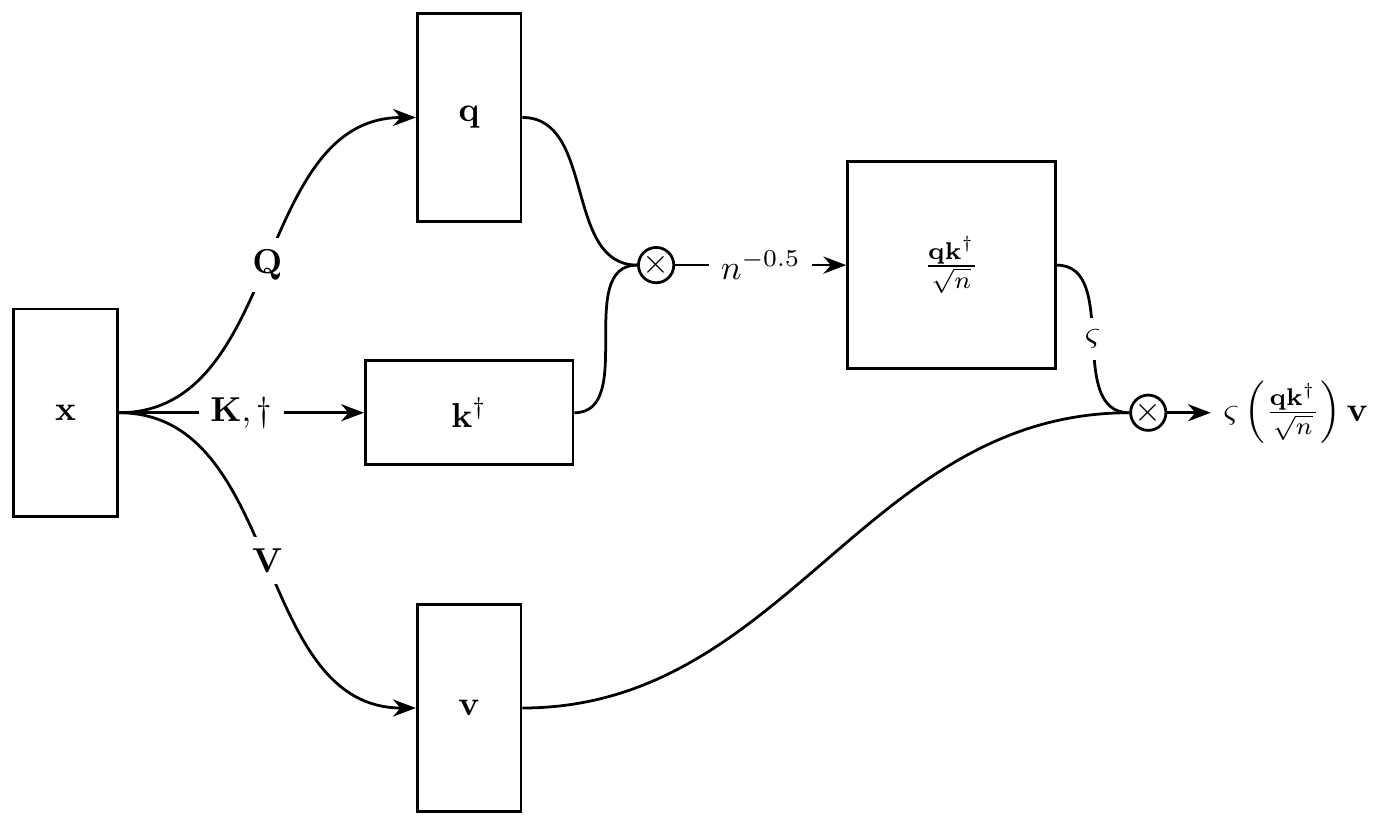}
    \caption[The self-attention mechanism]{An input $(\mathbf{x})$ is fed into a self-attention
    mechanism. The weights used to produce the query $(\mathbf{q})$, key $(\mathbf{k})$, and value
    $(\mathbf{v})$ matrices are learnt via backpropagation. Here the learnt weights
    are denoted as the capitalised versions of their child matrices.
    $\mathbf{q}$ and $\mathbf{k}$ are normalised and multiplied together, and a
    softmax nonlinearity ($\varsigma$) is applied. Finally, $\mathbf{v}$ is
    multiplied with output of the upper path and the final output is fed
    forward to the next neuronal layer. $\bigotimes$ denotes a matrix
    multiplication.}
    \label{fig_attention}
\end{figure}

Modern transformer architectures consist of a series of self-attention layers
interspersed with other layer types\footnote{
    In the original transformer formulation described in \citet{ref_aiayn},
    the network consisted of a connected `encoder' and `decoder' section much
    like a Seq2Seq model (Fig.~\ref{fig_seq2seq}). Later work has found this to
    be an unnecessary complication. For example, the generative pretrained
    transformer (GPT) 2 and 3 models
    \citep{ref_radford2019gpt2,ref_brown2020gpt3} consist of only decoder
    layers, and the bidirectional encoder representations from transformers
    (BERT) model consists of only encoder layers \citep{ref_bert}.
}.
Self-attention as described in \citet{ref_aiayn} is shown in
Fig.~\ref{fig_attention}. Intuitively, it captures the relationships between
quanta within a data input. To perform self-attention we first take an input
sequence 
\begin{equation*}
    \mathbf{x}
    =
    \begin{bmatrix}
        x_{1} & x_{2} & \cdots & x_{n} \\
    \end{bmatrix},
\end{equation*}
where $\mathbf{x}$ can be any sequence, such as a sentence, a variable star's time
series, or an unravelled galaxy image\footnote{
    One can go very general with this, as DeepMind demonstrated with their `Gato'
    transformer model \citep{ref_reed2022gato}. Gato can predict sequences
    for myriad tasks, from operating a physical robotic arm, to completing
    natural language sentences, to playing Atari games.
}. 
This sequence has a maximum length ($n$) that must be defined at
train time, but we can process shorter sequences by masking out any surplus
values so that they do not affect the loss.
Here we will follow the literature and refer to $[x_1, \ldots, x_n]$ as tokens.
As we can see in Fig.~\ref{fig_attention} the input is passed through a
trainable pair of weight matrices $\mathbf{Q}$ (or `query') and $\mathbf{K}$
(or `key'). The output matrices $\mathbf{q}$ and $\mathbf{k}^\dagger$ are then
multiplied together to yield
\begin{equation}
    (\mathbf{Q}\cdot\mathbf{x})(\mathbf{K}\cdot\mathbf{x})^\dagger
    = 
    \mathbf{q} \mathbf{k}^\dagger
    =
    \begin{bmatrix}
        Q_1 x_1 K_1 x_1  &  Q_1 x_1 K_2 x_2  &  \cdots  &  Q_1 x_1 K_n x_n \\
        Q_2 x_2 K_1 x_1  &  Q_2 x_2 K_2 x_2  &  \cdots  &  Q_2 x_2 K_n x_n \\
        \vdots  &  \vdots  &  \ddots  &  \vdots \\
        Q_n x_n K_1 x_1  &  Q_n x_n K_2 x_2  & \cdots  &  Q_n x_n K_n x_n
    \end{bmatrix}.
    \label{eqn_qkdagger}
\end{equation}
We can see that Eq.~\ref{eqn_qkdagger} describes the relationships between
tokens within $\mathbf{x}$. For example, if $x_1$ is similar semantically
to $x_2$, we would expect $Q_1 x_1 K_2 x_2$ and $Q_2 x_2 K_1 x_1$ to have a
high value.  We then normalise $\mathbf{q}\mathbf{k}^\dagger$ to mitigate
vanishing gradients\footnote{See Footnote~\ref{foot_explodinggradient}.},
and apply a softmax non-linearity so that the maximum weighting (or similarity)
is one, and the similarity values sum to unity.

Meanwhile, the input sequence $\mathbf{x}$ is passed through the neuronal layer
$\mathbf{V}$, resulting in a weighted representation $\mathbf{v}$:
\begin{equation*}
    \mathbf{V}\cdot\mathbf{x}
    =
    \mathbf{v}
    =
    \begin{bmatrix}
        V_1 x_1 &  V_2 x_2  &  \cdots  &  V_n x_n \\
    \end{bmatrix}.
\end{equation*}
$\mathbf{v}$ is multiplied with the similarity matrix 
$\varsigma (\mathbf{q}\mathbf{k}^\dagger/\sqrt{n})$. This process weighs similar
tokens within the sequence higher, increasing their relative importance in
later neuronal layers.

We will use an astronomical example to solidify our understanding
of the self-attention mechanism. Let us assume that our self-attention
mechanism is attending to a natural language caption describing a galaxy's
morphology that has been provided by a citizen scientist. The caption could be
something like:
\SetBlockThreshold{0}
\blockquote[]{\small $\mathbf{x} = $ `\emph{A barred galaxy with five spiral arms}',}
with each word acting as a separate token.
\begin{figure}[htbp]
    \centering
    \includegraphics{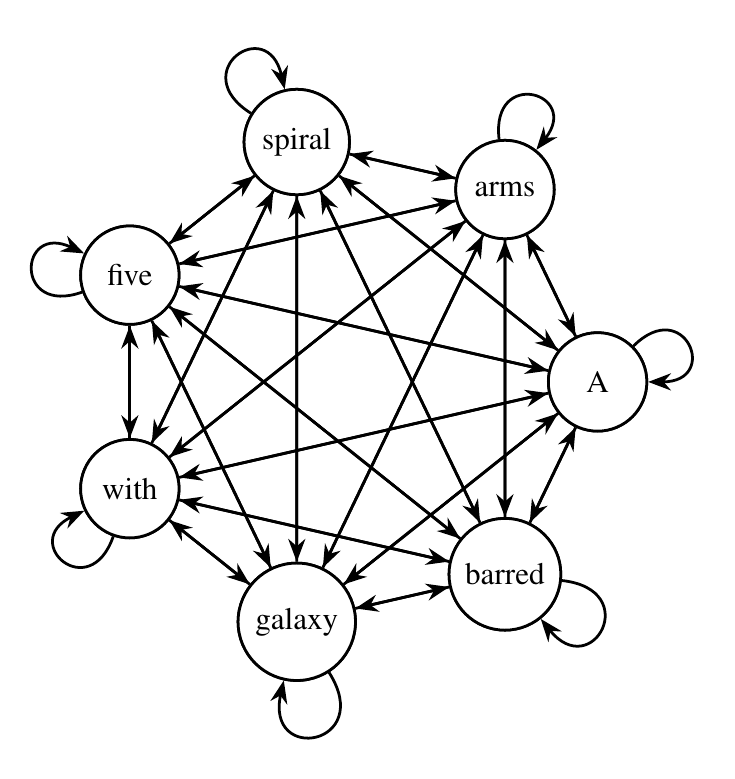}
    \caption{We can think of $\mathbf{q}\mathbf{k}^\dagger$ within
    self-attention as a graph of relationships between a prompt and itself. Each of the edges in this graph represents
    the weight shared between a pair of tokens in the input sequence.}
    \label{fig_selfattentiongraph}
\end{figure}
Let us imagine that we put this prompt into our self-attention mechanism:
\begin{equation*}
    (\mathbf{Q}\cdot\mathbf{x})(\mathbf{K}\cdot\mathbf{x})^\dagger = \mathbf{q} \mathbf{k}^\dagger =
    \bbordermatrix{
        & \text{A} & \text{barred} & \text{galaxy} & \text{with} & \text{five} & \text{spiral} & \text{arms} \cr
        \text{A}      & 0.7 & 0.2 & 0.1 & 0.0 & 0.0 & 0.0 & 0.0 \cr
        \text{barred} & 0.2 & 0.5 & 0.3 & 0.0 & 0.0 & 0.0 & 0.0 \cr
        \text{galaxy} & 0.1 & 0.3 & 0.5 & 0.1 & 0.0 & 0.0 & 0.0 \cr
        \text{with}   & 0.0 & 0.0 & 0.1 & 0.8 & 0.1 & 0.0 & 0.0 \cr
        \text{five}   & 0.0 & 0.0 & 0.0 & 0.1 & 0.6 & 0.3 & 0.0 \cr
        \text{spiral} & 0.0 & 0.0 & 0.0 & 0.0 & 0.3 & 0.5 & 0.2 \cr
        \text{arms}   & 0.0 & 0.0 & 0.0 & 0.0 & 0.0 & 0.2 & 0.8 
        }.
\end{equation*}
We can see that in the above matrix higher values have been assigned to pairs
of words that are more closely related within the sentence. For example, the
weight between `\emph{barred}' and `\emph{galaxy}' is relatively high (0.3), as
the term `\emph{barred}' describes a feature of galaxy.  Similarly, the weight
between `\emph{five}' and `\emph{spiral}' is also high (0.3), as these words together
define the number of spiral arms in the galaxy. Conversely, lower weights have
been assigned to word pairs that are less related, such as `\emph{A}' and
`\emph{with}' (0.0). As shown in Fig.~\ref{fig_selfattentiongraph}, one can
think of these relationships between tokens within our sequence as a learnt
mathematical graph\footnote{
    This view demonstrates that transformers can be thought of as a class of
    graph neural network---a network that is tasked with learning the
    relationships between nodes in a graph.  
    One can also approach this task with a feed-forward neural network
    \citep[\S\ref{sec_MLP};][]{ref_gori2005}, convolutional architecture
    \citep[\S\ref{sec_CNN};][]{ref_bruna2013,ref_kipf2017}, or with a
    recurrent architecture \citep[\S\ref{sec_LSTM};][]{ref_li2015}.
}.
Now that we have calculated $\mathbf{q}\mathbf{k}^\dagger$, we can use this
matrix to weigh our example sentence as shown in Fig.~\ref{fig_attention}.
This weighting gives the subsequent layers in our neural network an awareness
of the relationships between the tokens in our sequence.

\section{Astronomy's second wave of connectionism} \label{sec_cnnrnn_apps}

Compared to classical connectionist approaches\footnote{This
includes most MLP applications in astronomy, see \S\ref{sec_mlpinastro}.}
deep learning as outlined in \S\ref{sec_condeeplearn} does not require an
extraction of emergent parameters to train its models.  CNNs in particular are
well suited to observing raw information within image-based data.  Likewise,
RNNs are well suited to observing the full raw information within a time
series.  Astronomy is rich with both types of data, and in this section we will
review the history of the application of CNN, RNN, and transformer models to
astronomical data.

\subsection{Convolutional neural network applications}

It did not take long after \citet{ref_alexnet} established CNNs as the
\emph{de facto} image classification network for astronomers to take notice: in
\citeyear{ref_zhu2014} they were applied in the search for pulsars
\citep{ref_zhu2014} as part of an ensemble of methods.
\citet{ref_zhu2014} found that their ensemble was highly effective, with
100\% of their test set pulsar candidates being ranked within the top 961 of
the 90\,008 test candidates.  Shortly after, \citet{ref_hala2014} described
the use of one dimensional CNNs for a ternary classification problem. They find
that their model is capable of classifying 1D spectra into quasars, galaxies,
and stars to an impressive accuracy.  CNNs have been also been extensively used
in galaxy morphological classification.  First on the scene was
\citet{ref_dieleman2015}. They used CNNs to classify galaxy morphology
parameters as defined in the Galaxy Zoo dataset \citep{ref_galzoo} from
galaxy imagery. They observed their galaxies via the SDSS, and found a 99\%
consensus between the Galaxy Zoo labels, and the CNN classifications.
\citet{ref_huertas2015} showed that \citeauthor{ref_dieleman2015}'s CNN is
equally applicable to morphological classification of galaxies in the CANDELS
fields \citep{ref_koekemoer2011}. Likewise, \citet{ref_aniyan2017} showed that CNNs are capable of
classifying radio galaxies. The combined work of \citet{ref_dieleman2015},
\citet{ref_huertas2015}, and \citet{ref_aniyan2017} confirms that CNNs
are equally applicable to visually dissimilar surveys, with little-to-no
modification. Looking a little further afield, \citet{ref_wilde2022} used a
deep CNN model to classify simulated lensing events. They also applied some
interpretability techniques to their data,
using occlusion mapping \citep{ref_zeiler2014}, gradient class activation
mapping \citep{ref_selvaraju2016}, and Google's DeepDream to
prove that the CNN was indeed classifying via observing the gravitational
lenses. 
Alternative CNN models have also been used, such as the U-Net
(Fig.~\ref{fig_unet}). The U-Net was initially developed to segment biological
imagery \citep{ref_unet}. Its first use in astronomy was related:
\citet{ref_akeret2017} use a U-Net \citep{ref_unet} CNN to isolate via
segmentation, and ultimately remove, radio frequency interference from radio
telescope data. Likewise, \citet{ref_berger2019} used a three dimensional
U-Net \citep[V-Net;][]{ref_milletari2016} to predict and segment out galaxy
dark matter haloes in simulations, and \citet{ref_aragon2019} used a V-Net
to segment out the cosmological filaments and walls that make up the large
scale structure of the Universe.
\citet{ref_hausen2020} demonstrate that a U-Net is capable of performing
pixelwise semantic classification of objects in {\it HST}/CANDELS imagery, thus
proving that U-Nets are capable of useful work directly within large imaging
surveys, particularly in the deblending of overlapping objects, which is a
perennial challenge in deep imaging.
The U-Net in \citet{ref_lauritsen2021} is used
to superresolve simulated submillimetre observations. They found that the
U-Net could successfully do this when using a loss comprising of the L1 loss,
and a custom loss that measures the distance between predicted and ground
truth point sources.
\citet{ref_choma2018} were the first to demonstrate that graph convolutional
neural networks (GCNNs) are useful within astronomical context. They showed
that their 3D GCNN could classify signals from the IceCube neutrino
observatory, and found that it outperformed both a classical method, and a
standard 3D CNN.
\citet{ref_villanueva2021,ref_villanueva2022} demonstrated that EdgeNet---a
class of GCNN---can estimate halo masses when given the positions, velocities,
stellar masses, and radii of the host galaxies \citep{ref_wang2018graph}. The
authors also demonstrated that EdgeNet can estimate the halo masses of both
Andromeda and the Milky Way.
We must conclude from the studies described in this subsection that CNNs
are effective classifiers and regressors of image-based astronomical
data.

\subsection{Recurrent neural network applications}

RNNs were first applied in astronomy very close to home;
\citet{ref_aussem1994} predicted atmospheric seeing for observations from
ESO's Very Large Telescope, and the prediction of geomagnetic storms given data
on the solar wind was also explored in the mid-to-late 1990s and early 2000s
(\cite{ref_wu1996}, \cite{ref_lundstedt2002}, and other work from the same
group; \cite{ref_vassiliadis2000}). 

The first use of RNNs for classification in astronomy was carried out in a
prescient study by \citet{ref_brodrick2004}. They describe the use of an
RNN-like Elman network \citep{ref_elman1990}. Their RNN was tasked with the
search for artificially generated narrowband radio signals that resemble those
that may be produced by an extraterrestrial civilisation. They found that their
model had a test set accuracy of 92\%, suggesting that RNNs could be a useful
tool in the search for extraterrestrial intelligence.  More than a decade after
\citet{ref_brodrick2004}, \citet{ref_charnock2017} used an LSTM
(Fig.~\ref{fig_LSTM}) to classify simulated supernovae. They describe two
classification problems. One, a binary classification between type-Ia and non
type-Ia supernovae, and the other a classification between supernovae types
I, II, and III. For their best performing model they report an accuracy of more
than 95\% for their binary classification problem, and an accuracy of over 90\%
for their trinary classification. This study cemented the usefulness of RNNs
for classification problems in astronomy. \citet{ref_charnock2017} was
followed by numerous projects studying the use of RNNs for classification of
time series astronomical data. A non-exhaustive list of modern RNN use in
astronomy includes: stochastically sampled variable star classification
\citep{ref_naul2018}; exoplanet instance segmentation
\citep{ref_gonzalez2018}; variable star/galaxy sequential imagery
classification \citep{ref_carrascodavis2019}; and gamma ray source
classification \citep{ref_finke2021}.  We must conclude from these studies
that RNNs are effective classifiers of astronomical time series, provided
that sufficient data is available.

Of course, recurrent networks are not limited to classification; they can also
be used for regression problems.  First, \citet{ref_weddell2008}
successfully used an echo state network \citep{ref_jaeger2004} to
predict the point spread function of a target object in a wide field of view.
\citet{ref_capizzi2012} used an RNN to inpaint missing NASA Kepler time
series data for stellar objects. They found that their model could
recreate the missing time series to an excellent accuracy, suggesting that the
RNN could internalise information about the star it was trained on.  As
in the classification case, research into the use of RNNs for regression
problems picked up massively in the late 2010s, and here we will highlight a
selection of these studies that represent the range of RNN use cases.
\citet{ref_shen2017} used both an LSTM and an autoencoder based RNN to
denoise gravitational wave data, and \citet{ref_morningstar2019} used a
recurrent inference machine to reconstruct gravitationally lensed galaxies.
\citet{ref_liu2019} used an LSTM to predict solar flare activity. From these
studies, similarly to the classification case above, we can once again conclude
that RNNs are effective regressors of astronomical time series.

RNNs have also been used in cases that are a little more unconventional.
For example, \citet{ref_kugler2016} used an autoencoding RNN (specifically
an echo state network) to extract representation embeddings of variable main sequence stars.
They find that these embeddings capture some emergent properties of these
variable stars, such as temperature, and surface gravity, suggesting that
clustering within the embedding space could result in semantically meaningful
variable star classification. We will revisit this line of research when we
explore representation learning within astronomy in detail in
\S\ref{sec_athirdera}. An example of more drastic cross-pollination between
ideas within deep learning and those within astronomy is \citet{ref_smith2021}.
They use an encoder-decoder network comprising of a CNN encoder and RNN decoder to
predict surface brightness profiles of galaxies. This class of neural
network was previously used extensively within natural language image
captioning, and by treating surface brightness profiles as `captions' their model was capable
of prediction over $100\times$ faster than the previous classical, human-agent based method. 

\subsection{Transformer applications}

Although initially used for natural language, transformers have also been
adapted for use in imagery, first by \citet{ref_parmar2018imtrans}, and also
in \citet{ref_dosovitskiy2020vit}. To our knowledge, transformers have not yet been applied to
astronomical imagery, but they have started to find use in time-series astronomy.
\citet{ref_donosoolivia2022} used BERT \citep{ref_bert} to generate a
representation space for light curves in a self-supervised manner. 
\citet{ref_morvan2022} use an encoding transformer to denoise light curves 
from the Transiting Exoplanet Survey Satellite \citep[TESS;][]{ref_ricker2015},
and show that the denoising surrogate task results in an expressive embedding space.
\citet{ref_pan2022} also use a transformer model to analyse light curves for
exoplanets.  Transformers have taken the fields of natural language processing
and computer vision by storm (\S\ref{ch_conclusions}), and so if we extrapolate
from trends in other fields we expect to see many more examples of transformers
applied to astronomical use cases in the near future.  We will revisit the
transformer architecture in the context of foundation models \citep[][and
references therein]{ref_bommasani2021}, and their possible future astronomical
applications in \S\ref{ch_conclusions}.

\subsection{A problem with supervised learning}

Supervised learning requires a high quality labelled dataset to train a neural
network.  In turn, these datasets require labourious human intervention to create, and
so supervised data is in short supply. 
One can avoid this issue by prompting the deep learning model to
gather semantic information from entirely unlabelled data. This learnt semantic
information can then be accessed through a hidden descriptive `latent space', and
then used for downstream tasks like data generation, classification, and
regression.
Indeed, all of the networks described previously in this review can be
repurposed for non-supervised tasks, and in \S\ref{sec_gen_modelling} and
\S\ref{sec_representation} we will explore some deep learning frameworks that
do not require supervision.

\section{Deep generative modelling} \label{sec_gen_modelling}

\begin{figure}[htb]
    \begin{subfigure}{0.5\textwidth}
        \centering
        \includegraphics[width=\textwidth]{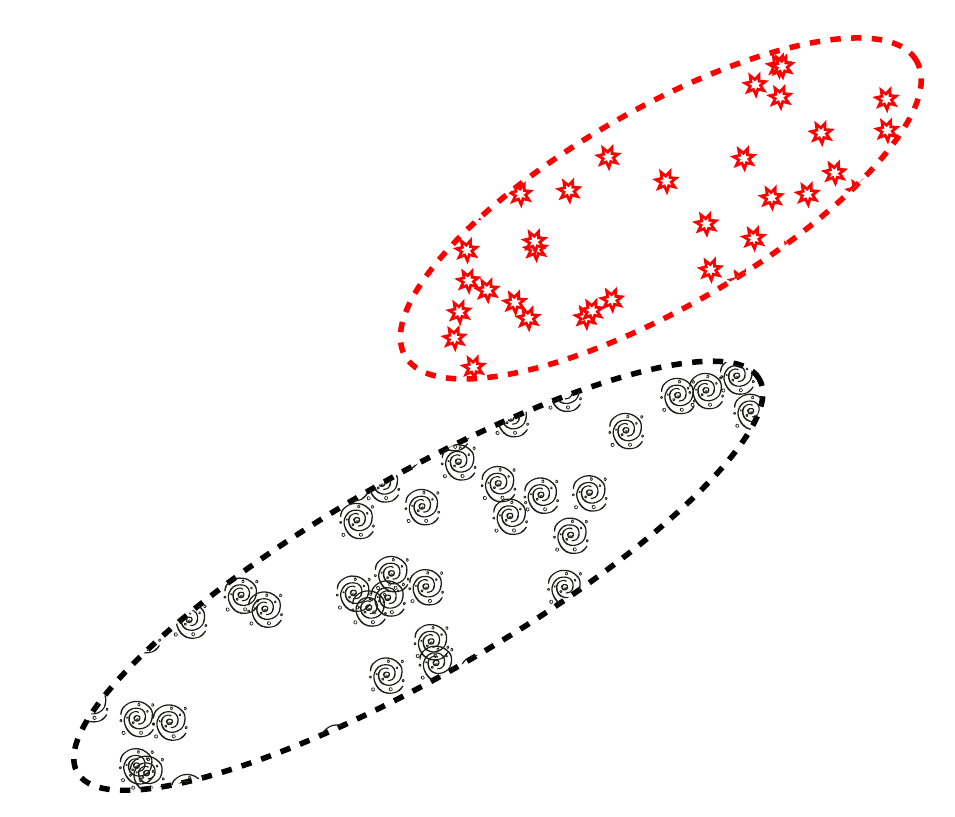}\hfill
    \end{subfigure}
    \begin{subfigure}{0.5\textwidth}
        \centering
        \includegraphics[width=\textwidth]{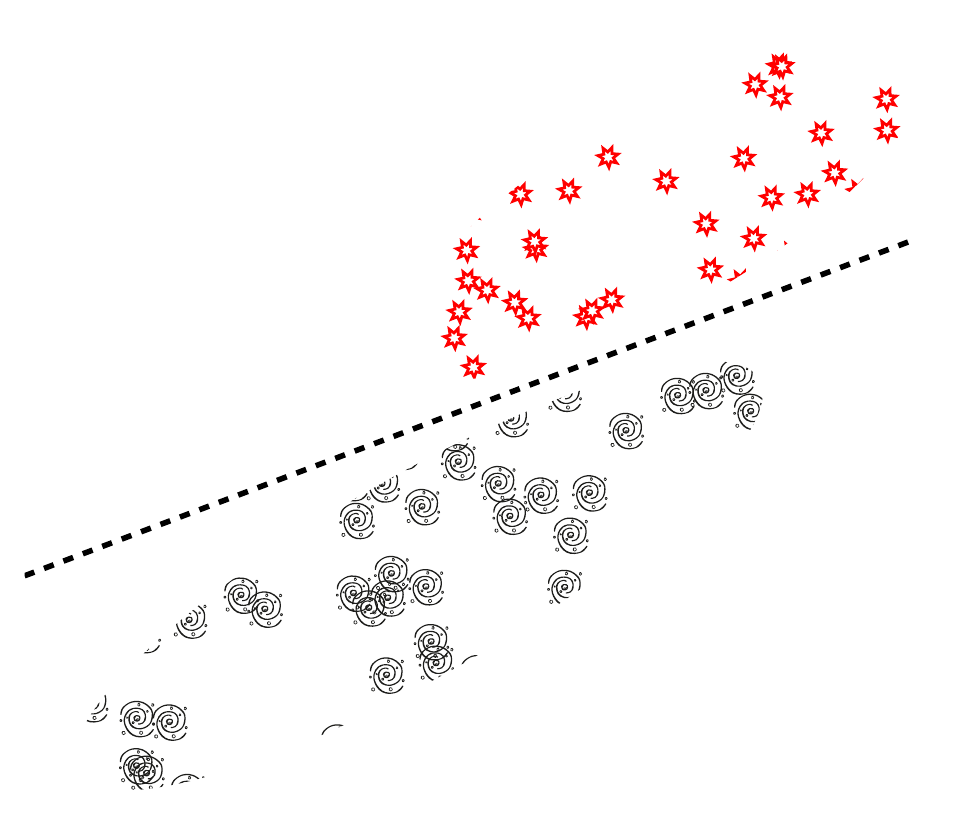}
    \end{subfigure}
    \caption[Discriminative and generative modelling]{Here we show 
    a possible latent space representation of a set of galaxies and a set of
    stars. A latent (or embedding) space is a compressed representation of a set
    of objects where similar objects are clustered closer together than dissimilar
    objects. While this space is often highly dimensional, here we project our
    latent space onto two dimensions for visualisation purposes.
    On the left we see a generative model attempting to learn the probability
    distributions of the latent representation of a dataset that contains a set
    of galaxies and a set of stars. On the right is a discriminative model,
    which is attempting to learn the boundary that separates the star and
    galaxy types.}
    \label{fig_generativemodelling}
\end{figure}

In this section we discuss generative modelling within the context of
astronomy. Unlike discriminative models, generative models explicitly learn the
distribution of classes in a dataset (Fig.~\ref{fig_generativemodelling}). Once
we learn the distribution of data, we can use that knowledge to generate new
synthetic data that resembles that found in the training dataset. 
In the following subsections we will explore in detail three popular forms of deep
generative model: the variational autoencoder (\S\ref{sec_VAE}); the generative
adversarial network (\S\ref{sec_GAN}); and the family of score-based (or
diffusion) models (\S\ref{sec_sbgm}). Finally, in \S\ref{sec_athirdera} we
discuss applications of deep generative modelling in astronomy.

\subsection{(Variational) autoencoders} \label{sec_VAE}

\begin{figure}[htbp]
    \centering
    \includegraphics{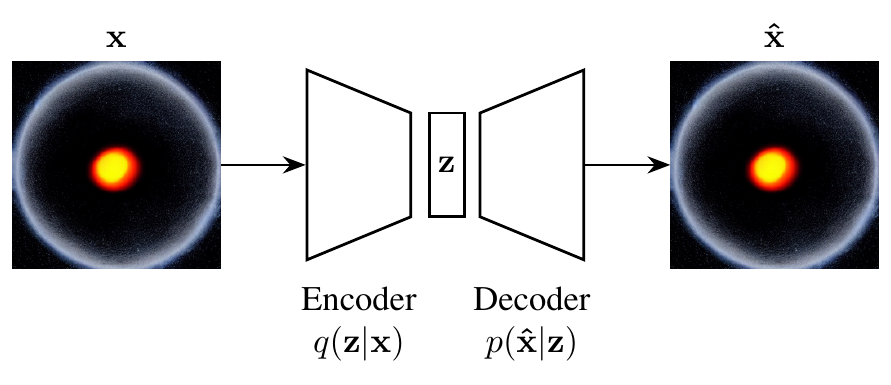}
    \caption[An autoencoder attends to an image of a black hole]{An autoencoder
    \citep{ref_rumelhart1986autoencoder} attends to an image of a black hole.
    $\mathbf{z}$ is a latent vector and $\mathbf{x}$ is a sample from a
    training set. The encoder, $q$ learns to encode the incoming data into a
    latent vector while the decoder $p$ takes as input $\mathbf{z}$ and attempts to
    recreate $\mathbf{x}$.}
    \label{fig_AE}
\end{figure}

Autoencoders have long been a neural network architectural staple.
In a sister paper to backpropagation's 
populariser, \citet{ref_rumelhart1986autoencoder} demonstrate backpropagation
within an autoencoder. Fig.~\ref{fig_AE} demonstrates the basic neural network
autoencoder architecture. An autoencoder is tasked with recreating some input
data, squeezing the input information ($\mathbf{x}$) into a bottleneck latent
vector ($\mathbf{z}$) via a neural network
$q(\mathbf{\mathbf{z}}\vert\mathbf{x})$. $\mathbf{z}$ is then expanded to an
imitation of the input data ($\mathbf{\hat{x}}$) by a second neural network
$p(\mathbf{\hat{x}}\vert\mathbf{z})$. The standard autoencoder is trained via a
reconstruction loss; $\mathcal{L}_R(\mathbf{x}, \mathbf{\hat{x}})$, where
$\mathcal{L}_R(\mathbf{x}, \mathbf{\hat{x}})$ measures the difference in
pixelspace between $\mathbf{x}$ and $\mathbf{\hat{x}}$.

\begin{figure}[htbp]
    \centering
    \includegraphics{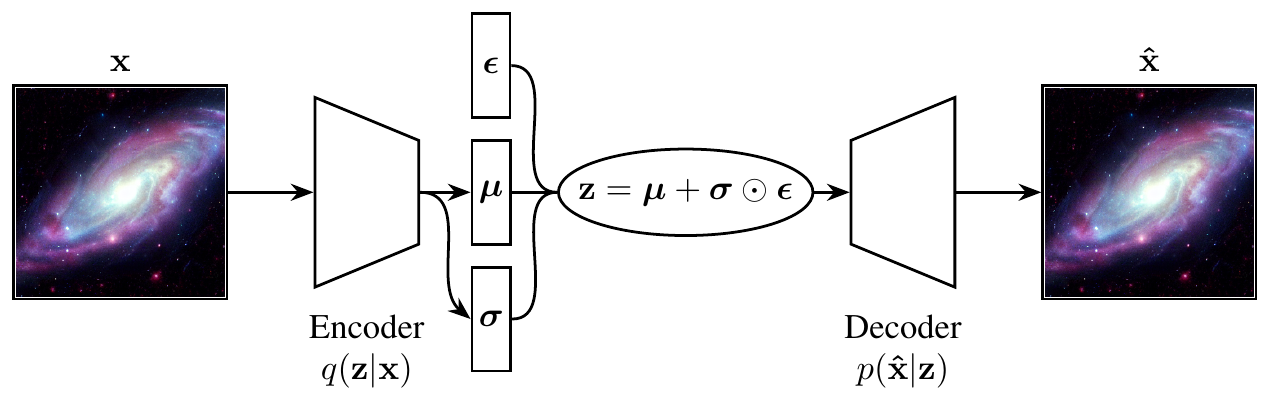}
    \caption[A variational autoencoder attends to an image of a spiral
    galaxy]{A variational autoencoder \citep{ref_vae} operates on a spiral
    galaxy. $\mathbf{z}$ is a latent vector and $\mathbf{x}$ is a sample from
    the training set. The encoder, $q$ learns to compress the incoming data
    into a latent vector that encodes the normal distribution. 
    The decoder $p$ takes as input $\mathbf{z}$ and
    attempts to recreate $\mathbf{x}$.}
    \label{fig_vae}
\end{figure}

Na\"ively, one would think that once trained, one could `just' sample a new
latent vector, and produce novel imagery via the decoding neural network
$p(\mathbf{\hat{x}}\vert\mathbf{z})$. We cannot do this, as autoencoders trained 
purely via a reconstruction loss have no incentive to produce a smoothly
interpolatable latent space. This means we can use a standard autoencoder to embed
and retrieve data contained in the training set, but cannot use one to generate
new data. To generate new data we require a smooth latent space, which
variational autoencoders (VAEs; Fig~\ref{fig_vae}) produce by design
\citep{ref_vae}.

A VAE differs from the standard autoencoder by enforcing a spread in each
training set samples' latent vector. We can see in Fig.~\ref{fig_vae} how this
is done; instead of directly predicting $\mathbf{z}$ the encoder $q$ predicts
two vectors, $\boldsymbol{\mu}$ and $\boldsymbol{\sigma}$.  $\mathbf{z}$ is
then sampled stochastically via the equation
\begin{equation}
    \mathbf{z} = \boldsymbol{\mu} + \boldsymbol{\sigma} \odot \boldsymbol{\epsilon},
\end{equation}
where $\odot$ is the Hadamard product, and $\boldsymbol{\epsilon}$ is noise
generated externally to the neural network graph\footnote{
    To avoid breaking the backpropagation chain the VAE injects noise
    via an external parameter, $\boldsymbol{\epsilon}$. This is described in
    \citet{ref_vae} as the `reparameterisation trick'.
}. This spread results in similar samples overlapping within the latent space,
and therefore we end up with a smooth latent space that we can interpolate
through. However, currently there is no incentive for the neural network to
provide a coherent, compact global structure in the latent space. For that we
require a regularisation term in the loss.  This regularisation is provided via
the Kullback-Leibler (KL) divergence, which is a measure of the difference
between two probability distributions. A standard VAE uses the KL divergence to
push the latent distribution towards the standard normal distribution,
incentivising a compact, continuous latent space. Hence, the final VAE loss
is a combination of the reconstruction loss and KL divergence:
\begin{equation}
    \mathcal{L}_{\text{VAE}} = \mathcal{L}_R (\mathbf{x}, \mathbf{\hat{x}}) + \text{KL}(q(\mathbf{x} | \mathbf{z}) \| \rho), \label{eqn_vae_prior}
\end{equation}
where $\rho$ is some prior. In a standard VAE $\rho =
\mathcal{N}(\mathbf{0},\mathds{1})$.

In practice, VAEs are able to generate smooth and coherent samples, as they
model the data distribution explicitly, which also means that we can perform
latent space arithmetic on the latent vector---such as interpolation,
reconstruction, and anonomly detection \citep{ref_vae}.  Their explicit
learning of the latent vector ($\mathbf{z}$) means that they can trivially be
repurposed for semi-supervised, self-supervised, and supervised downstream
tasks by manipulating $\mathbf{z}$ \citep{ref_regier2015deep,ref_spindler2020}.
However, the quality of samples generated by VAEs is lower than that of
generative adversarial networks or score-based generative models
\citep{ref_dosovitskiy2016}. This reduction in quality is due to the VAE's
simple posterior $q(\mathbf{z}|\mathbf{x})$, but one can
mitigate this shortcoming by iteratively approaching a more complex 
posterior\footnote{
    Interestingly, this iterative approximation is similar to the approach used
    in the training of score-based generative models and diffusion models
    \citep{ref_zhao2017}, and the similarities between the training methods of
    state-of-the-art in VAE models and SBGMs are striking. For example,
    the Vector-Quantised VAE, Very Deep VAE, and the Nouveau VAE all use a
    heirarchical architecture that iteratively injects latent codes that are
    used to produce finer and finer detail in the generated image
    \citep{ref_oord2017,ref_vahdat2020,ref_child2020}.
}.
To regularise the latent space, VAEs require an assumption of the prior
distribution which requires some knowledge of the dataset, although often this
is can be set as `just' a normal distribution as shown in
Eq.~\ref{eqn_vae_prior}.  

\subsection{Generative adversarial networks} \label{sec_GAN}

Generative adversarial networks \citep[GAN;][]{ref_gan} can be
thought of as a minimax game between two competing neural networks. If we
anthropomorphise we can gain an intuition for how a GAN learns: let us imagine
an art forger, and an art critic. The forger wants to paint paintings that
are similar to famous expensive works, and needs to fool the critic when selling
these paintings.  Meanwhile, the critic wants to ensure that no reproductions
are sold and so they need to accurately determine whether any painting is an
original or a reproduction.  At first, our forger is a poor painter, and so the
critic can easily identify our forger's works. However, the forger learns from
the critic's choices and produces more realistic paintings. As the forger's
paintings improve, the critic also learns better methods for detecting
forgeries. This minimax game incentivises the critic to keep improving their
classifications, and the forger to keep improving their painting.  If this
continues, we get to a point where the forger's works are indiscernible from
the real thing---the forger has learnt to perfectly mimic the dataset!  In
a GAN, we name the critic the discriminator ($D$), and we name the forger the
generator ($G$).

\begin{figure}[p]
    \begin{subfigure}[b]{\textwidth}
        \centering
        \includegraphics{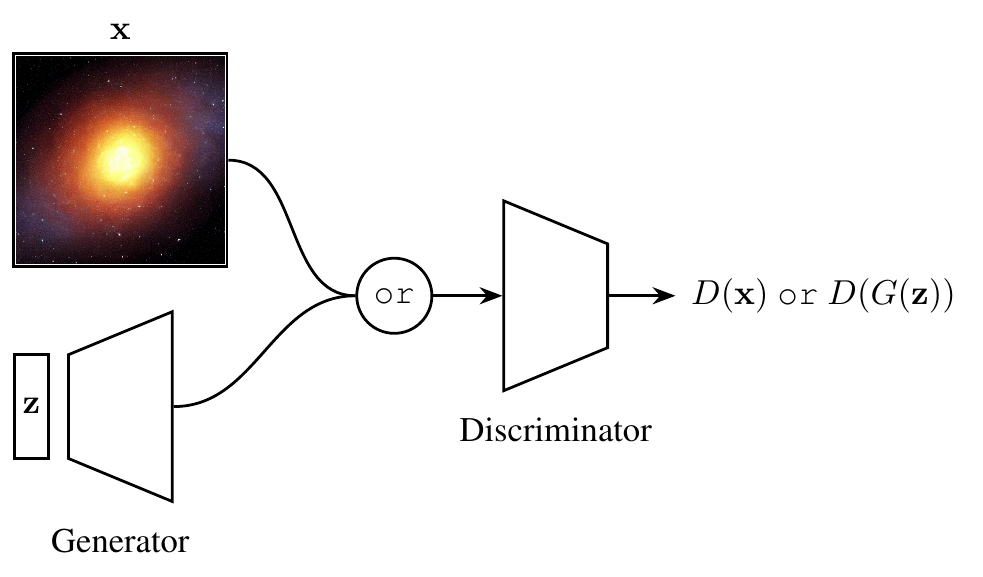}
        \caption[GAN]{A typical GAN according to
        \citet{ref_gan}. $\mathbf{z}$ is a noise vector, $\mathbf{x}$ is a
        sample from the training set. The discriminator learns to classify
        the incoming images as either fake or real, and the generator learns
        to fool the discriminator by producing realistic fakes.}
        \label{fig_GAN}
    \end{subfigure}
    \begin{subfigure}[b]{\textwidth}
        \centering
        \includegraphics[width=0.9\textwidth]{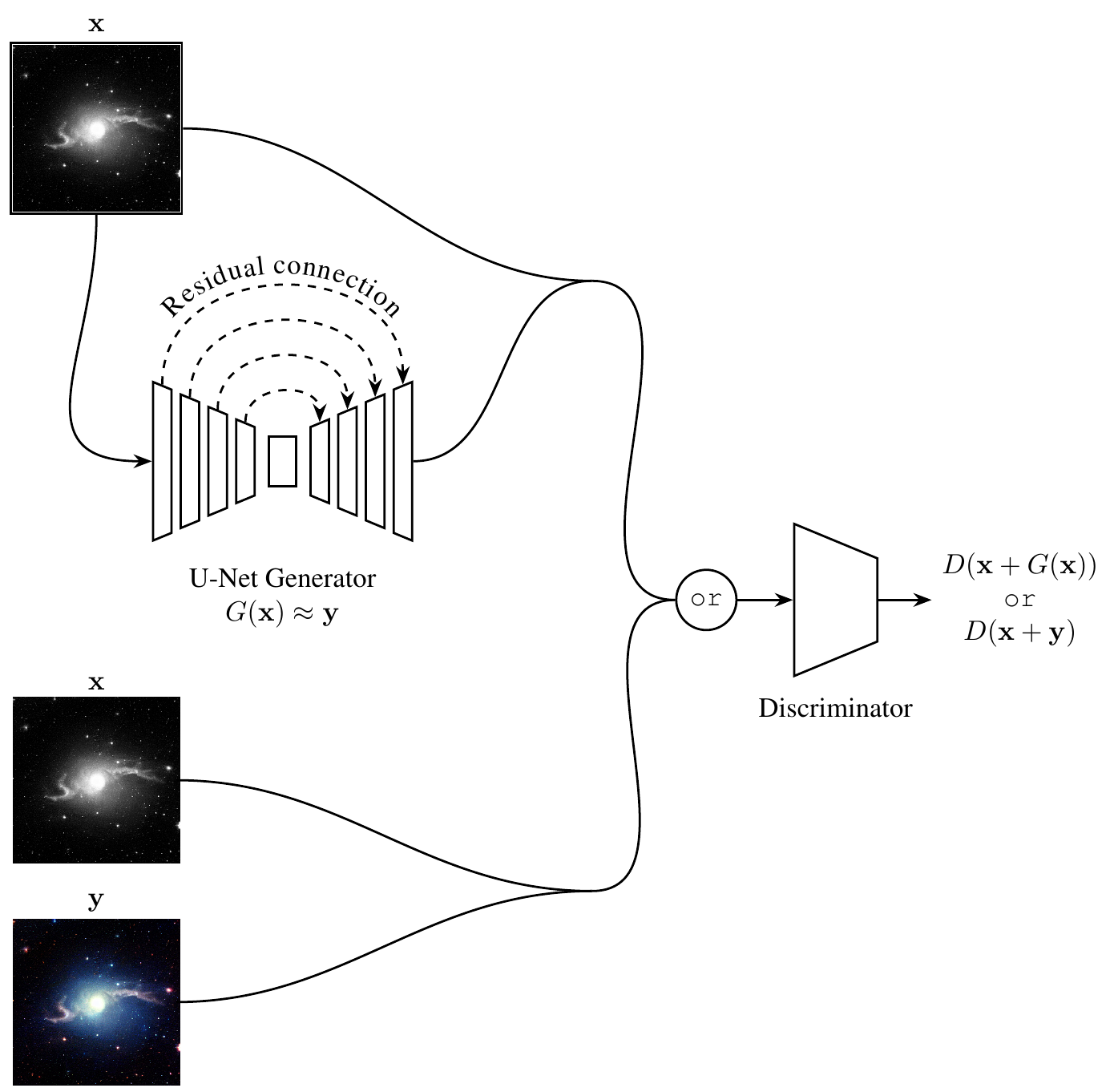}
        \caption[Pix2Pix]{A Pix2Pix-like model with a U-Net generator
        \citep{ref_pix2pix,ref_unet}.  The discriminator learns to classify the
        incoming image tuples as either fake or real.  Meanwhile, the generator
        learns to fool the discriminator by approximating the colourisation
        function mapping $\mathbf{x} \to \mathbf{y}$.  Line
        mergers denote channel-wise concatenations.}
        \label{fig_pix2pix}
    \end{subfigure}
    \caption{The GAN and Pix2Pix models.}
\end{figure}

In \citeauthor{ref_gan}'s original GAN formulation
(Fig.~\ref{fig_GAN}), $G$ and $D$ are neural networks (typically CNNs, although
other architectures can be used) that compete during training in a minimax game
where $G$ aims to maximise the probability of $D$
mispredicting that a generated datapoint is sampled from the real dataset. $G$
takes as input a randomly sampled latent vector $\textbf{z}$, and outputs a
synthetic datapoint $G(\textbf{z})$. $D$ takes either this synthetic datapoint,
or a real datapoint $\textbf{x}$, and outputs $D(G(\textbf{z}))$ or
$D(\textbf{x})$. This output is the probability that the datapoint is drawn
from the real dataset. 
To train the network we can write the GAN adversarial loss like so:
\begin{align*}
    \mathcal{L}_D &= -(\mathbb{E}_\textbf{x}[\log(D(\textbf{x}))] + \mathbb{E}_\textbf{z}[\log(1 - D(G(\textbf{z})))]),\\
    \mathcal{L}_G &= \mathbb{E}_\textbf{z}[\log(1 - D(G(\textbf{z})))],
\end{align*}
where here we attempt to minimise both $\mathcal{L}_D$ and $\mathcal{L}_G$. 
In practice we train the networks by alternating freezing the weights of $G$ and
backpropagating $\mathcal{L}_D$, and then freezing the weights of $D$ and
backpropagating $\mathcal{L}_G$ for each training batch. In this way the
networks' weights are updated to follow $\nabla_\mathbf{w}\mathcal{L}_G$ and
$\nabla_\mathbf{w}\mathcal{L}_D$ downwards until the distribution of
$G(\textbf{z})$ closely resembles that of the real dataset.  Once trained, $G$
can be used to generate entirely novel synthetic data that closely resembles
(but is not identical to) the training set data.

One can condition a GAN to guide the network towards a
desired output image \citep{ref_mirza2014}. To do this we alter the adversarial
loss so that it is conditioned on a label $\textbf{y}$:
\begin{align*}
    \mathcal{L}_D &= -(\mathbb{E}_{\mathbf{x}}[\log(D(\textbf{x} | \textbf{y}))] + \mathbb{E}_\mathbf{z}[\log(1 - D(G(\textbf{z} | \textbf{y})))]),\\
    \mathcal{L}_G &= \mathbb{E}_{\mathbf{z}}[\log(1 - D(G(\textbf{z} | \textbf{y})))].
\end{align*}
As an example, if we set $\mathbf{y}$ as the redshift of the galaxies in the
training set, we could use a conditional GAN to guide the network to generate
galaxies of a certain redshift. Furthermore, we are not restricted to
conditioning single values; GANs can also be conditioned on entire images. In
Fig.~\ref{fig_pix2pix} we see that the GAN adversarial loss can be used to
translate between image domains \citep{ref_pix2pix}. In
\citeauthor{ref_pix2pix}'s Pix2Pix model, the generator takes as input an image
$\mathbf{x}$, and attempts to produce a related image $\mathbf{y}$.  Meanwhile,
the discriminator attempts to discern whether the $(\mathbf{x},\mathbf{y})$
pair that it is given is sampled from the training set, or the generator.
Otherwise, Pix2Pix is trained in the same way as the standard GAN.

GANs are capable of generating high-quality, sharp, and realistic
samples \citep{ref_brock2018,ref_kang2023}.  They have long been a sweetheart
of the deep generative learning community, having been used for various
state-of-the-art applications, such as data embedding
\citep[e.g.][]{ref_cheng2016}, style transfer \citep[e.g.][]{ref_karras2018},
superresolution \citep[e.g.][]{ref_ledig2016}, and image inpaining and object
removal \citep[e.g.][]{ref_yu2018}.  Unfortunately however, GANs have some
downsides.  They are quite difficult to train; maintaining the balance between
the generator and discriminator networks is challenging and requires careful
finetuning \citep{ref_weng2019}. $G$ and $D$ must work in tandem and one cannot
overpower the other or learning will cease.  One of the most famous symptoms of
this imbalance is mode collapse, where $G$ only generates a limited variety of
samples that reliably fool $D$.  This instability during training makes it
quite a time-consuming task to find a stable network architecture if one is
designing a GAN themselves.  Finally, the GAN adversarial losses are relative
and so are not representative of the image quality.  This is not
the case for the VAE and SBGM families of models.  

\subsection{Score-based generative modelling and diffusion models} \label{sec_sbgm}

Diffusion models were introduced by
\citet{ref_sohldickstein2015} and were first shown to be capable of
producing high quality synthetic samples by
\citet{ref_ho2020}. Diffusion models are part of a family of generative deep learning
models that employ denoising score matching via annealed Langevin dynamic
sampling (first explored by \citet[][]{ref_hyvarinen2005,ref_vincent2011}.
More recent work can be found in
\citet[][]{ref_song2020,ref_ho2020,ref_ajm2020,ref_ajm2021,ref_song2021}).
This family of score-based generative models (SBGMs) can generate imagery of a
quality and diversity surpassing state of the art 
GAN models \citep{ref_gan}, a startling result
considering the historic disparity in interest and development between the two
techniques
\citep{ref_song2021,ref_nichol2021,ref_dhariwal2021,ref_ramesh2022dalle2}.
SBGMs can super-resolve images \citep{ref_kadkhodaie2020,ref_saharia2021},
translate between image domains \citep{ref_sasaki2021}, separate
superimposed images \citep{ref_jayaram2020}, and in-paint information
\citep{ref_kadkhodaie2020,ref_song2021}. 

\begin{figure}[htbp]
    \centering
    \includegraphics[width=\textwidth]{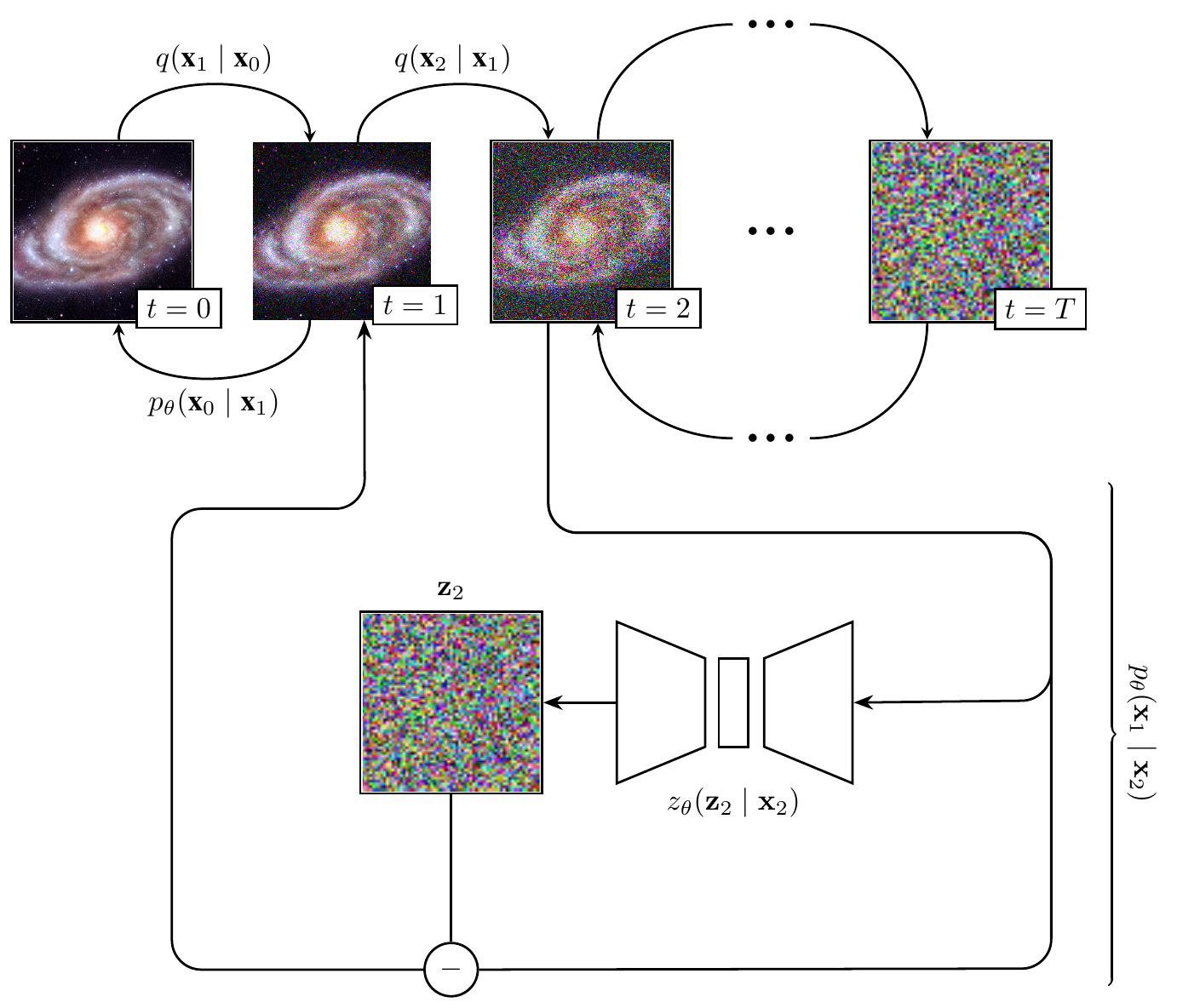}
    \caption[The Denoising Diffusion Probabalistic Model]{It is easy (and
    achievable without learnt parameters) to add noise to an image, but more
    difficult to remove it. Diffusion models attempt to learn an iterative
    removal process via training an appropriate neural network, 
    $p_\theta (\mathbf{x}_{t-1} \mid \mathbf{x}_t)$.}
    \label{fig_ddpm}
\end{figure}

Diffusion models define a diffusion process that projects a complex image domain space
onto a simple domain space.  In the original formulation, this diffusion
process is fixed to a predefined Markov chain $q(\mathbf{x}_t \mid
\mathbf{x}_{t-1})$ that adds a small amount of Gaussian noise with each step.
As Fig.~\ref{fig_ddpm} shows, this `simple domain space' can be noise sampled
from a Gaussian distribution $\mathbf{x}_T \sim
\mathcal{N}(\mathbf{0},\mathds{1})$.

\subsubsection{Forward process}

To slowly add Gaussian noise to our data we define a Markov chain
\begin{equation*}
    q(\mathbf{x}_{0 \ldots T}) = q(\mathbf{x}_0)\prod^T_{t=1} q(\mathbf{x}_t \mid \mathbf{x}_{t-1}),
\end{equation*}
where $\mathbf{x}_0$ is an image sampled from the training set.
The amount of noise added per step is controlled with a variance schedule
$\{\beta_t \in (0, 1)\}^T_{t=1}$:
\begin{equation}
    q(\mathbf{x}_t \mid \mathbf{x}_{t-1}) = \mathcal{N}(\mathbf{x}_t; \sqrt{1 - \beta_t}\,\mathbf{x}_{t-1},\,\beta_t\mathds{1}).
    \label{eqn_forwardbeta}
\end{equation}
This process is applied incrementally to the input image. Since we can define the 
above equation such that it only depends on $\mathbf{x}_0$ we can immediately
calculate an image representation $\mathbf{x}_t$ for any $t$
\citep{ref_ho2020}. If we define $\alpha_t = 1 - \beta_t$ and
$\bar{\alpha}_t = \prod^t_{i=1} \alpha_i$:
\begin{align}
    \mathbf{x}_t &= \sqrt{\alpha_t}\,\mathbf{x}_{t-1} + \sqrt{1 - \alpha_t}\,\mathbf{z}_{t-1}\notag\\
                 &= \sqrt{\alpha_t \alpha_{t-1}}\,\mathbf{x}_{t-2} + \sqrt{(1 - \alpha_t) + \alpha_{t} (1 - \alpha_{t-1})}\,\bar{\mathbf{z}}_{t-2}\notag\\
                 &= \sqrt{\alpha_t \alpha_{t-1} \alpha_{t-2}}\,\mathbf{x}_{t-3} + \sqrt{(1 - \alpha_t \alpha_{t-1}) + \alpha_{t} \alpha_{t-1} (1 - \alpha_{t-2})}\,\bar{\mathbf{z}}_{t-3}\notag\\
                 &= \ldots\notag\\
                 &= \sqrt{\bar{\alpha_t}} \mathbf{x}_{0} + \sqrt{1 - \bar{\alpha}_{t}}\mathbf{z},
    \label{eqn_niceproperty}
\end{align}
where $\mathbf{z}_t \sim \mathcal{N}(\mathbf{0}, \mathds{1})$ and
$\bar{\mathbf{z}}$ is a combination of Gaussians.  Plugging the above
expression into Eq.~\ref{eqn_forwardbeta} removes the $\mathbf{x}_{t-1}$
dependency and yields
\begin{equation}
    q(\mathbf{x}_t \mid \mathbf{x}_0) = \mathcal{N}(\mathbf{x}_t; \sqrt{\bar{\alpha}_t}\,\mathbf{x}_0,\,(1-\bar{\alpha}_t)\mathds{1}).
    \label{eqn_ddpm_practice}
\end{equation}

\subsubsection{Reverse process}

Diffusion models attempt to reverse the forward process by applying a Markov chain with
learnt Gaussian transitions. These transitions can be learnt via an
appropriate neural network, $p_\theta$:
\begin{align*}
    p_\theta(\mathbf{x}_{0 \ldots T}) &= p(\mathbf{x}_T)\prod^T_{t=1} p_\theta(\mathbf{x}_{t-1} \mid \mathbf{x}_t),\\
    p_\theta(\mathbf{x}_{t - 1} \mid \mathbf{x}_t) &= \mathcal{N}(\mathbf{x}_{t - 1}; \boldsymbol{\mu}_\theta(\mathbf{x}_t, t),\boldsymbol{\Sigma}_\theta(\mathbf{x}_t, t)).
\end{align*}
While $\mathbf{\Sigma}_\theta(\mathbf{x}_t, t)$ can be learnt\footnote{See for
example \citet{ref_nichol2021}.}, the \citet{ref_ho2020} formulation fixes $\mathbf{\Sigma}_\theta$ to
an iteration-dependent constant $\sigma_t^2 \mathds{1}$, where
\mbox{$\sigma^2_t = 1 - \alpha_t$}.

By recognising that diffusion models are a restricted class of hierarchical 
VAE\footnote{
    Denoising autoencoders (\S\ref{sec_VAE}) have an interesting
    relationship with score-based generative (or diffusion) models. As a
    taster, \citet{ref_turner2021} reframe diffusion models as a class of
    hierarchical denoising VAE, and \citet{ref_dieleman2022} show through a
    brief derivation that diffusion models optimise the same loss as a
    denoising autoencoder.
}, 
we see that we can train $p_\theta$ by optimising the evidence lower bound
\citep[ELBO, introduced in][]{ref_vae} that can be written as a
summation over the Kullback-Leibler divergences at each iteration
step\footnote{
    See Appendix~B in \citet{ref_sohldickstein2015} and
    Appendix~A in \citet{ref_ho2020} for the full derivation.  
}:
\begin{align}
    \mathcal{L}_{\text{ELBO}} &= \mathbb{E}_q \Bigl[ D_\text{KL}(q(\mathbf{x}_T \mid \mathbf{x}_0) \| p(\mathbf{x}_T)) + \notag
                                \\&\hspace{5em}\sum_{t > 1} D_\text{KL}(q(\mathbf{x}_{t - 1} \mid \mathbf{x}_t, \mathbf{x}_0) \| p_\theta(\mathbf{x}_{t - 1} \mid \mathbf{x}_t))
                                + \log p_\theta(\mathbf{x}_0 \mid \mathbf{x}_1)\Bigr].
                                \label{eqn_elbo_ddpm}
\end{align}
In the \citet{ref_ho2020} formulation, the first term in
Eq.~\ref{eqn_elbo_ddpm} is a constant during training and the final term is
modelled as an independent discrete decoder. This leaves the middle summation.
Each summand can be written as
\begin{equation}
    \mathcal{L}(\boldsymbol{\mu}_t, \boldsymbol{\mu}_\theta) = \frac{1}{2 \sigma_t^2} \| \boldsymbol{\mu}_t(\mathbf{x}_t, \mathbf{x}_0) - \boldsymbol{\mu}_\theta(\mathbf{x}_t, t) \|^2,
    \label{eqn_lossmu_ddpm}
\end{equation}
where $\boldsymbol{\mu}_\theta$ is the neural network's estimation of the forward process
posterior mean $\boldsymbol{\mu}_t$. In practice it would be preferable to predict the noise
addition in each iteration step ($\mathbf{z}_t$), as $\mathbf{z}_t$ has a distribution that by
definition is centred about zero, with a well defined variance.  To this end we
can define $\boldsymbol{\mu}_\theta$ as
\begin{equation}
    \boldsymbol{\mu}_\theta(\mathbf{x}_t, t) = \frac{1}{\sqrt{\alpha_t}} \left(\mathbf{x}_t - \frac{1 - \alpha_t}{\sqrt{1 - \bar{\alpha}_t}} \mathbf{z}_\theta(\mathbf{x}_t, t)\right),
    \label{eqn_mu_ddpm}
\end{equation}
and by combining Eqs.~\ref{eqn_lossmu_ddpm} and \ref{eqn_mu_ddpm} we get
\begin{align}
    \mathcal{L}(\mathbf{z}_t, \mathbf{z}_\theta) &= 
        \frac{1}{2 \sigma_t^2} \Biggl\| \frac{1}{\sqrt{\alpha_t}} \left(\mathbf{x}_t - \frac{1 - \alpha_t}{\sqrt{1 - \bar{\alpha}_t}} \mathbf{z}_t\right) -
         \frac{1}{\sqrt{\alpha_t}} \left(\mathbf{x}_t - \frac{1 - \alpha_t}{\sqrt{1 - \bar{\alpha}_t}} \mathbf{z}_\theta(\mathbf{x}_t, t)\right) \Biggr\|^2\notag\\
         &= \frac{(1 - \alpha_t)^2}{2 \sigma_t^2 \alpha_t (1 - \bar{\alpha}_t)} \| \mathbf{z}_t - \mathbf{z}_\theta(\mathbf{x}_t, t) \|^2.
         \label{eqn_complexloss_ddpm}
\end{align}

\citet{ref_ho2020} empirically found that a simplified version of the loss
described in Eq.~\ref{eqn_complexloss_ddpm} results in better sample quality.
They use a simplified version of Eq.~\ref{eqn_complexloss_ddpm} as their
loss, and optimise to predict the noise required to reverse a forward process
iteration step:
\begin{equation}
    \mathcal{L}(\mathbf{z}_t, \mathbf{z}_\theta) = \| \mathbf{z}_t - \mathbf{z}_\theta (\textbf{x}_t,\,t) \|^2,\quad\text{where}~\mathbf{x}_t = \sqrt{\bar{\alpha}_t} \mathbf{x}_0 + \sqrt{1 - \bar{\alpha}_t}\mathbf{z}_t.
    \label{eqn_loss_ddpm}
\end{equation}
By recognising that $\mathbf{z}_t = \sigma_t^2 \nabla_{\mathbf{x}_t} \log q(\mathbf{x}_t
\mid \mathbf{x}_{t - 1})$, we see that Eq.~\ref{eqn_loss_ddpm} is equivalent to
denoising score matching over $t$ noise levels \citep{ref_vincent2011}.  This
connection establishes a link between diffusion models and other SBGMs \citep[such
as][]{ref_song2019,ref_song2020,ref_ajm2020}.  

To run inference for the reverse process, one progressively removes the predicted
noise $\mathbf{z}_\theta$ from an image. The predicted noise is weighted
according to a variance schedule:
\begin{equation*}
    \mathbf{x}_{t-1} = \frac{1}{\sqrt{\alpha_t}} \left(\mathbf{x}_t - \frac{1 - \alpha_t}{\sqrt{1 - \bar{\alpha}_t}}\,\mathbf{z}_\theta(\mathbf{x}_t, t)\right) + \boldsymbol{\sigma}_t \mathbf{z}.
\end{equation*}
If we take $p(\mathbf{x}_T) \sim \mathcal{N}(\mathbf{x}_T; \mathbf{0},
\mathds{1})$, we can use $p_\theta$ to generate entirely novel data that are
similar---but not identical to---those found in the training set. 

In practice, diffusion models are trained by sampling an integer value of $t
\sim \mathcal{U}(1, T)$, where $T$ is a large value typically in the 1000s.
We then use Eq.~\ref{eqn_ddpm_practice} to sample an image $\mathbf{x}_t$ that
has had noise added to it $t$ times. The model then attempts to predict the
exact noise required to reverse a forward iteration time step---that is, the
output of a neural network\footnote{
    Typically a U-Net; see \S\ref{sec_LSTM} for more detail.
}
of the form $z_\theta (\mathbf{z}_{t}|\mathbf{x}_{t-1})$. As shown in
Fig.~\ref{fig_ddpm}, we can estimate $\mathbf{x}_t$ by removing the
predicted noise from $\mathbf{x}_{t-1}$.  To optimise the model $\mathbf{z}_t$
is compared via Eq.~\ref{eqn_loss_ddpm} to the actual noise required to reverse
the forward iteration, and this is the loss that is reduced during training.
For a detailed astronomical example with code we direct the reader to
\citet{ref_smith2022}.

\subsubsection{Denoising diffusion implicit models}

\citeauthor{ref_ho2020}'s diffusion model performs inference at a rate orders of
magnitude slower than single shot generative models like the VAE
(\S\ref{sec_VAE}) or the GAN (\S\ref{sec_GAN}). This is because diffusion models need to
sequentially reverse every step in the forward process Markov Chain. Reducing the inference
time for diffusion models is an active area of research
\citep{ref_ajm2021,ref_luhman2021,ref_watson2022}, and here we will review
one proposed
solution to the problem; the denoising diffusion implicit model
\citep[DDIM;][]{ref_song2020ddim}.

\begin{figure}[h]
    \centering
    \includegraphics[angle=90,width=\textwidth]{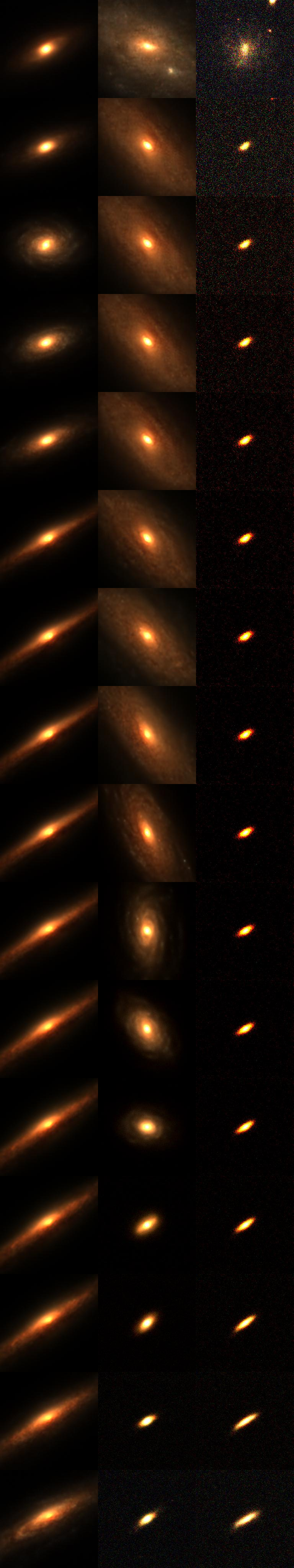}
    \caption[DDIM latent space interpolation]{Meaningful latent space interpolation
    via a DDIM model \citep{ref_song2020ddim,ref_smith2022}. This property
    comes `for free' with most other generative models, however the denoising diffusion probabilistic model
    \citep{ref_ho2020} requires a tweak to its sampling scheme
    (Eq.~\ref{eqn_ddim_loss}).}
    \label{fig_ddim_latentinterp}
\end{figure}

\citet[][\S\S3-4]{ref_song2020ddim} propose the following reparameterisation of Eq.~\ref{eqn_niceproperty}:
\begin{align*}
     \mathbf{x}_{t-1} &= \sqrt{\bar{\alpha}_{t-1}}\,\mathbf{x}_0 + \sqrt{1 - \bar{\alpha}_{t - 1} - \sigma_t^2}\,\mathbf{z}_{\theta}^{(t)} + \sigma_t \mathbf{z}_t\\
                     &= \sqrt{\bar{\alpha}_{t-1}}\,\underbrace{\left(\frac{\mathbf{x}_t - \sqrt{1- \bar{\alpha}_t}\,\mathbf{z}^{(t)}_\theta}{\sqrt{\bar{\alpha}_t}}\right)}_{\mathbf{x}_0 \text{ prediction}} + \underbrace{\sqrt{1 - \bar{\alpha}_{t - 1} - \sigma_t^2}\,\mathbf{z}^{(t)}_{\theta}}_{\text{vector towards }\mathbf{x}_t} + \underbrace{\sigma_t \mathbf{z}_t}_{\text{noise}},
\end{align*}
where $(t)$ is noted as a superscript to denote the output of the neural
network $z_\theta$ at timestep $t$.  Intuitively, the first term can be thought
of as the prediction of the input image $\mathbf{x}_0$, given an iteration step
$t$. The second term can be thought of as a vector from $\mathbf{x}_{t-1}$
towards the current iteration step image $\mathbf{x}_t$. The third term is
random noise. If we substitute in $\mathbf{x}_t$ from Eq.~\ref{eqn_loss_ddpm}
we make this intuition explicit:
\begin{equation*}
    \mathbf{x}_{t - 1} = \sqrt{\bar{\alpha}_{t-1}}\,\mathbf{x}_0 + \sqrt{1 - \bar{\alpha}_{t - 1} - \sigma_t^2}\,\frac{\mathbf{x}_t - \sqrt{\bar{\alpha}_t}\,\mathbf{x}_0}{\sqrt{1 - \bar{\alpha}_{t}}} +\sigma_t \mathbf{z}.
\end{equation*}
If we then set $\sigma_t = 0$, we remove the noise dependency and the forward process becomes deterministic:
\begin{equation}
    q_{\text{DDIM}} (\mathbf{x}_{t - 1} \mid \mathbf{x}_t, \mathbf{x}_0) = \sqrt{\bar{\alpha}_{t-1}}\,\mathbf{x}_0 + \sqrt{1 - \bar{\alpha}_{t - 1}}\,\frac{\mathbf{x}_t - \sqrt{\bar{\alpha}_t}\,\mathbf{x}_0}{\sqrt{1 - \bar{\alpha}_{t}}}.
    \label{eqn_ddim_loss}
\end{equation}
This means that DDIMs can deterministically map to and from the latent space,
and so inherit all the benefits of this property. For example, two objects
sampled from similar latent vectors share high level properties, latent space
arithmetic is possible, and we can perform meaningful interpolation within this
space. We demonstrate DDIM latent space interpolation in
Fig.~\ref{fig_ddim_latentinterp}.

We can also subsample every $\tau$ number of steps at inference time, where
$\tau$ is a set of evenly spaced steps between $0$ and $T$, the maximum number
of steps in the forward process:
\begin{equation}
    q_{\text{DDIM}} (\mathbf{x}_{\tau_{i - 1}} \mid \mathbf{x}_{\tau_i}, \mathbf{x}_0) = \sqrt{\bar{\alpha}_{t - 1}}\,\mathbf{x}_0 + \sqrt{1 - \bar{\alpha}_{t - 1}}\,\frac{\mathbf{x}_{\tau_i} - \sqrt{\bar{\alpha}_t}\,\mathbf{x}_0}{\sqrt{1 - \bar{\alpha}_t}}.
\end{equation}
As shown in \citet{ref_song2020ddim} this results in acceptable generations
with a $T/\tau$ inference speed up.

SBGMs have emerged as a promising alternative to GANs, VAEs, and other
generative models, showcasing their ability to generate high-quality samples
with a level of detail comparable to that of the previous state-of-the-art
\citep{ref_dhariwal2021,ref_nichol2021,ref_ramesh2022dalle2}. One of the key
advantages of SBGMs is how easy they are to train; they do not inherit any of
the instability issues that plague GANs.  However, SBGMs do have their share of
weaknesses. For instance, the SBGM sampling process is computationally
expensive and slow.  This is because generating a single sample requires a pass
through a learnt Markov chain (Fig.~\ref{fig_ddpm}), which can limit their
practicality in certain applications.  Finally, diffusion models and other
SBGMs have not been as extensively explored in the deep learning literature as
VAEs and GANs (although this is changing fast!).  This leaves their
applicability across various domains still under investigation.  

\section{Representation learning} \label{sec_representation}

Self-supervised\footnote{
    A model that employs self-supervised learning is one that obtains a
    supervisory signal from the data itself. `Self-supervised learning' as a
    descriptor has largely superseded the older term `unsupervised learning'.
    This is because the older term suggests that there is no supervisory signal
    at all---but the signal is there, just not explicitly defined by a human
    expert!
}
representation learning has recently exploded in popularity,
with a slew of models being developed in rapid succession
\citep[e.g.][]{ref_simclr,ref_simclrv2,ref_byol,ref_moco,ref_mocov2,ref_durkan2020}.
At its core, representation learning attempts to produce semantically
meaningful compressed representations (or embeddings) of complex highly
dimensional data.  Aside from simply being a compression device, these
embeddings can also be taken and used in downstream tasks, like clustering,
anomaly detection, or classification.

In this section we will describe two approaches to representation learning that are
popular within astronomy. The first approach uses contrastive learning as defined
by the SimCLR model. The second approach
defines and uses a `surrogate task' (such as autoencoding or next value
prediction) to train a deep learning model, and extracts semantically
meaningful representations from the subsequent trained network.

\subsection{Contrastive learning} \label{sec_simclr}

\begin{figure}[htbp]
    \centering
    \begin{subfigure}[b]{0.5\textwidth}
        \centering
        \includegraphics[width=\textwidth]{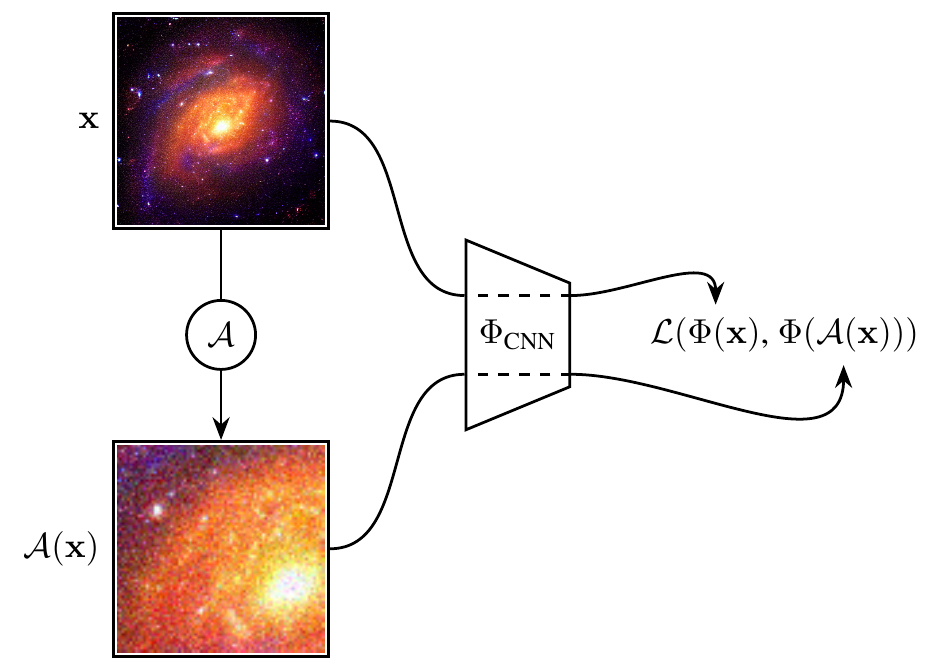}
        \caption{Possible application to imagery.}
    \end{subfigure}\hfill
    \begin{subfigure}[b]{0.5\textwidth}
        \centering
        \includegraphics[width=\textwidth]{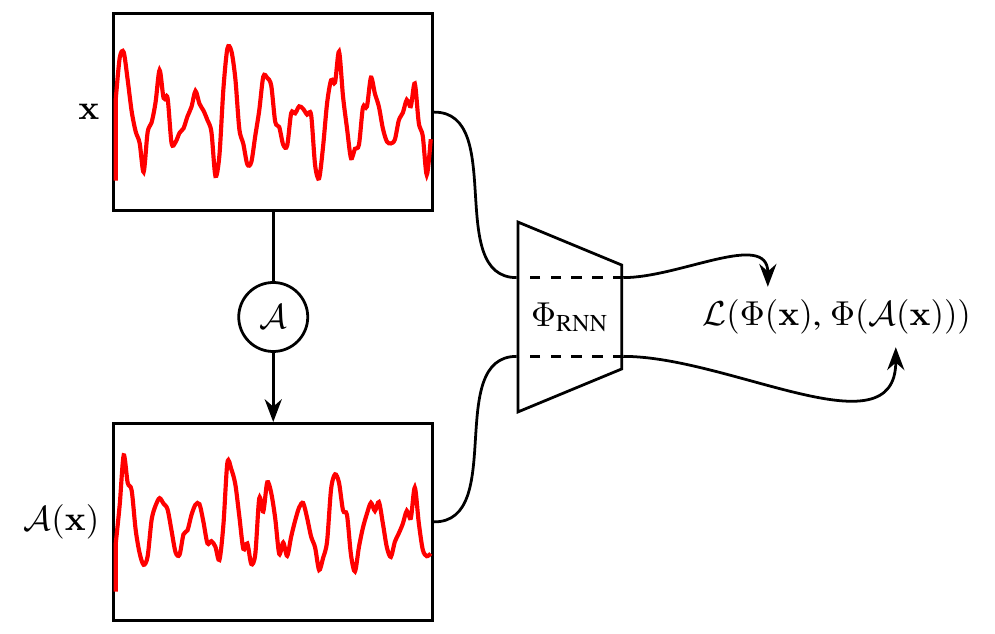}
        \caption{Possible application to sequential data.}
    \end{subfigure}
    \caption[A simple contrastive learning model]{
        A simple contrastive learning model is applied to both imagery and
        sequential data. $\mathcal{A}$ is an augmentation pipeline. For imagery,
        $\mathcal{A}$ could consist of random crops, noise addition, and colour
        jitter. For sequential data, $\mathcal{A}$ could consist
        of noise addition, stochastic temporal shifting, and random data
        deletion. $\Phi$ is a function approximator that projects inputs onto
        an embedding space. $\Phi$ is typically a neural network: when
        processing imagery, $\Phi$ could take the form of a CNN, and when
        processing sequential data $\Phi$ could be an RNN.  The loss $\mathcal{L}$
        measures the distance between the embeddings $\Phi(\ve{x}) = \mathbf{z}_i$ and
        $\Phi(\mathcal{A}(\ve{x})) = \mathbf{z}_j$, and we train by attempting to minimise
        this distance while maximising the distance between dissimilar samples.}
    \label{fig_simclr}
\end{figure}

Fig.~\ref{fig_simclr} describes a simple contrastive learning model similar to
SimCLR \citep{ref_simclr}. This model takes as input a sample $\mathbf{x}$ from
the training set, and augments it to produce $\mathcal{A}(\mathbf{x})$. This
augmentation is performed in such a way that $\mathcal{A}(\mathbf{x})$ shares
enough semantically meaningful data with $\mathbf{x}$ to belong to the same
class. In the contrastive learning literature $(\mathbf{x},
\mathcal{A}(\mathbf{x}))$ is known as a positive pair.  This positive pair is
passed to a Siamese neural network $\Phi$, which projects the high dimensional
input data onto a lower dimensional `embedding space'. All other training set
samples are assumed to belong to a different class to $\mathbf{x}$, and so can
be combined with $\mathbf{x}$ to produce `negative pairs'.  
Once we produce some embeddings we need to define a loss that clusters similar
samples together, while simultaneously pushing away dissimilar samples.
\citet{ref_hadsell2006} propose such a loss---the maximum margin contrastive
loss:
\begin{equation*}
    \mathcal{L}(\mathbf{z}_i, \mathbf{z}_j) = 
        \delta_{y_iy_j}\,d(\mathbf{z}_i, \mathbf{z}_j) +
        (1 - \delta_{y_iy_j}) \max(0, m - d(\mathbf{z}_i, \mathbf{z}_j)),
\end{equation*}
where $\delta$ is the Kronecker delta, $\mathbf{z}_i$ and $\mathbf{z}_j$ are
embedding vectors\footnote{
    All embeddings in this subsection are normalised.
}, $y_i$ and $y_j$ are the class labels for the embedding vectors, and $m$ is
the margin. $d$ is a `distance metric' (such as for example the L1 loss) that
reduces to zero in the case where its inputs are identical. If $\mathbf{z}_i$
and $\mathbf{z}_j$ are a positive pair, the loss pulls the embeddings closer,
and if they are a negative pair the loss pushes the embeddings away from each
other. The margin imposes an upper distance bound on dissimilar embeddings. 

While useful, the maximum margin contrastive loss does not take
into account the embedding space beyond the pair it is attending to in each
training step. This limitation ultimately results in a less expressive
embedding space. The triplet loss \citep{ref_chechik2010} solves this issue
by taking into account the broader embedding space and simultaneously
attracting a positive pair while repulsing a negative pair
with each training step:
\begin{equation}
    \mathcal{L}(\mathbf{z}_i, \mathbf{z}_j, \mathbf{z}_k) = 
        \max(0,\,d(\mathbf{z}_i, \mathbf{z}_j) - d(\mathbf{z}_i, \mathbf{z}_k) + m)
    \label{eqn_triplet_loss}
\end{equation}
where $\mathbf{z}_k$ is a sampled from a different class to $\mathbf{z}_i$,
and $\mathbf{z}_j$ is sampled from the same class as $\mathbf{z}_i$.

\begin{figure}[htbp]
    \centering
    \begin{subfigure}[b]{0.45\textwidth}
        \centering
        \includegraphics{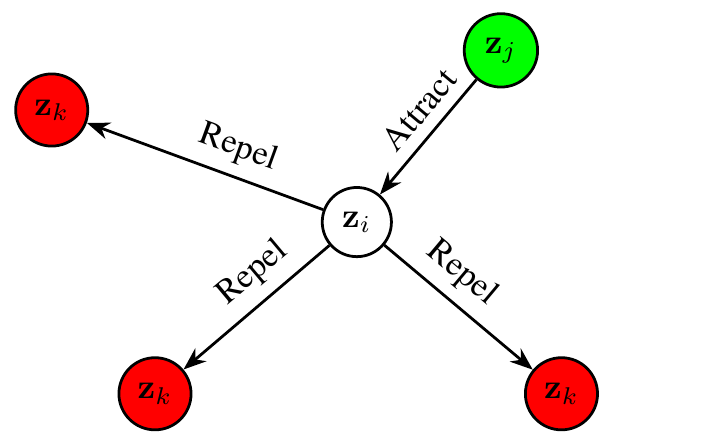}
        \caption[Triplet loss]{The triplet (Eq.~\ref{eqn_triplet_loss}) and \mbox{NT-Xent}
                               (Eq.~\ref{eqn_ntxent}) losses simultaneously
                               incentivise attraction between embeddings sampled
                               from the same class ($\mathbf{z}_i$ and
                               $\mathbf{z}_j$), and repulsion between embeddings
                               sampled from different classes
                               ($\mathbf{z}_i$ and $\mathbf{z}_k$).}
        \label{fig_contrastive_loss}
    \end{subfigure}
    \hfill
    \begin{subfigure}[b]{0.52\textwidth}
        \centering
        \includegraphics{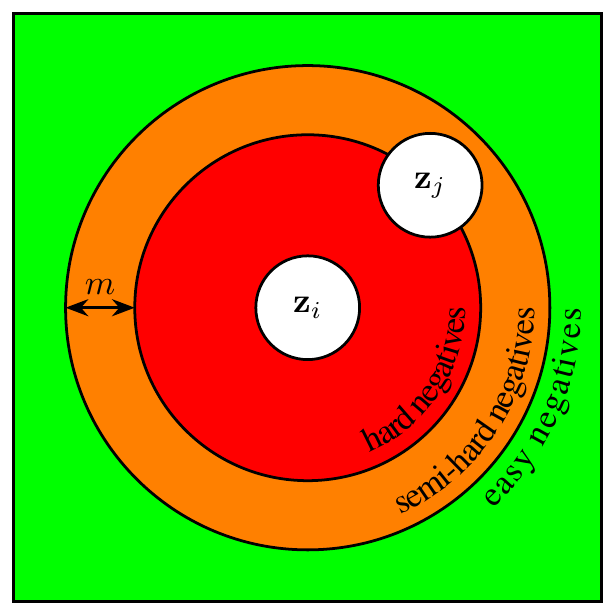}
        \caption[Types of negative embeddings]{Types of negative embeddings.
            $\mathbf{z}_i$ and $\mathbf{z}_j$ form a positive embedding pair.
            If a negative is closer than the current positive it is considered
            a hard negative, if it lies within the margin it is considered a
            semi-hard negative, and if it is beyond the margin it is considered an
            easy negative.}
        \label{fig_negtypes}
    \end{subfigure}
    \caption[More information on self-supervised embeddings]{More information
        on self-supervised embeddings. Fig.~\ref{fig_contrastive_loss} depicts
        the inner workings of the triplet and NT-Xent losses, and
        Fig.~\ref{fig_negtypes} shows the three possible negative embedding
        types as described in the literature.}
\end{figure}

If we study Eq.~\ref{eqn_triplet_loss} we see that it is possible to
generalise our loss even further, taking into account an arbitrary number
of negative samples. The normalized temperature-scaled cross entropy
loss \citep[NT-Xent;][]{ref_sohn2016} does precisely this:
\begin{equation}
    \mathcal{L}(\mathbf{z}_i, \mathbf{z}_j) = 
                       - \text{log}\left(\frac
                             {\exp(d(\mathbf{z}_i, \mathbf{z}_j)/\mathcal{T})}
                             {\sum^{2N}_{k=1}(1 - \delta_{ik}) \exp(d(\mathbf{z}_i, \mathbf{z}_k)/\mathcal{T})}
                         \right),
    \label{eqn_ntxent}
\end{equation}
where $\mathbf{z}_i$ and $\mathbf{z}_j$ are a positive embedding pair, and
$\mathbf{z}_i$ and $\mathbf{z}_k$ are a negative pair.  $\mathcal{T}$ is a
`temperature' hyperparameter introduced in \citet{ref_simclr} to help the model
learn from hard negatives (negatives closer to the anchor than the comparison
positive, see Fig.~\ref{fig_negtypes}).

\subsection{Learning representations via a surrogate task}

One can also learn representations via a surrogate task. A surrogate task is
any task that is unrelated to the network's final use. However, in the process
of learning to perform the surrogate task, the network learns what is
important, and what is unimportant about data within the training set. This
information can then be extracted in the form of learnt representations. If the
surrogate task is general enough, these representations will contain useful
semantic information about the items in the dataset, and can then be used for
downstream applications.

\begin{figure}[htbp]
    \centering
    \begin{subfigure}[b]{\textwidth}
        \centering
        \includegraphics{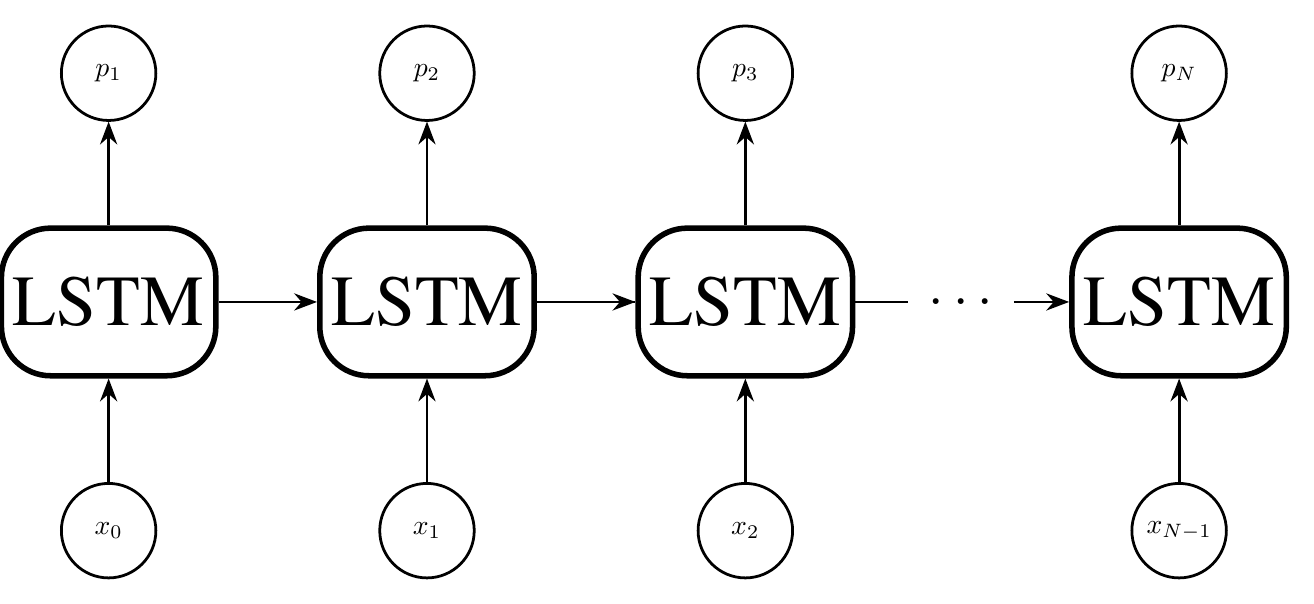}
        \caption{While training we feed in the galaxy rotation curve, and
        predict the next observation in its sequence.}
    \end{subfigure}

    \vspace{1em}

    \begin{subfigure}[b]{\textwidth}
        \centering
        \includegraphics{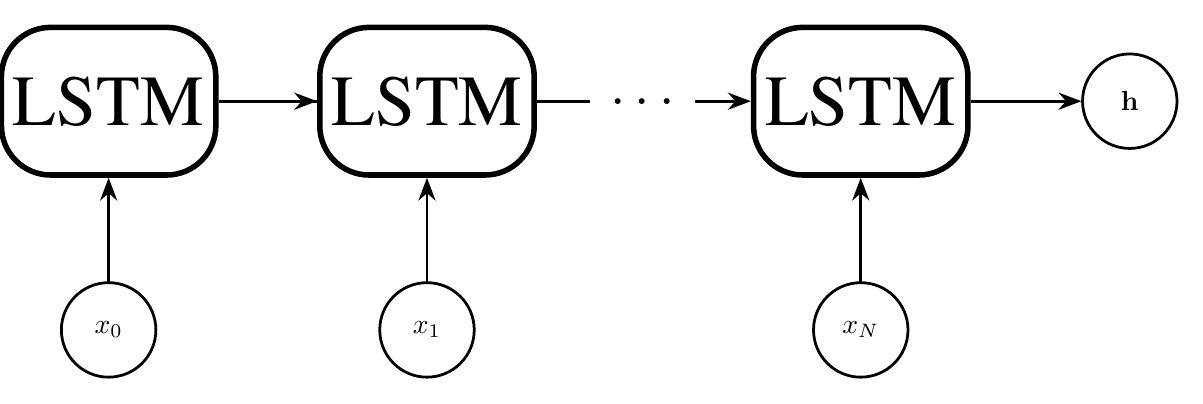}
        \caption{While inferring we feed in the full galaxy rotation curve,
        and extract the LSTM hidden state as a compressed representation embedding 
        of the curve. Otherwise, we ignore whatever output (i.e.
        $\{p_1,\ldots,p_N\}$) the LSTM generates.}
    \end{subfigure}
    \caption[A surrogate task for extracting rotation curve representations]{A
    hypothetical surrogate task for extracting rotation curve representations is shown.
    $\{x_0,\ldots,x_N\}$ is a set of observations from a galaxy rotation curve,
    in order of radial distance from the galactic centre. $\{p_1,\ldots,p_N\}$
    is the LSTM's corresponding set of predictions for $\{x_1,\ldots,x_N\}$.
    $\mathbf{h}$ is the LSTM hidden state vector.  See Fig.~\ref{fig_LSTM} for
    more about the internal workings of the LSTM.}
    \label{fig_lstm_surrogatetask}
\end{figure}

Let us concretise this process by revisiting an example that we previously
discussed in \S\ref{sec_rnn}. Let us imagine we have a large set of galaxy
rotation curves that we want to extract embeddings from. We could train an LSTM
model (Fig.~\ref{fig_lstm_surrogatetask}) on the task of
predicting the next item in the rotation curve, with the model only having
access to the previous items in the profile. Once the LSTM model is trained on
this task, we can feed in a full, new rotation curve, and repurpose the final
hidden state as a representative embedding. Note that this set up does not rely
on any external labels, only on the rotation curve itself\footnote{
    This self-supervised training set up is similar to that used to train
    autoregressive foundation models. These models will be explored in 
    detail in \S\ref{sec_foundationmodels}.
}.

We can generate embeddings via an autoencoding task. Again, let us use an
astronomical example to specify this, and say that we want to extract embeddings
from a set of galaxy observations. We could repurpose a
variational autoencoder for this, training it as normal as described in
\S\ref{sec_VAE}. However, once the model is trained we would discard the decoding
part of the network, and only consider the encoder. To generate embeddings we would
then simply pass in our galaxy images to the trained encoder. The same process can
be carried out by a GAN (\S\ref{sec_GAN}). In 
the GAN case, we would discard the generator after training, and use the
discriminator's penultimate layer outputs as our embeddings.

Supervised networks can also be used to generate embeddings. If a network has been
trained in a supervised manner to classify or regress data, it will have learnt
some properties about that data that helps it to carry out its task. We
can access these learnt representations by taking the outputs from a trained
network's penultimate layer as an embedding\footnote{
    Interestingly, this process is used in the calculation for the Fr\'echet
    Inception Distance (FID) \citep{ref_heusel2017,ref_seitzer2020}. The
    FID acts as a measurement of the visual similarity between two datasets.
    The FID works by taking the penultimate layer representations from a
    trained Inception-v3 model \citep{ref_szegedy2016} for each dataset and
    calculating the distance between them. 
}.

\section{Astronomy's third wave of connectionism} \label{sec_athirdera}

Since its astronomical debut in the mid-2010s 
\citep{ref_regier2015deep}\footnote{
    Also compare its companion paper \citep{ref_regier2015celeste}.
},
deep generative modelling has become a popular subfield within astronomical
connectionism. This popularity is driven by its inherent scalability; the lack
of a need for labelled data allows the methods to be repurposed for any dataset
that might be at hand. Self-supervised connectionism has been around for longer
\citep[i.e.][]{ref_serra1993},
but again has recently exploded in popularity due to its usefulness in
wrangling enormous unlabelled datasets. This section is split into two major
parts. We will first outline the history of deep astronomical generative
modelling in \S\ref{sec_genapp}, and the history of astronomical representation
learning will be discussed in \S\ref{sec_repapp}. Although representation
learning is the explicit goal for only the studies described in
\S\ref{sec_repapp}, it must be stressed that representations \emph{can also be
extracted from all the deep generative models described in
\S\ref{sec_genapp}.}

\subsection{Deep astronomical generative modelling} \label{sec_genapp}

Capturing genuine astronomical data demands accurate knowledge of telescope
behaviour, equipment features, environmental factors during observations, and
data reduction techniques. These complex steps are often tailored to individual
observation sets. However, there's an alternative to classical simulation:
leveraging examples from a specific survey allows for the development of a
data-driven method to simulate not only the astronomical signal but also the
data's inherent characteristics. In addition to this, deep learning models trained to replicate
astronomical observations are much cheaper to run than classical simulation,
and so can rapidly generate massive amounts of data. Data that can then be used
for astronomical pipeline prototyping at scale, aiding the development of new
analysis methods, and for dataset augmentation.  Data-driven simulation is made
possible via the power of deep generative models, and this section describes
the history of their use within astronomy.

Seminally, \citet{ref_regier2015deep} proposed the use of a VAE to model
galaxy observations.  They trained their network on downscaled $69 \times 69$
crops of galaxies from a SDSS-sampled dataset containing 43\,444 galaxies.
They trained their network in the same way as described in \S\ref{sec_VAE}, and
find that the network is capable of generating galaxies similar to those found
in the training set. They also find that their network produces semantically
meaningful embeddings, noting that their galaxies are clustered by orientation
and morphological type. This same line of enquiry was followed by
\citet{ref_ravanbakhsh2016}, who showed that VAEs could be used to generate
galaxies conditionally. \citet{ref_ravanbakhsh2016} also pioneered the use
of GANs to generate galaxy imagery. \citet{ref_spindler2020} used a VAE
combined with a Gaussian mixture model prior (see Eq.~\ref{eqn_vae_prior} and
accompanying text) to generate and cluster galaxy images into morphological
types. While the previous studies in this paragraph used images with relatively
small pixel dimensions in their training set, \citet{ref_fussell2019} and
\citet{ref_holzschuh2022} demonstrated that GANs are capable of generating
large high fidelity galaxy observations.  \citet{ref_fussell2019} achieved
this with a stacked GAN architecture \citep{ref_zhou2016cvpr}, and
\citet{ref_holzschuh2022} use the related StyleGAN architecture
\citep{ref_karras2018} to the same end.  \citet{ref_bretonniere2021} use a
flow-based model\footnote{
    Flow-based models have not been discussed in detail in this review,
    but see \citet{ref_weng2018flow} for a magisterial introduction to the
    subject.
} 
\citep{ref_germain2015,ref_papamakarios2017} to conditionally simulate
galaxy observations. They found that their approach could produce more accurate
simulations than the previous analytical approach, at the cost of inference time.
Relatedly, \citet{ref_smith2022} use a diffusion model to generate large high
fidelity galaxies. They trained their network on two datasets comprised of
galaxies as observed by the Dark Energy Spectroscopic Instrument
\citep[DESI;][]{ref_desi}.  One, a set of $306\,006$ galaxies catalogued in the
SDSS Data Release 7 \citep{ref_sdss,ref_sdss_dr7,ref_wilman2010}, and the other
a set of $1962$ late-type galaxies, as catalogued in the Photometry and
Rotation curve OBservations from Extragalatic Surveys
\citep[PROBES;][]{ref_stone2019} dataset.  PROBES contains well resolved
galaxies that exhibit spiral arms, bars, and other features characteristic of
late-type galaxies. They found that their model produces galaxies that are both
qualitatively and statistically indistinguishable from those in the training set,
proving that diffusion models are a competitive alternative to the more
established GAN and VAE models for astronomical simulation.  From all of these
studies we can conclude that deep generative models can internalise a model
capable of physically and morphologically describing galaxies.

Generative models have also been used to simulate astronomical data on larger
scales.  In a use-case tangential to galaxy generation, \citet{ref_sagan}
show that a Spatial-GAN \citep{ref_jetchev2016} can simulate arbitrarily
wide field surveys. They train on the Hubble eXtreme Deep Field, and find that
galaxies `detected' within their model's synthetic deep fields are statistically
indistinguishable from the real thing. 
Cosmological simulations have also been explored,
with \citet{ref_rodriguez2018} using a GAN to generate cosmic web
simulations at pace, and \citet{ref_mustafa2019} generating weak lensing
convergence maps at a pace faster than classic simulations.
Beyond GANs,
\citet{ref_remy2020}\footnote{
    This preliminary work has been subsequently extended in
    \citet{ref_remy2022}.
} 
trained a SBGM on simulated maps from MassiveNus
\citep{ref_liu2018}, and found that their model was capable of replicating these maps.
They also demonstrated that their model was capable of producing a likely
spread in the posterior predictions.
Finally, they demonstrate that a SBGM is capable of predicting the mass
map of the real Hubble Cosmic Evolution Survey (COSMOS) field
\citep{ref_scoville2007}.

The image domain translation abilities of GANs in a Pix2Pix-like formulation
\citep[][also see Fig.~\ref{fig_pix2pix}.]{ref_pix2pix} is particularly useful
in astronomy. \citet{ref_schawinski2017} demonstrated this use first by
training a Pix2Pix-like model to denoise astronomical data. They trained their
network on 4550 galaxies sampled from SDSS. The galaxies were convolved to
increase the seeing, and speckle noise was added. The GAN was tasked with
reversing this process.  They found that their method outperformed both blind
deconvolution, and Lucy-Richardson deconvolution.
Generative models are also capable of separating sources, as
\citet{ref_stark2018} demonstrate by using a Pix2Pix model to deblend a quasar's
point source emission from the extended light of its host galaxy.  \citet{ref_reiman2019}
use a similar model to \citet{ref_stark2018} to deblend overlapping
galaxies.  

At the time of writing there are only three examples of score-based (or
diffusion) modelling in the astronomy literature
\citep{ref_remy2020,ref_smith2022,ref_remy2022}\footnote{
    Since the first posting of this review there have been several workshop
    papers presented at the 36th Conference on Neural Information Processing
    Systems (NeurIPS 2022) on the application of SBGMs to astronomical
    problems \citep[e.g.][]{ref_adam2022,ref_karchev2022,ref_mudur2022}.
    Here we will highlight a particularly neat example of diffusion model
    application: \citet{ref_karchev2022} tackle the inverse problem of
    strong-lensing source reconstruction and prove that a denoising diffusion
    restoration model \citep[DDRM;][]{ref_kawar2022} inference scheme alongside
    an off-the-shelf `AstroDDPM' model \citep{ref_smith2022} can restore
    galaxies that have been through a lensing process.  Remarkably, they
    achieved this \emph{without any retraining or finetuning of the original
    AstroDDPM model}, demonstrating that generalist pretrained score-based
    models like that described in \citet{ref_smith2022} can easily be
    repurposed for seemingly out-of-distribution downstream tasks. We will
    revisit the idea of pretrained models that can be repurposed for downstream
    tasks when we discuss `foundation' models in \S\ref{ch_conclusions}.
}.  
It is surprising that these studies are the only examples of score-based
modelling in astronomy, as SBGMs produce generations that rival that of
state-of-the-art GAN models, without drawbacks present in other models (like
blurring in the case of VAEs, or mode collapse and training instability in the
case of GANs). SBGMs also have some natural uses in astronomical data
pipelines.  For example, an implementation similar to \citet{ref_sasaki2021}
could be used for survey-to-survey photometry translation similarly to
\citet{ref_buncher2021}. The source image separation model described in
\citet{ref_jayaram2020} has the obvious application as an astronomical object
deblender \citep[i.e.][]{ref_stark2018,ref_reiman2019,ref_arcelin2021}.  To
summarise, SBGMs are ripe for exploitation by the astronomical community and we
hope to see much interest in this area in the coming years.

\subsection{Self-supervised astronomical representation learning} \label{sec_repapp}

In \citeyear{ref_serra1993}, \citet{ref_serra1993} proposed using an
autoencoder to learn embeddings for stars as observed by the Two Micron
Galactic Survey \citep{ref_calbet1993}. They first proved that their
autoencoder model worked better than principle component analysis (PCA) on
the toy problem of separating Gaussian distributions, and they then showed
that their model also outperformed the classic PCA method on real data.
More than twenty years later, \citet{ref_graff2014}\footnote{ 
    See Footnote~\ref{ftn_astrofoundation} for commentary of this study in the
    context of astronomical foundation models.
}
showed that autoencoders are also capable of capturing the properties of galaxies as
described in the Mapping Dark Matter Challenge \citep{ref_kitching2015} by
demonstrating  that embeddings extracted from their autoencoder were beneficial
for computing the ellipticities of their galaxies as a downstream task.
We are not limited to imagery; \citet{ref_yang2015} show that an autoencoder can
learn representations that can then be used to train a neural network for the
downstream task of estimating stars' atmospheric parameters, and
\citet{ref_tsang2019} demonstrate that an autoencoder can generate embeddings
that can then be used to classify variable star light curves.  From these
studies we must conclude that neural networks trained via a surrogate task are
capable of learning semantically meaningful embeddings across astronomical
domains.

Very recently there has been work applying self-supervised
contrastive learning models to galaxy image clustering.
\citet{ref_hayat2021} trained SimCLR \citep{ref_simclr} on multi-band
galaxy photometry from the SDSS
\citep{ref_sdss}. They show that the resulting embeddings capture
useful information by directly using them in a training set for a galaxy
morphology classification model, and a redshift estimation model.
Similarly, \citet{ref_sarmiento2021} trained SimCLR on integral field
spectroscopy data captured from galaxies in the Mapping Nearby Galaxies at
Apache Point Observatory survey \citep[MaNGA;][]{ref_manga}. Again,
they find that SimCLR produces semantically meaningful embeddings.
\citet{ref_slijepcevic2022} demonstrate that the `Bootstrap Your Own Latent'
\citep[BYOL;][]{ref_byol}\footnote{
    A contrastive learning framework that unlike SimCLR does not use negative
    samples to learn an embedding space.  
} contrastive learning model is capable of learning
semantically meaningful representations of radio galaxies. Their model is
trained on 100\,000 Radio Galaxy Zoo galaxies, and inference is run on the
1256 galaxy strong Mirabest dataset \citep{ref_porter2020}. They find that
embeddings derived from their model are semantically meaningful, suggesting
that self-supervised methods are transferable between disparate surveys.
These studies show that contrastive learning is applicable to 
imagery, further study will be required to demonstrate its effectiveness with
other types of astronomical data, such as time series and volumetric data. 

\section{Foundation models: a fourth astroconnectionist wave? } \label{ch_conclusions}

This review has shown thus far that deep learning has found wide use in
astronomy, a use predicated on the availability of enormous amounts of
computational power and data. This section looks to the future and predicts
an outcome if astronomy continues to follow in the footsteps of other applied
deep learning fields. In short, we predict and argue that astronomical
connectionism will likely see the removal of expertly crafted deep learning
models, to be replaced with an all encompassing `foundation' model. In
\S\ref{sec_foundationmodels} we explore what foundation models are, and their
context within deep learning.  \S\ref{sec_datamoats} then contextualises these
models within astronomy, and suggests actions we can take as a community to
realise an astronomical foundation model. Finally, \S\ref{sec_practical_foundation}
demonstrates as a thought experiment a state-of-the-art use-case for an
astronomical foundation model
and explores other theoretical and practical uses and implications within (and beyond) astronomy.

\subsection{Foundation models} \label {sec_foundationmodels}

Since its inception, connectionism has followed a path of greater compute
and greater generality \citep{ref_sutton2019,ref_gwern2022}. In that time,
human crafted biases have fallen by the wayside, to be replaced with models and
techniques that learn directly from data.  \citet{ref_sutton2019}
exemplifies this process via the field of speech recognition:
\blockquote{\small
    \emph{In speech recognition, there was an early competition, sponsored by DARPA [Defense Advanced Research Projects Agency], in
        the 1970s. Entrants included a host of special methods that took advantage of
        human knowledge---knowledge of words, of phonemes, of the human vocal tract,
        etc. On the other side were newer methods that were more statistical in nature
        and did much more computation, based on hidden Markov models (HMMs). Again, the
        statistical methods won out over the human-knowledge-based methods. This led to
        a major change in all of natural language processing, gradually over decades,
        where statistics and computation came to dominate the field. The recent rise of
        deep learning in speech recognition is the most recent step in this consistent
        direction. Deep learning methods rely even less on human knowledge, and use
        even more computation, together with learning on huge training sets, to produce
        dramatically better speech recognition systems. As in [computer Go and
        computer chess], researchers always tried to make systems that worked the way
        the researchers thought their own minds worked---they tried to put that
        knowledge in their systems---but it proved ultimately
        counterproductive, and a colossal waste of researcher's time, when,
        through Moore's law, massive computation became available and a means
        was found to put it to good use.
    }}
We are seeing this principal play out once again through a new paradigm
shift in deep learning, where even the underlying neural network architecture does not
matter. Previously, neural networks were adapted for a specific domain via
inductive biases injected by researchers, such as convolutions for computer
vision, and recurrence for language processing. Now we are seeing transformer
networks \citep[see \S\ref{sec_transformers} and ][]{ref_aiayn} competing\footnote{
    For now! It may be that network architecture does not matter all that much
    at scale, and that any sufficiently large neural network is adequate. If
    this is true, we will see the simplest (and most scalable) architectures
    win out. Although this theory has not yet been rigorously tested, we are
     currently seeing rumblings that suggest that this is the case
    \citep[e.g. the section `\emph{Transformers are not special}'
    in][]{ref_porby2022}. 
    \citet{ref_bo2021} stands as a particularly notable example of this
    hypothesis, showing that an attention-free RNN is capable of matching the
    performance of a similarly-scaled transformer network.  Also see
    Footnote~\ref{ftn_mlpresurgence} for commentary on the performance
    capabilities of MLPs and transformers.  
}
in all deep learning domains applied or otherwise: from language processing
\citep{ref_bert,ref_brown2020gpt3}\footnote{
    These models are collectively known in the literature as large language
    models, or LLMs.
}
to computer vision \citep{ref_parmar2018imtrans,ref_dosovitskiy2020vit} to
graph learning \citep{ref_kim2022} to protein folding
\citep{ref_jumper2021} to astronomy
\citep{ref_donosoolivia2022,ref_morvan2022,ref_pan2022}. The transformer's
versatility allows us to take a model trained on one task and apply it to a
similar yet different task, a process known as transfer learning. For example,
we could train a model on the `surrogate' task of predicting the next word in a
sequence, and then apply that model to a similar yet different task of
predicting the answer to a geography question. In this example the first model
is known as a `foundation' model, and the downstream model is derived from it.
This set up brings with it some useful advantages. For example, if the
foundation model is improved, all downstream tasks also see improvement.
Therefore, the need for only one model allows researchers to pool their efforts
in a way not possible when resources are split between many projects.

To train a foundation model, we first need to define a surrogate task. As
labelled datasets are expensive, and raw data is relatively cheap, the easiest and most
scalable way to do this is via self-supervised learning\footnote{
    For more on self-supervised learning see \S\ref{sec_representation}.
}.
Self-supervised learning does not require a human to provide a labelled dataset
for training.  Instead, the supervisory signal is generated automatically from
the raw data.  For example, in the context of astronomy this task could be
predicting a masked value in a variable star's light curve
\citep{ref_donosoolivia2022}. Another task could be using an autoencoder
(\S\ref{sec_VAE}) to replicate a galaxy observation \citep{ref_spindler2020}.
A further task could be training within a self-supervised framework, like
contrastive learning (\S\ref{sec_simclr}).  The important thing about
self-supervised learning is that it does not require annotated data. This means
that we can leverage vast reserves of raw data (such as textbooks, scraped
Internet text, raw imagery, etc.).

\begin{figure}[htb]
    \includegraphics[width=\textwidth]{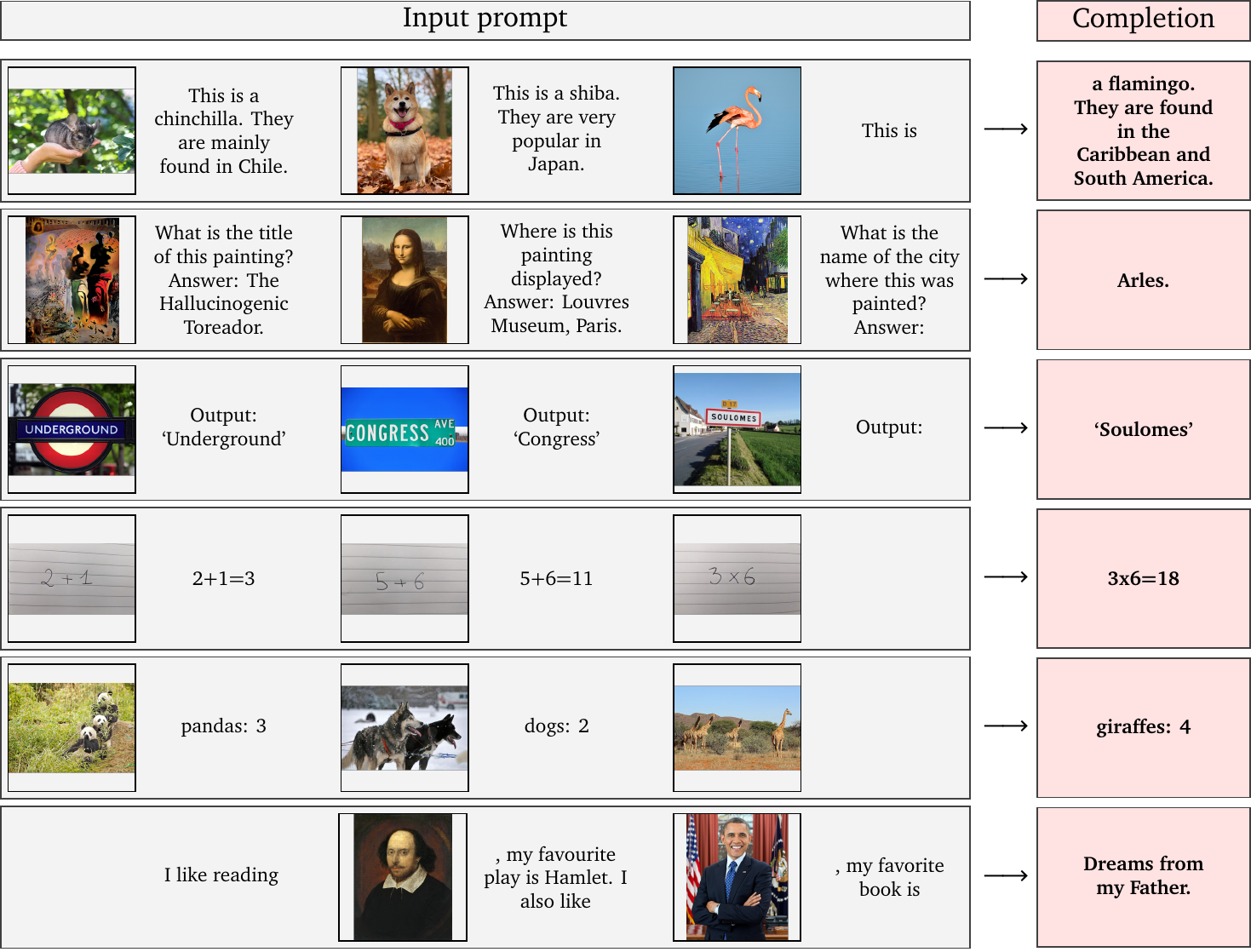}
    \caption[A demonstration of the generality of DeepMind's Flamingo
    model]{Flamingo is a foundation model that is capable of understanding
    images within the context of natural language. Here we see some examples
    of Flamingo's emergent abilities. This figure is adapted from Fig.~1 in
    \citet{ref_alayrac2022flamingo}.}
    \label{fig_flamingo}
\end{figure}

Very large models trained on vast amounts of data demonstrate surprising
emergent behaviour. For instance, GPT-3 \citep{ref_brown2020gpt3} is a
175B
parameter model that can be `prompted' to perform a novel task (see
Fig.~\ref{fig_flamingo} for more on prompting foundation models). This ability
was not shown at all in GPT-3's older, smaller 1.5B parameter sibling
\citep{ref_radford2019gpt2}. Furthermore, a meta-study described in
\citet{ref_wei2022emergent} found that larger models suddenly `unlock' abilities
such as arithmetic, translation, and understanding of figures of speech once
they reach a certain scale. These findings suggest that architectural changes
are not required beyond scaling to perform many tasks in natural
language processing \citep{ref_gwern2022,ref_chowdhery2022palm}.  In
Fig.~\ref{fig_flamingo} we see some results from
\citet{ref_alayrac2022flamingo}, a model comprising of an LLM,
and an image encoder. In this figure we can see that the model is capable 
of arithmetic, reading, counting, and has a broad knowledge (albeit not `understanding') of art, geography
and zoology\footnote{
    Interestingly, the authors of Flamingo first assumed that Flamingo's prediction of the
    species range of its eponymous bird was incorrect: flamingos are found in the
    Caribbean, South America, Africa, Europe, and South Asia. However, they
    later realised that the picture in Fig.~\ref{fig_flamingo} is of an
    \emph{American} flamingo, which is specifically found in the Caribbean and
    South America, so the network was right after all! See
    the reddit thread for the full context \citep{ref_flamingoreddit}.
}, and
literature.  This model is comprised of a ResNet variant
\citep{ref_resnet,ref_brock2021} to encode imagery, and the Chinchilla LLM
\citep{ref_hoffman2022} to encode and generate text. Chinchilla (and
therefore Flamingo) was trained with the surrogate task of predicting the next
word in a text sequence, and so none of the emergent properties stated above were
explicitly optimised for.

In the next subsection, we will state and explain the need for an astronomical
foundation model\footnote{
    \citet{ref_walmsley2022foundation} explore in a preliminary study a
    `galaxy foundation model' trained on Galaxy Zoo labels, and
    corresponding paired galaxy observations. They find that their pretraining
    is beneficial for training a network that performs a downstream task.
    However, the idea has been around for far longer that that; possibly the first
    demonstration of an astronomical foundation model was described eight years
    earlier in \citet{ref_graff2014}.  \citet{ref_graff2014} demonstrated that
    embeddings learnt with their autoencoding SkyNet network can be
    used for downstream tasks, but they do not use the moniker `foundation
    model' to describe SkyNet as the term had not yet been invented!
    Notably, neither study trains a model of the scale required to exhibit
    emergent properties or task generalisability.   These `blessings of scale'
    require data and compute at a level that has not yet been seen within
    astronomical connectionism. \label{ftn_astrofoundation}
}, not only for astronomy's sake, but also for the sake of
openness in deep learning research.

\subsection{Scaling laws and data moats} \label{sec_datamoats}

\citet{ref_hoffman2022} suggested an update to the foundation model
scaling law first proposed in \citet{ref_kaplan2020}. Their scaling law
equation relates the size of a neural network model and the training dataset
size to the minimum achievable loss. Mathematically, the equation is
\begin{equation}
    \mathcal{L}_{\min}(N, D) = \underbrace{\frac{A}{N^\alpha}}_{\text{parameter term}} + \underbrace{\frac{B}{D^\beta}}_{\text{data term}} + \underbrace{E}_{\text{dataset entropy}},
    \label{eqn_scalingloss}
\end{equation}
where $E$ is a constant that represents the lowest possible loss, given a
particular training dataset. $N$ is the number of trainable parameters within
the neural network, and $D$ is the size of the dataset in tokens (see
\S\ref{sec_transformers} for more about tokenisation). We can see that when we
have an infinitely large model trained on an infinitely large dataset (i.e. 
$N = D = \infty$), the only term remaining is the `dataset entropy' constant, $E$.
We can therefore only reduce the loss by increasing the size of our model, or
the size of our training set.

After fitting Eq.~\ref{eqn_scalingloss}, \citet{ref_hoffman2022} find
\begin{equation*}
    \mathcal{L}_{\min}(N, D) = \frac{406.4}{N^{0.34}} + \frac{410.7}{D^{0.28}} + 1.69.
\end{equation*}
If we then plug in $N$ and $D$ for a selection of real foundation models we
arrive at Fig.~\ref{fig_scalinglaw}. We can see in Fig.~\ref{fig_scalinglaw}
that the model size term for real foundation models is far lower than the
dataset size term. This means that an increase in dataset size has the
potential to reduce the minimum loss by a far larger amount than a larger
model would. Therefore, an obvious next step to improve these foundation models 
further is by increasing their dataset size.

\begin{figure}[htbp]
    \centering
    \input{images/conclusions/scaling_laws3.pgf}

    \begin{tabular}{lrrrr|r} \\ \toprule
        Model & $N$ & $D$ & $A/N^\alpha$ & $B/D^\beta$ & $\mathcal{L}_{\min}$ \\ \midrule
        LaMDA \citep{ref_thoppilan2022lamda} & 137B & 168B & 0.066 & 0.295 & 2.051\\
        GPT-3 \citep{ref_brown2020gpt3} & 175B & 300B & 0.061 & 0.251 & 2.002\\
        Gopher \citep{ref_rae2021gopher} & 280B & 300B & 0.052 & 0.251 & 1.993\\
        MT-NLG \citep{ref_smith2022mtnlg} & 530B & 270B & 0.041 & 0.259 & 1.990\\
        Chinchilla \citep{ref_hoffman2022} & 70B & 1.4T & 0.083 & 0.163 & 1.936\\
        PaLM \citep{ref_chowdhery2022palm} & 540B & 780B & 0.042 & 0.192 & 1.924\\
        \bottomrule
    \end{tabular}
    \caption[A slack in the minimum losses of foundation models]{A comparison
    between the minimum losses of a selection of foundation models.  The table
    above shows the number of parameters in a model ($N$), the number of tokens
    within that model's training set ($D$), and their corresponding calculated
    emergent terms from Eq.~\ref{eqn_scalingloss}. Here we use
    \citet{ref_hoffman2022} to source values for $A$, $\alpha$, $B$, and $\beta$. The
    minimum loss for each model according to \citet{ref_hoffman2022} is
    shown as $\mathcal{L}_{\min}$.  The contour plot shows the emergent
    parameters $B/D^\beta$ and $A/N^\alpha$ plotted against each other for our
    models. The closer the models' scatterpoints are to the bottom left, the
    lower their minimum loss value.
    }
    \label{fig_scalinglaw}
\end{figure}

The largest dataset \citep[MassiveText-English;][]{ref_hoffman2022} in the
comparison shown in Fig.~\ref{fig_scalinglaw} amounts to $1.4$T tokens.
However, this dataset is proprietary, being only available to researchers
employed by Google.  The largest public text dataset available at
the time of writing is The Pile \citep{ref_thepile}, with a total size of
${\sim}260$B tokens. We could increase the size of these datasets by
indefinitely scraping text data from the surface web, but this data tends to be
of low quality. Also, we have already exhausted some important high quality
data reserves, like fundamental research papers, and open source code
\citep{ref_friel2022}. We also have to ask ourselves: what happens when
generative models start to create data \emph{en masse}, and dump it indiscriminately
onto the Internet?  If a significant proportion of text in a dataset scraped
from the Internet is generated via an LLM, training on it will cause unforeseen
issues and may ultimately result in a model with worse performance. We must
therefore ensure that the data is not generated by a deep generative model.  In
addition to all this, the academy and the public at large will never have
access to the vast reserves of data contained in the deep web administered by
ByteDance, Google, Meta, Microsoft, and other tech giants.  For all these
reasons, we will need to think outside the box if we want to mine new high
quality data.

Enter the multimodal foundation model. \citet{ref_reed2022gato}\footnote{
    Earlier work from \citet{ref_kaiser2017} also demonstrated a deep
    learning model that could learn from disparate tasks, however Gato is the
    first model that achieves this while staying within a single deep learning
    paradigm.
}
demonstrated that a large transformer neural network is capable of learning
many tasks, from playing Atari, to captioning images, to chatting, to operating
a real robot arm. The model shares weights across all tasks, and decides at
inference time from context which task to predict.  Importantly,
\citet{ref_reed2022gato} find that their model follows the same scaling laws
as other foundation models, and so multimodal foundation models have the same
hunger for data that we see in Fig.~\ref{fig_scalinglaw}. Even
more astonishingly, \citet{ref_aghajanyan2023} find that a foundation model
trained on concatenated independent datasets significantly outperforms
separately trained unimodal models once the neural networks reach a certain
scale. We can therefore augment our text datasets
with high quality, publicly available astronomical data.

The Vera Rubin Observatory's 189 16~megapixel CCDs will observe 1000 science frames per night while conducting LSST
\citep{ref_lsst}.  This amounts to $3\times10^{12}$ pixels per night, or
approximately 12B tokens a night if we use the same tokenising scheme as
\citeauthor{ref_dosovitskiy2020vit}'s vision transformer
\citep{ref_dosovitskiy2020vit}. After only one year of observing, the LSST
will have produced $4.4$T tokens of raw data, larger than even the
MassiveText-English dataset\footnote{Of course, the reduced, useful data will be
far smaller than our raw estimate here. The motivation behind this calculation is to
show that even a single astronomical survey rivals the largest text dataset in
size. A compilation of all useful astronomical data would certainly dwarf any
contempory text dataset, whether public or proprietary.}. This data, and other astronomical
data like it, could be compiled into a very large open dataset similar to EleutherAI's
Pile \citep{ref_thepile}. This dataset would provide a way for academics
employed outside of Big Tech to train and research very large foundation
models.  Compiling a dataset like this would be difficult for a single
relatively underresourced research group, but it could be accomplished via
bazaar style open development \citep{ref_raymond1999}. We have already seen
this development model succeed in large open source projects, the most famous
of which is the Linux kernel. This development model has also been shown to
work within the field of deep learning by EleutherAI
\citep[e.g.][]{ref_thepile,ref_gptneox,ref_vqganclip}, and with HuggingFace's
BigScience initiative \citep{ref_scao2022}.  Once
compiled, we must ensure that progress is kept in the open, and that the data
is not simply absorbed into proprietary datasets---to do this we must give our
dataset a strong (viral) copyleft style licence. 

Once the dataset is compiled all we need for training are some self-supervised
surrogate tasks for our `astrofoundation' model to attempt. These tasks could
include predicting the next observation in a variable star's time sequence,
predicting the low surface brightness profile of a galaxy, predicting a
galaxy's morphological parameters, or simply generating the next crop in a
sequence of observations\footnote{
    This is essentially training the model to act as a physics simulator. Viewing
    foundation models as world simulators is not unprecedented. This
    perspective has already been explored in the simulation of thousands of
    `social simulacra' within a model online community
    \citep{ref_park2022}, and with the simulation of participants in classic
    (i.e. Milgram's shock experiment, the Ultimatum Game) and novel
    psychological studies \citep{ref_aher2022}. 
}. 
As we will explore in the next subsection, these surrogate tasks do not need to be at all related to the downstream
tasks we will eventually use our model for.
Once trained, our astrofoundation model will inherit all the interesting
properties that LLMs enjoy, such as few to zero-shot generation and other
emergent behaviours. 

\subsection{The practical implications and uses of an astrofoundation model} \label{sec_practical_foundation}

This section explores the wider implications of a hypothetical astrofoundation
model (\S\ref{sec_theory}), as well as some practical astronomical uses
(\S\ref{sec_practice}).  In \S\ref{sec_newsim} we highlight one possible
downstream task that would be useful in astronomy; a conditional generative
model for astronomical simulation.

\subsubsection{Democratising foundation models} \label{sec_theory}



\begin{figure}[htbp]
    \input{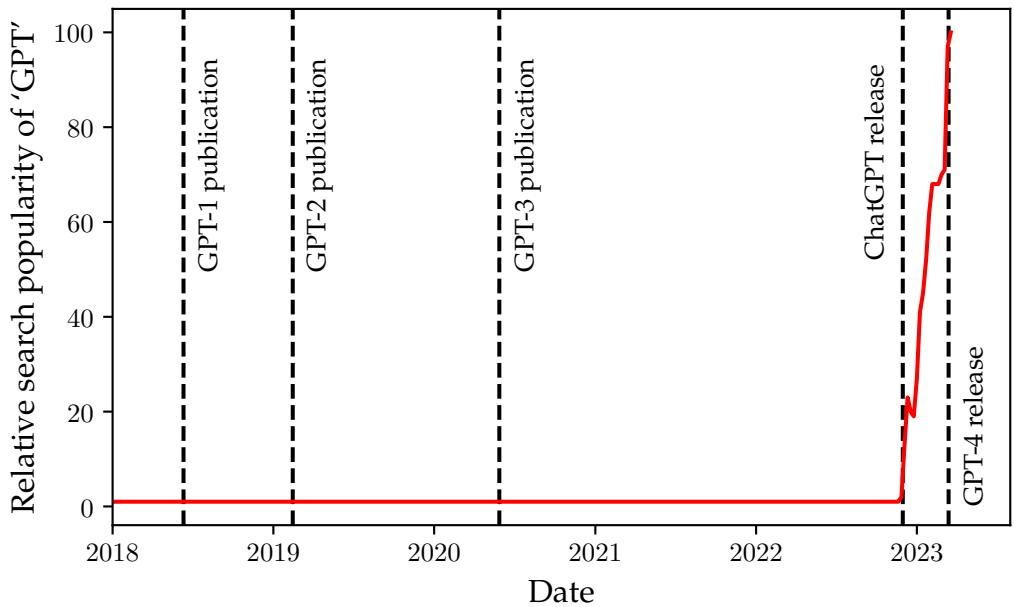}
    \caption[The stunning popularity of the GPT family of models.]
    {Here we show the relative Google search popularity for the
    term `GPT'. We can see a huge increase in searches for GPT when the ChatGPT
    model was launched for public use (and surprisingly little increase in
    searches when the GPT-1, GPT-2, and GPT-3 papers were released!)
    \citep{ref_radford2018gpt1,ref_radford2019gpt2,ref_brown2020gpt3,ref_openai2023gpt4}.
    This data is taken from Google Trends (\url{https://trends.google.co.uk/}).
    }
    \label{fig_gptpop}
\end{figure}

\noindent The spring of 2023\footnote{
    While we revisited this subsection for our review rebuttal.
} 
has brought with it a shift in the global zeitgeist's
attention towards foundation models in general, and the GPT family of large
language models in particular.  Leading the charge is OpenAI's ChatGPT, whose
release has become a very public advertisement of the abilities that large
language models possess (Fig.~\ref{fig_gptpop}). While impactful, we note that
ChatGPT is `just' a web interface wrapper for versions of GPT-3 and GPT-4 that
have been finetuned using human feedback
\citep{ref_knox2008,ref_ziegler2019}. ChatGPT's popularity therefore suggests
that there is a lot of latent general interest in deep learning and foundation
models, and that this interest can be realised through a convincing public
demonstration. Fully open development and dissemination of these models is
perhaps the most public demonstration there is. And we have indeed seen that
the release of open source foundation models leads to an explosion of
innovation and interest\footnote{ 
    This is a specific example of the more general rule that `bazaar' (public
    from conception) style open development outcompetes the `cathedral' model
    (closed until release, or in this case closed even after release)
    on an equal playing field \citep{ref_raymond1999}.  
}.
One particular example is the release and impact of the `large language model
[from] Meta AI' \citep[LLaMA;][]{ref_touvron2023}. The LLaMAs are a collection
of an open source LLMs, and the largest LLaMA has a comparable performance to
GPT-3.  Since LLaMA's release, an entire ecosystem of projects have spun up
that use the model in innovative and interesting ways
\citep[e.g.][]{ref_alpaca2023,ref_vicuna2023,ref_koala2023,ref_beeching2023stackllama}.
A similar story occurred in 2022 when StabilityAI released an open
text-to-image diffusion model based on latent diffusion
\citep{ref_rombach2021}. The following flurry of activity far outstripped the
progress OpenAI made with their competing closed source DALL-E 2 model
\citep{ref_ramesh2022dalle2,ref_stability2023}.  We believe that a similar
explosion of innovation to that seen with the release of the LLaMA and Stable
Diffusion models would lay in store for astronomy if an open astronomical
foundation model is developed and marketed effectively.

\begin{figure}[htbp]
    \input{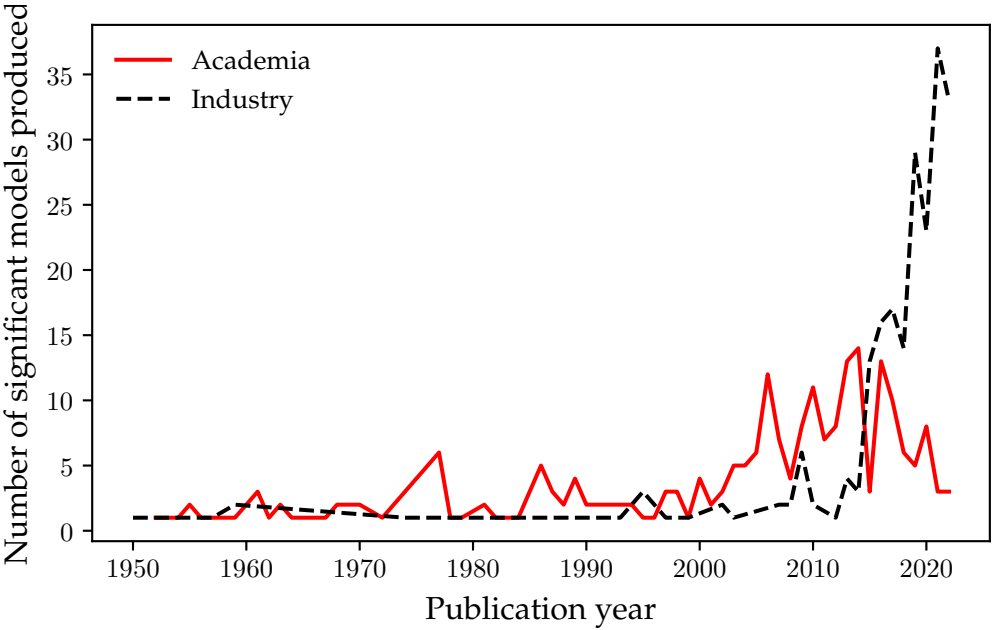}
    \caption[]
    {
    Here we show the number of highly cited, state-of-the-art, or
    historically significant works produced per year within academia and
    industry. This data is taken from \citet{ref_sevilla2021}.
    }
    \label{fig_indvsac}
\end{figure}

In mid-March GPT-4 was released \citep{ref_openai2023gpt4}. Its accompanying
`Technical Report' contains no detail on the model's architecture, training set
size, or training routine\footnote{
    Although if we extrapolate from the historical trend of LLM development,
    OpenAI's general research culture and direction, and the time GPT-4 takes
    to run inference, we could arrive at the conclusion that GPT-4 is
    essentially a scaled up `GPT-3' model that follows a Chinchilla-optimal
    scaling law (\S\ref{sec_datamoats}).
}.
The unashamed release of a closed model is quite a worrying development for a
field that has historically been built on open source and open research. Of
most concern is industrial actors within this space closing up shop as a
reaction to the open/closed model prisoner's dilemma set by OpenAI. As
Fig.~\ref{fig_indvsac} shows, industry has produced the lion's share of
impactful deep learning models since the mid-2010s; if future developments are
kept hidden due to commercial pressure we will see a concentration of talent
and innovation locked away behind industry's closed doors. 
Furthermore, the latest developments in foundation modelling have the potential
to significantly impact the global economy and workforce through pervasive
automation  \citep{ref_bommasani2021,ref_eloundou2023}.  
As automation increases, the concentration of power, expertise, and economic
clout within large industrial actors will weaken the economic bargaining
position of those that do not have access to these technologies.
This could result in a societal equilibrium where fewer and fewer people have
access to economic and social opportunity. This is an equilibrium that
\citet{ref_brynjolfsson2022} memetically dubs the `\emph{Turing Trap}':
\blockquote{
    \small\emph{A fully automated economy could, in principle, be structured to redistribute the benefits
    from production widely, even to those who are no longer strictly necessary for value creation.
    However, the beneficiaries would be in a weak bargaining position to prevent a change in the
    distribution that left them with little or nothing. They would depend precariously on the
    decisions of those in control of the technology. This opens the door to increased concentration of
    wealth and power.
}}
To avoid this trap we must collectively work towards making foundation
models---and by proxy the latest fruits of automation---available to all. 
A copyleft foundation model trained on a copyleft dataset (such as our
hypothetical astronomical foundation model) would go some way towards reducing
the growing technological inequality between Big Tech and wider society.

With the above discussion in mind, we would like to revisit our brief analysis
in \S\ref{sec_datamoats} and restate and emphasise the pressing need for an
independent, verifiable, completely open, and strong copyleft licenced
alternative to the closed foundation models controlled by OpenAI, Microsoft,
Anthropic, Google, and other Big Tech conglomerates. While expensive, the
compute is fairly easy to source---the paramount issue is that foundation
models require a huge amount of data to train them effectively. These models
are usually trained via a large amount of high quality publicly unavailable
textual data that is locked within the deep web.  Fortunately however,
\S\ref{sec_datamoats} shows that a large amount of useful multimodal data can
be easily sourced from astronomical observations. We can therefore conclude
this subsection on a positive note---astronomy is ideally poised to play an
outsized role in the democratisation of foundation models.  

\subsubsection{Possible astronomical use-cases} \label{sec_practice}

In this subsection we outline some possible exciting astronomical uses for our
astrofoundation model. Before we dive in, we must state that here we only skim
the surface of this technology's potential, and we hope that---as evidenced by
the LLaMA and Stable Diffusion ecosystems (\S\ref{sec_theory})---there will be
many more use-cases that we have not discussed here that would emerge from
community involvement. We divide this subsection into two parts. The first part
talks about how a foundation model could aid outreach, citizen science, and
cross-disciplinary collaboration, and the second part discusses how the model
could aid astronomical research.

\paragraph{Collaboration, citizen science, and outreach.}

By providing a common platform for generating simulations and analysing data, a
neural network-based astrofoundation model would ease and facilitate
collaboration between researchers in previously disparate fields. In addition
to this, any improvement in the underlying technology could easily be
integrated into field-specific (or field-agnostic) foundation models that could
be used for tasks that previously needed years of specialist training to
operate.  One example specific to astronomy is astronomical simulation. A
physically aware astrofoundation model could be used to simulate and
interrogate simulated astronomical events in much the same way that classical
simulations do now
\citep{ref_springel2018,ref_vogelsberger2020,ref_angulo2022}.
\S\ref{sec_newsim} describes in detail one framework that could facilitate such
a model.

The multi-modal training of neural networks lets us make connections between
data modes that would be impossible or difficult with current methods.  As just
one example, let us consider citizen science. In a citizen science project like
Galaxy Zoo \citep{ref_galzoo}, citizen scientists are asked to label
astronomical objects with quantitative labels. This can be an unintuitive
process for someone untrained in astronomy.  An astronomical foundation model
that has an awareness of natural language would allow participants to describe
astronomical objects using their own words. This would reduce the need for
specialised training and therefore would increase the accessibility of these
projects.  One could imagine a new Galaxy Zoo-like project where citizen
scientists provide natural language descriptions of galaxy morphologies. The
foundation model could then process and analyse these descriptions, which would
eventually contribute to a more comprehensive understanding of galaxy
evolution\footnote{Work is already being done to realise this. For example,
\citet{ref_bowles2023} propose a semantic natural language labelling scheme for
the Galaxy Zoo evolutionary map of the universe project.}.

A foundation model with astronomical knowledge could be used to develop
chatbots capable of engaging students, educators, and the general public in
conversations about astronomy. These chatbots could answer questions, provide
explanations, or even suggest personalized learning resources based on the
user's interests and prior knowledge. This would widen and democratise access
to astronomical knowledge, and this easy access to astronomical knowledge could
enthuse and help to recruit the next generation of astronomers.  Foundation
models can already act as tutors, and commercial actors are currently working
in this space; the most notable examples being `Duolingo Max' which provides
users a personalised chatbot for foreign language learning, and Khan Academy's
`Khanmigo' which provides students a personal tutor for their courses. Both
Duolingo Max and Khanmigo are paid offerings powered by OpenAI's GPT-4 API
\citep{ref_openai2023gpt4}, and so an open astronomical foundation model would
provide wider access than a closed GPT-$N$ model that has been prompted to
become astronomically-aware.

\paragraph{Augmenting research.}

While the foundation model is necessarily trained on existing data, its ability
to identify patterns and relationships within the data can lead to new
knowledge discovery, and a more efficient way to process data that previously
was difficult or time consuming. As discussed previously in
\S\S\ref{sec_gen_modelling}--\ref{sec_athirdera},  an astroconnectionist could
use the foundation model to generate embeddings for a set of astronomical
objects. Like we discussed in \S\S\ref{sec_gen_modelling}--\ref{sec_athirdera},
these embeddings could be used for downstream astronomical tasks, or could be
placed into visualisation pipelines like the t-distributed stochastic neighbour
embedding method \citep{ref_hinton2002,ref_van2008}. Since the astronomical
foundation model would be multimodal, a researcher could combine the embeddings
of multiple datasets generated from entirely different instruments, giving them
a birds-eye view of their data that would currently be difficult to achieve.
We can also use the foundation model's emergent abilities to our advantage; as
shown in Fig.~\ref{fig_flamingo} we could use few-shot learning and prompt the
trained model with a few example pairs of inputs. For instance, we could use
pairs of input galaxy observations and corresponding output surface brightness
profiles \citep{ref_smith2021}.  If the astronomical foundation model is a
few-shot learner (and is aware of a similar input output pairing within its
training data), it would identify that the researcher wants to calculate the
surface brightness profiles of new galaxies.  The researcher would then use the
prompted model as a surface brightness profile extractor, sidestepping the need
for a specialised analytical or deep learning model for such a task.  This
process is not limited to this example---it would work for any input output
pair within a mode that the foundation model is aware of. Even better, this
process would require no retraining of the foundation model, it would only
require the few-shot prompt at inference time.

Autonomous agents are no longer science fiction; task-driven autonomous agents
powered by the simulacra of a foundation model are capable of 
solving very general tasks when given only a high-level prompt by a human
operator \citep{ref_park2023,ref_nakajima2023}.  One could therefore imagine a
semi-automated research pipeline, where an autonomous agent with astronomical
knowledge is given access to a set of astronomical data through an API.  The
agent would be prompted with a high-level research goal (such as `\emph{find
something interesting and surprising within this dataset}'), and would then
take steps to achieve this task.  These steps could include querying research
papers for a literature review, searching a large multi-modal astronomical
dataset to find data that supports a theory, evoking and discussing its
findings with additional simulacra, or spinning up simulations to test a
hypothesis \citep{ref_liang2023}.  While the agent operates in the background,
the human researcher would be able to provide high-level interpretation of the
results, and would be a steady hand providing guidance and refinement of a more
general research direction.  In this way, an astronomical foundation model
would provide the tools to make all astronomers the principal investigator of
their own powerful `AI lab'.

\subsubsection{A new class of simulation} \label{sec_newsim}

We would like to end this subsection with a tangible
application of our hypothetical astrofoundation model; a conditional
generative model for astronomical simulation in the spirit of recent work on
text-to-image modelling \citep[i.e.][]{ref_rombach2021,ref_saharia2022}.
If we train an unconditional generative model, we cannot control its output at
inference time. This is an issue if we want to generate specific classes of
observations to train models for downstream tasks, such as redshift estimation,
or galaxy type classification.  To achieve a model capable of generating
specific classes, one could simply train a conditional generative model of the
form
\begin{equation}
    G_\phi(\hat{\mathbf{x}} \mid \mathbf{z}, \mathbf{y}), \label{eqn_congen}
\end{equation}
where $\hat{\mathbf{x}}$ is a generated image, $\mathbf{z}$ is some
noise that acts to capture all detail not encoded in $\mathbf{y}$, and
$\mathbf{y}$ is a conditioning vector. As an example, $\mathbf{y}$ could
contain a galaxy's redshift or morphological type.  However, this means that
we must be very specific when choosing $\mathbf{y}$.
Multimodal modelling provides us the means to sidestep this fundamental issue,
and lets us play with fuzzy inputs.

\begin{figure}[htbp]
\setlength{\tabcolsep}{6.0pt}
\captionsetup[subfigure]{labelformat=empty}
\setlength{\tabcolsep}{6.0pt}
\centering
\begin{tabular}{ccc}
    \begin{subfigure}[t]{0.30\textwidth} \centering \includegraphics[width=\textwidth]{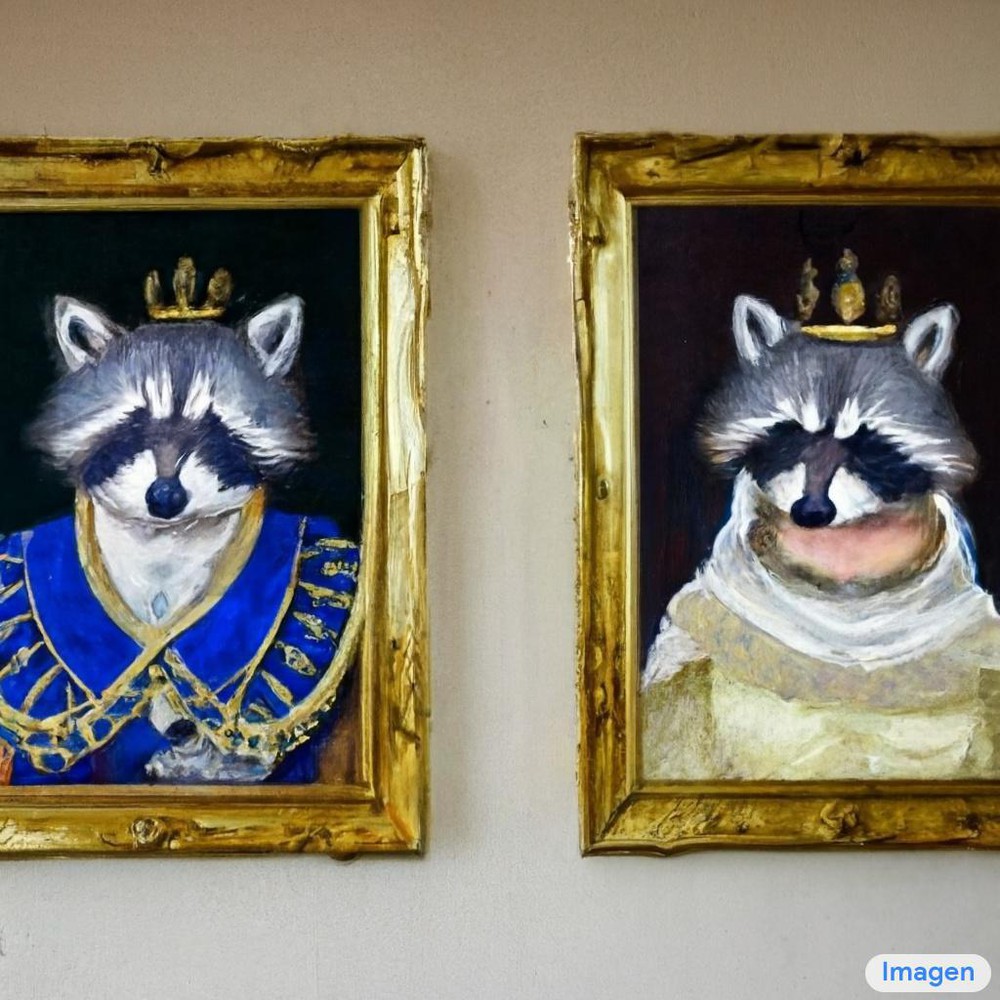} \caption{\scriptsize{A wall in a royal castle. There are two paintings on the wall. The one on the left a detailed oil painting of the royal raccoon king. The one on the right a detailed oil painting of the royal raccoon queen.}} \end{subfigure} &
    \begin{subfigure}[t]{0.30\textwidth} \centering \includegraphics[width=\textwidth]{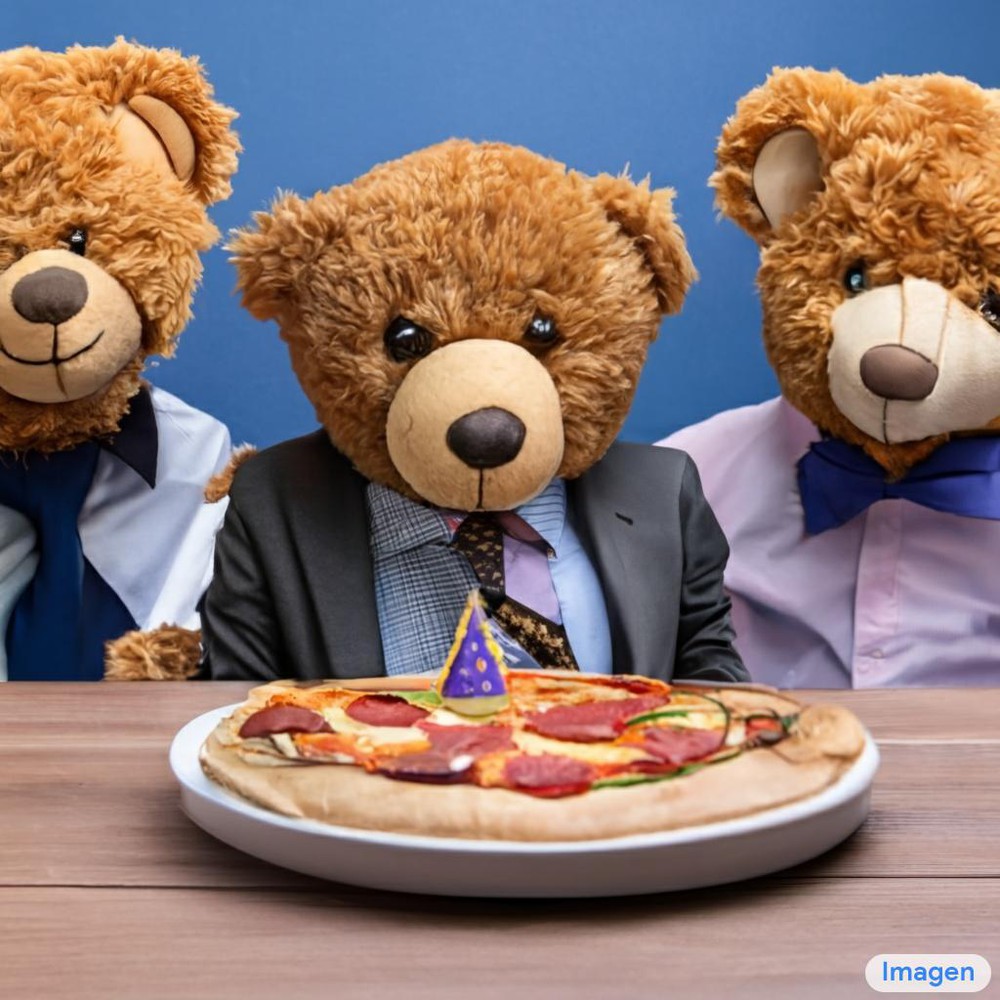} \caption{\scriptsize{A group of teddy bears in suit in a corporate office celebrating the birthday of their friend. There is a pizza cake on the desk.}} \end{subfigure} &
    \begin{subfigure}[t]{0.30\textwidth} \centering \includegraphics[width=\textwidth]{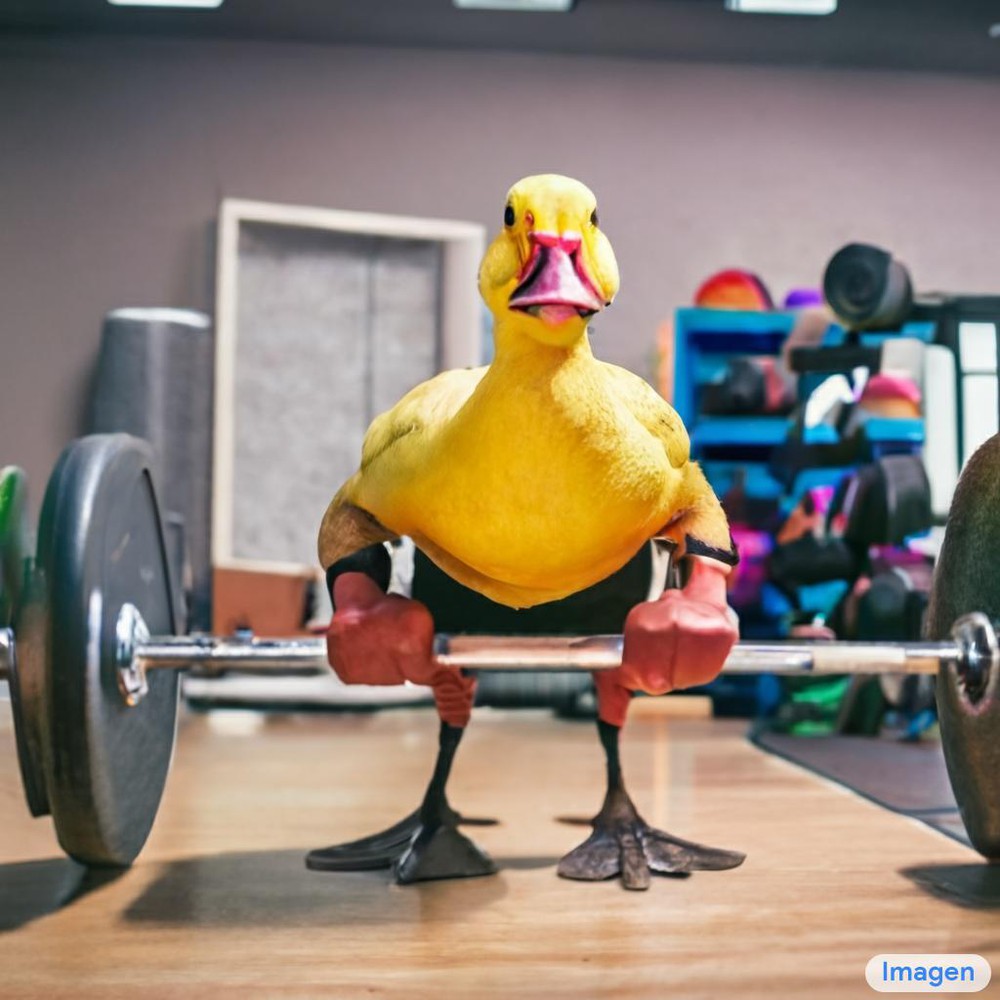} \caption{\scriptsize{An angry duck doing heavy weightlifting at the gym.}} \end{subfigure}\\ 
    \addlinespace
    \begin{subfigure}[t]{0.30\textwidth} \centering \includegraphics[width=\textwidth]{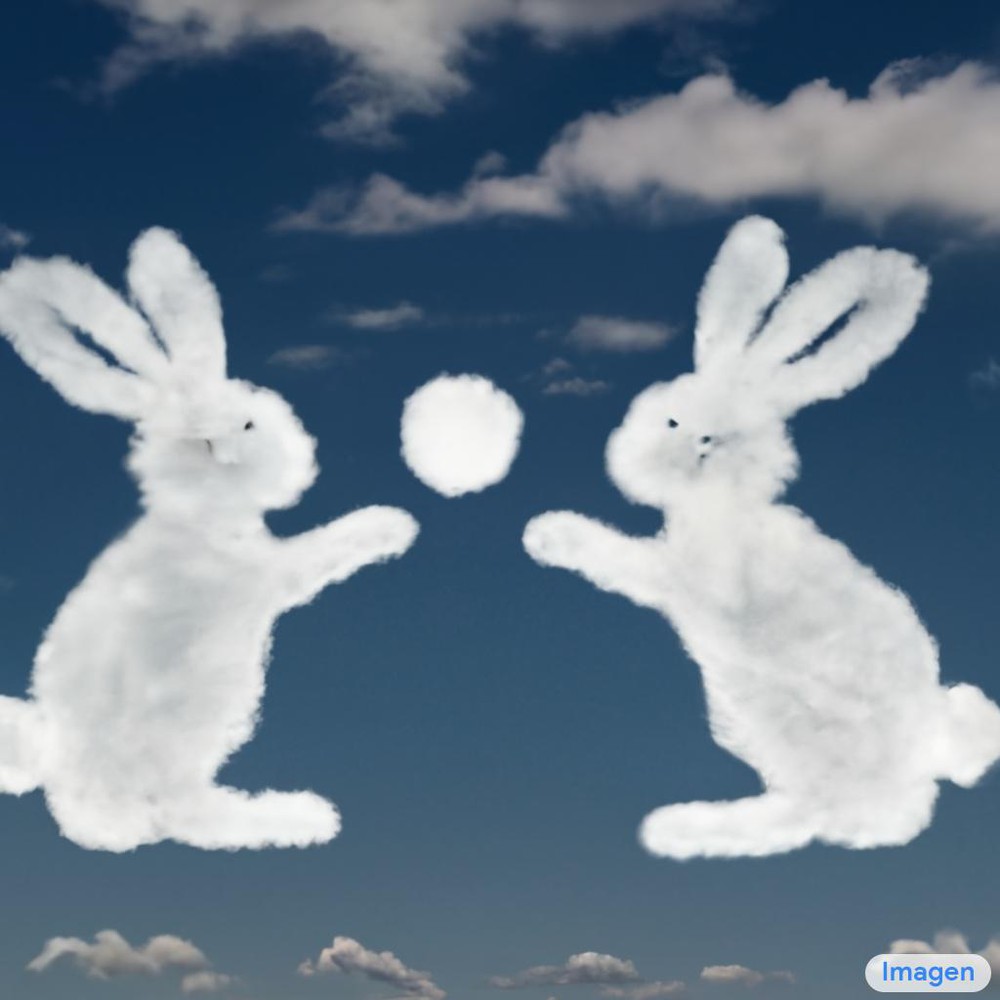} \caption{\scriptsize{A cloud in the shape of two bunnies playing with a ball. The ball is made of clouds too.}} \end{subfigure} &
    \begin{subfigure}[t]{0.30\textwidth} \centering \includegraphics[width=\textwidth]{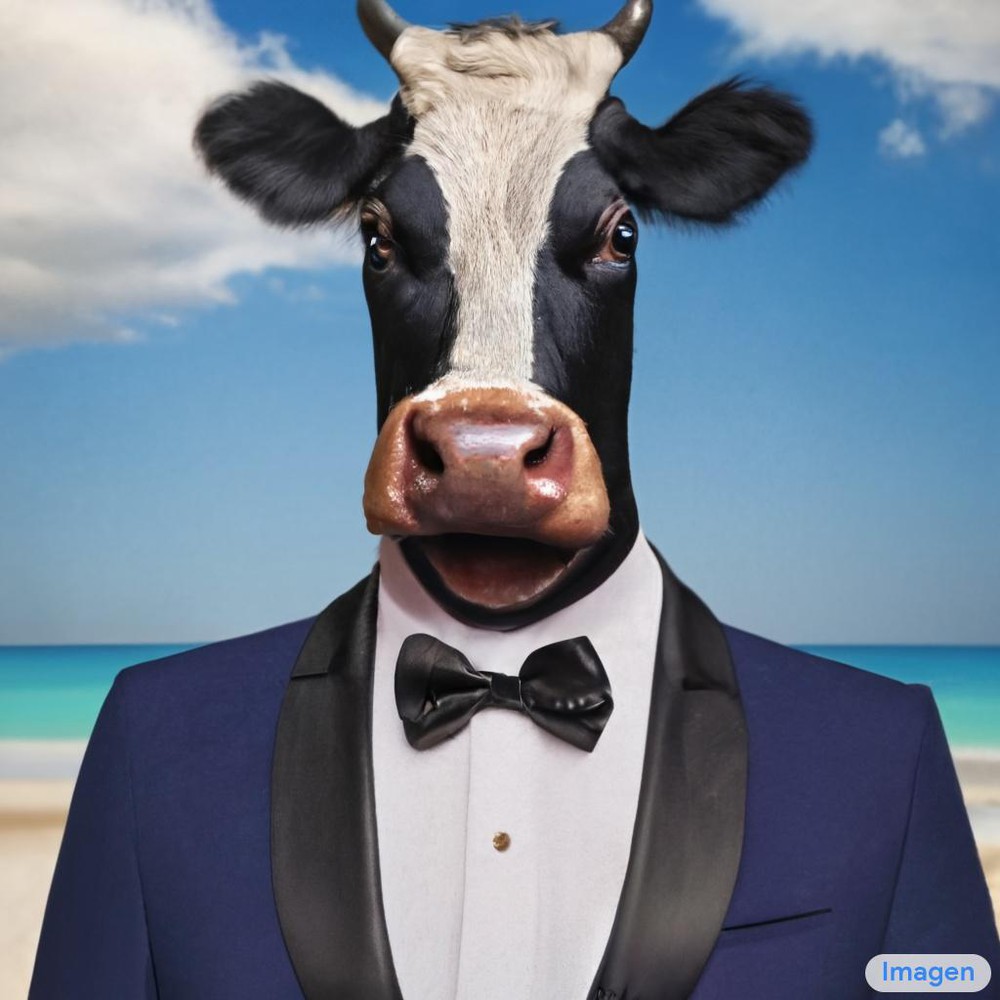} \caption{\scriptsize{A photo of a person with the head of a cow, wearing a tuxedo and black bowtie. Beach wallpaper in the background.}} \end{subfigure} &
    \begin{subfigure}[t]{0.30\textwidth} \centering \includegraphics[width=\textwidth]{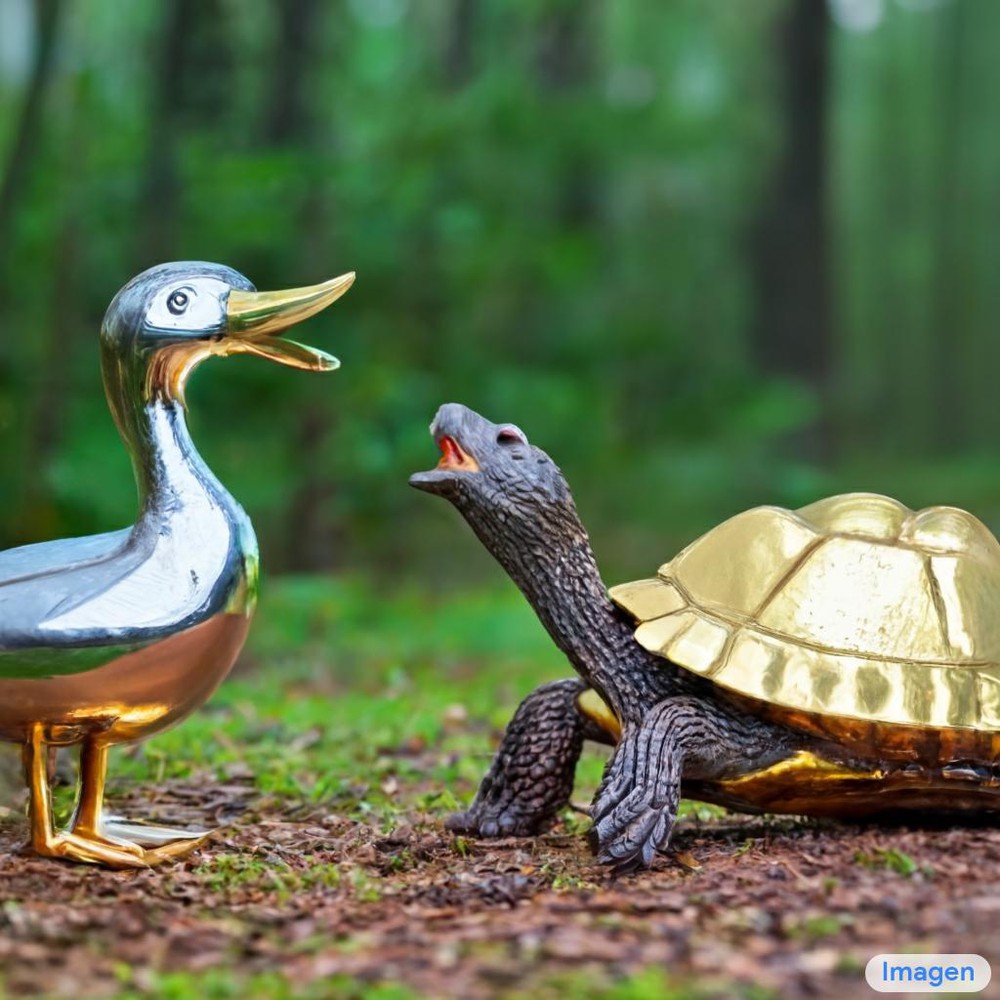} \caption{\scriptsize{A chrome-plated duck with a golden beak arguing with an angry turtle in a forest.}} \end{subfigure} \\
\end{tabular}
    \caption[Imagen, a recent text to image translator model]{Select
    $1024\times1024$ Imagen samples generated from text inputs. Below each
    image is its corresponding conditioning text. Figure adapted from
    Fig.~A.2 in \cite{ref_saharia2022}.}
    \label{fig_imagen}
\end{figure}%

\begin{figure}[htbp]
    \centering
    \includegraphics{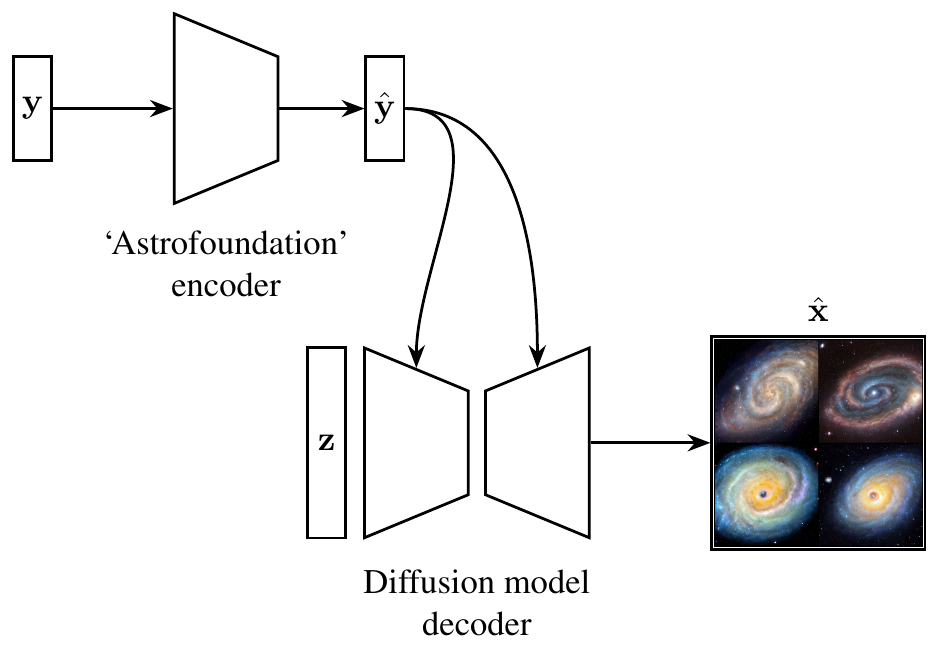}
    \caption[Simulation via an astronomical foundation model]{An Imagen-like
    model uses a frozen foundation model to encode text, and then uses that
    encoding to condition a cascaded diffusion model of the form
    $G_\phi(\hat{\mathbf{x}} \mid \mathbf{z}, \hat{\mathbf{y}})$
    \citep{ref_ho2021,ref_saharia2022}.  Here we see one possible
    realisation of this type of model in astronomy.  $\mathbf{y}$ is some kind
    of descriptive vector that can be paired with a ground truth image. For
    example, $\mathbf{y}$ could be the surface brightness profile of a galaxy,
    or the summary statistics of a variable star light curve, or some
    cosmological parameters. In general, $\mathbf{y}$ could be any vector that
    the astrofoundation model understands. $\hat{\mathbf{y}}$ is $\mathbf{y}$'s
    projected latent space equivalent. Since we do not need to train the
    foundation model here, training cost is far lower than for an equivalent
    end-to-end trained model.}
    \label{fig_astroimagen}
\end{figure}

As a thought experiment let us consider Google's recent `Imagen' model\footnote{
    Naturally, no implementation is provided by Google. However, there is a
    fantastic MIT licenced implementation of Imagen provided by Phil Wang and
    others (\url{https://github.com/lucidrains/imagen-pytorch}), 
    and StabilityAI has a similar trained open source model released under the
    name `Stable Diffusion'
    (\url{https://github.com/Stability-AI/stablediffusion}).
}, 
and imagine how it could be repurposed for an astronomical use case
\citep[Figs.~\ref{fig_imagen} and
\ref{fig_astroimagen};][]{ref_saharia2022}. Imagen is a combination of a frozen LLM
\citep[specifically T5-XXL;][]{ref_tay2021} and a cascaded diffusion model
\citep[][also see \S\ref{sec_sbgm}]{ref_ho2021}. The LLM acts as a language
encoder, and then passes its generated latent space representations onto the
diffusion model as a conditioning vector.  If we were to replace the frozen LLM
with an `astrofoundation' model (see \S\ref{sec_foundationmodels} and
\S\ref{sec_datamoats}), we could leverage astronomy's fundamentally multimodal
nature. For example, if our astrofoundation model were trained to understand
the Galaxy Zoo 2 (GZ2) morphological classifications
\citep{ref_willett2013}, we could take the GZ2 descriptors as $\mathbf{y}$
and their corresponding galaxy pair as $\mathbf{x}$ and train on those.

Once trained, our astronomical Imagen model could generate synthetic galaxies
that resemble the real galaxy observations that it was trained on. However,
unlike an unconditional astronomical simulator, this model would be capable of
generating galaxies that specifically resemble a real galaxy that shares the
conditioning set of GZ2 parameters!

Unlike the conditional model described by Eq.~\ref{eqn_congen}, an
astrofoundation type model allows us to be creative with the conditioning
vector. For example, we could run the model in reverse to generate
representations that refer to a very specific astronomical object, and then
generate many more objects of that `class' with injected features like
satellite occlusion, a specific instrument response function, a specific
redshift, etc.  \citep[see work on `textual inversion' by][]{ref_gal2022}.
These simulations would enable researchers to create tailored datasets for
various research purposes, such as studying particular galaxy types,
morphologies, or cosmological phenomena.  We could even create a `Galaxy Zoo'
type dataset that asks citizen scientists to describe galaxy morphology via
natural language (\S\ref{sec_practice}).  This is possible since the encoding
foundation model does not fundamentally care about which form the caption
takes. This approach would cut down on citizen scientist training cost due to
natural language's inherent intuitiveness.  Furthermore, as inference-time
generation is relatively cheap, a model like the one described in this section
would allow astronomers to explore and test hypotheses and scenarios more
rapidly than they could if they used a classical simulation.  

\section{Connectionism's caveats}

Thus far in this review we have been very optimistic about astronomical
connectionism's potential. However, this does not mean that connectionism is
without its pitfalls. \S\ref{sec_practical_pitfalls} outlines some practical
downsides of astronomical connectionism, and discusses how a practitioner can mitigate
them. Due to its importance, we dedicate \S\ref{sec_carbon} to the discussion
of climate change and carbon emissions, and illustrate connectionism's impact
with a case study on the carbon emissions of modern large language and
foundation models.

\subsection{Possible practical pitfalls} \label{sec_practical_pitfalls}

As illustrated in Fig.~\ref{fig_scalinglaw}, deep learning has an insatiable
hunger for data. Acquiring and labelling data for the training of deep learning
models can be extraordinarily expensive and time-consuming. The savvy
astroconnectionist could mitigate this problem through self-supervised or
generative learning that does not require labelled data, and then repurposing
learnt embeddings for more specialised
downstream tasks\footnote{
    This process is also known as `transfer learning'.
}
(see \S\S\ref{sec_gen_modelling}--\ref{ch_conclusions}).
Related to this, rare or entirely unexpected astronomical events and
phenomena\footnote{
    Such as Green Bean Galaxies \citep{ref_schirmer2013}, or SETI events akin
    to the `\emph{Wow!}' signal \citep{ref_kraus1994}.
}
are by definition poorly sampled within any training data, and so a deep
learning model will have difficulty generalising and internalising these
events. One solution is using an anomaly detection method to find these
rare phenomena. We direct the reader to \citet{ref_pang2021} for an excellent
recent review of anomaly detection techniques.

Very large deep learning models can be expensive to train and run inference
with. Some astronomical applications, such as detecting transient events,
require real-time processing of large volumes of data.  The computational
complexity of deep learning models can pose challenges for their deployment in
these time-sensitive scenarios. In that case it may be preferable to employ a
fast, simple, classical technique, or to use a smaller deep learning model.

Astronomical data can be observed via a variety of different instruments (or
simulations), and the final output data can be processed by any number of
post-processing pipelines. These pipelines each have their own characteristics,
idiosyncrasies, and foibles, and so can appear very different when propagated
through a deep neural network.  Also, the distribution of known celestial
objects within a survey may be influenced by observational biases or historical
interests, and so careful inspection of datasets is required to ensure that
they are representative for the desired use-case. In addition to care, an
astroconnectionist might employ domain adaptation techniques to ensure that
their datasets are representative for their downstream tasks
\citep{ref_wang2018}. Finally, as we explored in \S\ref{ch_conclusions}, it may
even be enough to simply train a very large deep learning model on a collection
of datasets \citep{ref_aghajanyan2023}, but this approach is currently out of
reach for the average researcher.

The perennial criticism of deep learning is---of course---interpretability.  As
deep learning models are highly parametrised it is difficult to understand why
they arrive at a certain behaviour or decision.  There are many ways to
sidestep this issue, and this paragraph will briefly outline some developments
in this direction that might be of use to a practitioner.  Perhaps the gold
standard for interpretability is a neural network walking the user through its
`thought' process step-by-step with natural language, as a human would do.
Large language foundation models can do this, and this ability comes `for free'
with a sufficiently large model and dataset \citep{ref_wei2022}. Unfortunately
however, no such foundation model currently exists that also has a deep
knowledge of astronomy (\S\ref{ch_conclusions}) so we must be a little more
creative. Attentional mapping can be used to show which features the deep
learning model are attending to when producing an output, and this attentional
mapping can be depicted as a heat-map over our data. Attentional mapping can be
generated in several ways; for example, we could use a mechanism like we
discussed in \S\ref{sec_transformers} to highlight the most useful parts of an
input datum for the model to predict or generate its output. One can also use
class activation mapping \citep{ref_zhou2016cvpr} to trace the outputs of a
fully convolutional neural network back to its inputs to see which parts of an
input image are used in a prediction. Occlusion mapping (and other perturbation
techniques) can be used to visualise attention for all architectures. Occlusion maps require us to 
occlude parts of an input datum and in turn allow us observe how that affects
the output prediction \citep{ref_zeiler2014}.  We can also apply certain
statistical methods to deep learning models to gain an insight into their
inner workings. Stochastic neural networks trained within the Bayesian paradigm
(or `Bayesian neural networks') can be used to estimate the uncertainty in
neural network predictions \citep{ref_wang2020}. One does not need to have
prior knowledge of the dataset when training a Bayesian neural network; neural
networks can make use of approximate Bayesian computation techniques like
likelihood-free inference to estimate the posterior \citep{ref_tavare1997}.  
Besides these methods, many other deep interpretability pipelines are in
use---far more than we have space to go over here---and so we highly recommend
\citet{ref_ras2020} for a general and extensive overview of the field of
explainable deep learning.

\subsection{Connectionism's carbon crisis} \label{sec_carbon}

The training of deep learning models in general requires a considerable amount
of energy, and it is only natural that the training of ultra-large foundation
models significantly ups the ante. In this section we illustrate
connectionism's hunger for energy by estimating the total carbon footprint
created in the training of the GPT-3\footnote{
    We would compare GPT-4, but OpenAI has neglected to disclose
    any information regarding the training routine of the network in their
    `Technical Report' \citep{ref_openai2023gpt4}.
} and PaLM foundation models
\citep{ref_brown2020gpt3,ref_chowdhery2022palm}. 

Let us start with the eminent GPT-3 model. Unfortunately, the total energy cost
is not stated in \citet{ref_brown2020gpt3} but we can make a ballpark estimate
using information from that work.  GPT-3 was trained on a high performance
computing cluster containing $N=10\,000$ NVIDIA V100 chips, and required a
total $\Sigma=3.14\times10^{23}$~FLOPs to train to completion
\citep{ref_brown2020gpt3}. A single V100 has a throughput of
$C=2.8\times10^{13}$~FLOPS for half precision floats and so we can estimate
GPT-3's total training time in datacentre-seconds as
\begin{equation*}
    \frac{\Sigma}{C \cdot N} = \frac{3.14\times10^{23}}{2.8\times10^{12} \cdot 10^4} = 1.12\times10^6\,\text{s},
\end{equation*}
which is approximately 311 hours. We know the thermal design power of a single
V100 chip is 300\,W and so we can safely assume a lower bound on the datacentre
power usage as 3000\,kW. Therefore, we estimate the total power consumed while
training GPT-3 as
\begin{equation*}
    3000 \cdot 311 =  933\,000\,\text{kWh}.
\end{equation*}
The emissions per kWh of the datacentre where GPT-3 was trained is
0.429~kg~CO$_2$e~kWh$^{-1}$ \citep{ref_patterson2021}, leaving us with a
total emission of around 400\,000\,kg\,CO$_2$e\footnote{
    We must keep in mind that this estimate is a lower limit. We do not
    include CPU power, cooling, or any other overheads in our calculation,
    never mind the cost to do a full hyperparameter sweep!
}.

\begin{figure}[htbp]
    \centering
    \input{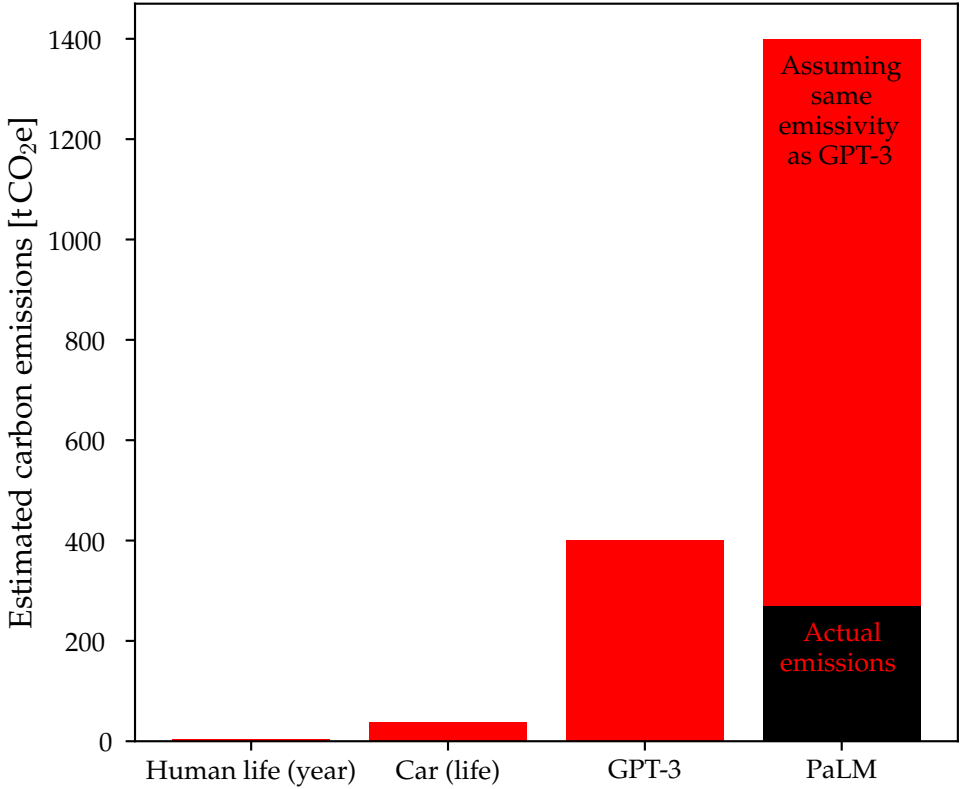}
    \caption{Here we contextualise the huge carbon footprints generated when
    training foundation models. The average person's yearly carbon footprint is
    estimated as 4750\,kg\,CO$_2$e using data from
    \citet{ref_friedlingstein2022}, and the car lifetime emissions is
    38\,504\,kg\,CO$_2$e assuming a Mercedes-Benz C 300 d model
    \citep{ref_buberger2022}.}
    \label{fig_carbon}
\end{figure}

However, GPT-3 is already years old; so we will also estimate the energy used
when training Google's state-of-the-art `PaLM' foundation model.
\citet{ref_chowdhery2022palm} state:
    `\emph{We trained PaLM-540B on 6144 TPU v4 chips for 1200 hours and 3072 TPU v4
    chips for 336 hours including some downtime and repeated steps\ldots [We found a]
    378.5\,W measured system power per TPU v4 chip\ldots}'
We can therefore calculate PaLM's total energy usage as 
\begin{equation*}
    378.5 \cdot (6144 \cdot 1200 + 3072 \cdot 336) \approx 3\,180\,000\,\text{kWh}.
\end{equation*}
If PaLM was trained on the same datacentre as GPT-3 (i.e.\ at an emissivity of
0.429~kg~CO$_2$e~kWh$^{-1}$) it would have emitted a staggering
1\,400\,000\,kg\,CO$_2$e---quadruple the average person's lifetime carbon
footprint \citep{ref_friedlingstein2022} and approaching the annual emission of
some small countries. Luckily, the datacentre that PaLM was trained on was far
greener than that used by OpenAI, and PaLM actually produced
$\sim$270\,000\,kg\,CO$_2$e \citep{ref_chowdhery2022palm}, although this is
still rather large.  We contextualise our calculated footprints visually in
Fig.~\ref{fig_carbon}.

PaLM's contribution to Fig.~\ref{fig_carbon} demonstrates the importance of
choosing and using datacentres that run on clean energy sources when training
deep learning models, and make efficient use of heat output (e.g.\ through
recovery systems). Besides this, researchers can also take care when optimising
their neural network models to reduce their carbon footprint. For instance by
choosing hyperparameters through a more efficient manual or randomised search,
instead of via a brute force method \citep{ref_bergstra2012}.  As stated in
\citet{ref_energy_ml_model} researchers can also combat redundant retraining of
models (and thus unnecessary energy usage) by ensuring that fully trained
models, data, and code are released under an open licence. The publishing of a
fully trained model's energy usage, computation requirements, and carbon
footprint also allows downstream researchers to determine whether replication
of a work is economically and environmentally viable. Calculating one's energy
usage in the spirit of openness does not have to be difficult: we have been
using the excellent and user-friendly `Machine Learning CO$_2$ Impact
Calculator' in our own work to calculate and publish the carbon footprint of
our models \citep{ref_energy_calculator}. A recommendation of this review is
that an environmental impact statement should become standard practice in
journal articles, conference presentations and proceedings when deep learning
models (or any HPC-heavy research for that matter) is used. 

\section{Final comments, or how we learnt to stop worrying and
love astronomy's Big Data Era} \label{sec_finalcomments}

\begin{figure}[thbp]
    \centering
    \input{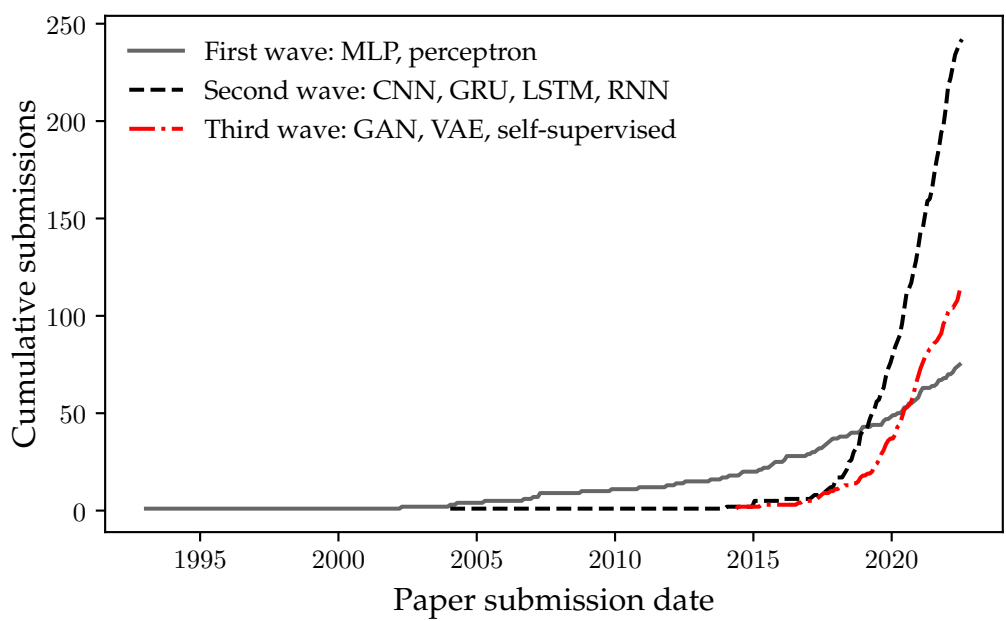}
    \caption[Three distinct waves within astronomical connectionism]{Here we
    see the number of arXiv:astro-ph submissions whose titles or abstracts
    match the terms given in the legend. We can see three distinct `waves'. The
    first corresponds to studies that use MLPs
    (\S\ref{sec_AN}-\S\ref{sec_mlpinastro}), the second corresponds to studies
    that use `deep learning' methods that injest raw data
    (\S\ref{sec_CNN}-\S\ref{sec_cnnrnn_apps}) and the third corresponds to
    studies that use generative or self-supervised models
    (\S\ref{sec_gen_modelling}-\S\ref{sec_athirdera}). The raw data is in the
    public domain, and is available at
    \url{https://www.kaggle.com/Cornell-University/arxiv}.}
    \label{fig_waves}
\end{figure}

\noindent To repeat our introductory statement: in every field that deep learning has infiltrated we have seen a
reduction in the use of specialist knowledge, to be replaced with knowledge
automatically derived from data. We have already seen this process play out in many disparate fields from
computer Go \citep{ref_silver2016}, to protein folding \citep{ref_jumper2021},
to natural language processing \citep{ref_brown2020gpt3}, to computer vision
\citep{ref_dosovitskiy2020vit}. This process is already well known within the
deep learning community as `\emph{The Bitter Lesson,}' a precept that is
summarised by the quote:
\blockquote{\small
    \emph{The biggest lesson that can be read from 70 years of AI research is that
    general methods that leverage computation are ultimately the most effective,
    and by a large margin.} \citep{ref_sutton2019}\\\vspace{-1em}
}
There is no reason to believe that
astronomy is fundamentally different. Indeed, within this review we have seen a
narrative pointing to this conclusion (Fig.~\ref{fig_waves}). Initial work on
MLPs within astronomy required manually selected emergent properties as input
\citep[e.g.][]{ref_angel1990,ref_odewahn1992}. With the advent of CNNs and
RNNs, these manually selected inputs gave way to raw data ingestion
\citep[e.g.][]{ref_dieleman2015,ref_charnock2017}.  Now we are seeing the
removal of human supervision altogether with deep learning methods inferring 
labels and knowledge directly from the data \citep[e.g.][]{ref_spindler2020,ref_morvan2022}.
Ultimately, if astronomy follows in the footsteps of other applied deep learning fields,
we will see the removal of expertly crafted deep learning models, to be replaced
with finetuned versions of an all-encompassing `foundation' model
\citep{ref_bommasani2021}.  This process is by no means a bad thing; the
removal of human bias in the astronomical discovery process allows us to find
`unknown unknowns' through serendipity
\citep{ref_sarmiento2021,ref_donosoolivia2022}. Likewise,
the ability to leverage data allows us to directly generate and interrogate
realistic yet synthetic observations, sidestepping the need for an expensive
and fragile classical simulation \citep{ref_sagan,ref_smith2022}.

Astronomy's relative data wealth gives us the opportunity to form a symbiotic
relationship with the cutting edge of deep learning research, an increasingly
data hungry field \citep{ref_gwern2022,ref_friel2022}. Many ultra-large
datasets in machine learning are proprietary, and so the astronomical community
has the opportunity to step in and provide a high quality multimodal public
dataset. In turn, this dataset could be used to train an astronomical
`foundation' model that can be used for state-of-the-art downstream tasks (such
as astronomical simulation, see \S\ref{sec_newsim}). Finally, following recent
developments in connectionism \citep{ref_brown2020gpt3,ref_hoffman2022} most
astronomers lack the resources to train models on the cutting edge of the
field. If astronomy is to have any chance of keeping up with the Big Tech
goliaths, we must follow the examples of EleutherAI and HuggingFace and pool
our resources in a grassroots-style open source fashion
(\S\ref{ch_conclusions}). We leave this as a challenge for the community.

\vskip1pc

\ethics{
    This article is a review and so no ethical approval is required.
}

\dataccess{
    This article has no additional data.
}

\competing{
    There are no competing interests.
}

\funding{MJS acknowledges support from the Alan Turing Institute by way of the Turing Enrichment Scheme.
    JEG acknowledges funding from the Royal Society and the Science and Technology Facilities Council.
}

\ack{
    The authors would like to thank Hans-Martin Adorf for providing access to
    his manuscripts. The galaxy icon shown in Fig.~\ref{fig_generativemodelling}
    is by Agata Kuczmi\'nska and is reproduced here under the CC-BY-4.0 licence.
    We would like to thank Connor Stone, Micah Bowels, and the anonymous
    reviewers for their helpful comments and suggestions on the first draft of
    this manuscript.
}

\disclaimer{
    This study is an adaptation of work presented in Chapters 1 and 5 of MJS's
    PhD thesis \citep{ref_smith2022thesis}. GPT-4 wrote the abstract. 
}

\pagebreak

\newrefcontext[sorting=nyt]
\printbibliography

\end{document}